\documentclass[onecolumn,sort&compress,numbers]{els-mrw} % For numbered
\usepackage{amsmath,amssymb,amsfonts,amsthm,makeidx,graphicx}
\usepackage{txfonts}
\usepackage{helvet}
\usepackage{hyperref}

\newcommand{\epm}[2]{
 \raisebox{-0.5ex}{\shortstack[l]{$\scriptstyle+#1$\\$\scriptstyle-#2$}}}

\newcommand{\bea}{\begin{eqnarray}}
\newcommand{\eea}{\end{eqnarray}}
\newcommand{\beq}{\begin{equation}}
\newcommand{\eeq}{\end{equation}}

\newcommand{\mm}{meson-antimeson}
\newcommand{\mmm}{\mm\ mixing}
\newcommand{\mmc}{Meson-antimeson}
\newcommand{\mmmc}{\mmc\ mixing}
\newcommand{\bb}{\ensuremath{B\!-\!\Bbar\,}}
\newcommand{\bbm}{\bb\ mixing}
\newcommand{\bbs}{\ensuremath{B_s\!-\!\Bbar{}_s\,}}
\newcommand{\bbms}{\bbs\ mixing}
\newcommand{\bbd}{\ensuremath{B_d\!-\!\Bbar{}_d\,}}
\newcommand{\bbmd}{\bbd\ mixing}
\newcommand{\bbq}{\ensuremath{B_q\!-\!\Bbar{}_q\,}}
\newcommand{\bbmq}{\bbq\ mixing}
\newcommand{\dd}{\ensuremath{D\!-\!\Dbar\,}}
\newcommand{\ddm}{\dd\ mixing}
\newcommand{\kk}{\ensuremath{K\!-\!\Kbar\,}}
\newcommand{\kkm}{\kk\ mixing}
\newcommand{\dg}{\ensuremath{\Delta \Gamma}}
\newcommand{\dm}{\ensuremath{\Delta M}}
\newcommand{\BdorBdbar}{\raisebox{5.7pt}{$\scriptscriptstyle(\hspace*{4.6pt})$}
  \hspace*{-5.5pt}\!\Bbar_{d}}
\newcommand{\BsorBsbar}{\raisebox{5.7pt}{$\scriptscriptstyle(\hspace*{4.6pt})$}
  \hspace*{-5.5pt}\!\Bbar_{s}}
\newcommand{\DorDbar}{\raisebox{5.7pt}{$\scriptscriptstyle(\hspace*{4.6pt})$}
  \hspace*{-5.5pt}\!\Dbar}
\newcommand{\KorKbar}{\raisebox{5.7pt}{$\scriptscriptstyle(\hspace*{4.6pt})$}
  \hspace*{-5.5pt}\!\Kbar}

\newcommand{\eq}[1]{Eq.~(\ref{#1})}
\newcommand{\eqsand}[2]{Eqs.~(\ref{#1}) and (\ref{#2})}
\newcommand{\eqsto}[2]{Eqs.~(\ref{#1}--\ref{#2})}
\newcommand{\sgn}{\mbox{sign}\,}
\newcommand{\Dbar}{\bar{D}}
\newcommand{\Bbar}{\bar{B}}
\newcommand{\Kbar}{\bar{K}}
\newcommand{\Mbar}{\bar{M}}
\newcommand{\no}{\nonumber}
\newcommand{\nn}{\nonumber \\}
\newcommand{\ov}[1]{\overline{#1}}
\newcommand{\lt}{\left}
\newcommand{\rt}{\right}

\newcommand{\mev}{\,\mbox{MeV}}
\newcommand{\gev}{\,\mbox{GeV}}
\newcommand{\tev}{\,\mbox{TeV}}
\newcommand{\fig}[1]{Fig.~\ref{#1}}
\newcommand{\ds}{\displaystyle}
\newcommand{\imag}{\mathrm{Im}\,}
\newcommand{\real}{\mathrm{Re}\,}

\newcommand{\lqcd}{\Lambda_{\rm QCD}}
\newcommand{\bra}[1]{\ensuremath{\langle #1 |}}
\newcommand{\ket}[1]{\ensuremath{| #1 \rangle }}

\newcommand{\e}{\epsilon}

\newcommand{\gtf}{\ensuremath{\Gamma (M(t) \rightarrow f )}}
\newcommand{\gbtf}{\ensuremath{\Gamma (\Mbar{}(t) \rightarrow f )}}
\newcommand{\gtfb}{\ensuremath{\Gamma (M(t) \rightarrow \bar{f} )}}
\newcommand{\gbtfb}{\ensuremath{\Gamma (\Mbar{}(t) \rightarrow \bar{f} )}}

\newcommand{\gtfcp}{\ensuremath{\Gamma (M(t) \rightarrow f_{\rm CP} )}}
\newcommand{\gbtfcp}{\ensuremath{\Gamma (\Mbar{}(t) \rightarrow f_{\rm CP} )}}

\newcommand{\gtfs}{\ensuremath{\Gamma (M(t) \rightarrow f_{\rm fs} )}}
\newcommand{\gbtfs}{\ensuremath{\Gamma (\Mbar{}(t) \rightarrow f_{\rm fs} )}}
\newcommand{\gtfbs}{\ensuremath{\Gamma (M(t) \rightarrow \bar{f}_{\rm fs} )}}
\newcommand{\gbtfbs}{\ensuremath{\Gamma (\Mbar{}(t) \rightarrow 
                      \bar{f}_{\rm fs} )}}

\newcommand{\guntf}{\ensuremath{\Gamma  [f,t] }}
\newcommand{\guntfs}{\ensuremath{\Gamma  [f_{\rm fs},t] }}

\newcommand{\guntfbs}{\ensuremath{\Gamma  [\bar f_{\rm fs},t] }}

\newcommand{\sv}[1]{\begin{pmatrix} #1 \end{pmatrix}}

\begin{document}

%%%%%%%%%%%%%%%%%%%%%%%%%%%%%%%%%%%%%%%%%%%%%%%%%%%%%%%%%%%%%%%%
%% the following items are mandatory: 
%% - title
%% - author names
%% - affiliation details
%% - abstract
%% - keywords

% \chapter{Article title (template for all chapters in Sections 1-5:
%   General Concepts, Hadron Physics, EW Physics, Neutrino Physics,
%   BSM)}\label{chap1}
\chapter{Meson-antimeson mixing}\label{chap1}

%% All author names and affiliations, and email address for corresponding author
\author[1]{Ulrich Nierste}%

\address[1]{\orgname{Karlsruhe Institute of Technology (KIT)},
    \orgdiv{Institute for Theoretical Particle Physics (TTP)},
    \orgaddress{Wolfgang-Gaede-Stra\ss e 1, 76131 Karlsruhe, Germany}}

% preprint no.:  
%\articletag{Chapter Article tagline: update of previous edition, reprint.}

\maketitle

%%%%%%%%%%%%%%%%%%%%%%%%%%%%%%%%%%%%%%%%%%%%%%%%%%%%%%%%%%%%%%%%
%% the following item is mandatory: 
%% 100-150 word summary of the chapter
\begin{abstract}[Abstract]
  Meson-antimeson transitions are flavor-changing neutral current
  processes in which the strangeness, charm, or beauty quantum number
  changes by two units. In the Standard Model (SM) these transitions
  originate from box diagrams with two $W$ bosons. They permit the
  preparation of time-dependent, oscillating quantum states which are
  superpositions of a meson and its antimeson. By studying their decays
  one gains information on both the \mmm\ amplitude itself and the decay
  amplitude involved, in particular one can measure complex phases
  quantifying the violation of charge-parity ($CP$) violation.  I
  present a comprehensive overview on the topic, starting with
  phenomenological presentations of \kk, \bbd, \bbs, and \ddm\ and their
  impact on the formulation of the SM.  Highlights are the discovery of
  the violation of $CP$ and other discrete symmetries, the prediction of
  the charm quark and its mass, the prediction of a heavy top quark, and
  the confirmation of the Kobayashi-Maskawa mechanism of $CP$ violation.
  Further sections cover the theoretical formalism needed to describe
  \mmm\ and to calculate observables in terms of the fundamental
  parameters of the SM or hypothetical theories of new physics. I
  discuss the unitarity triangle of the Cabibbo-Kobayashi-Maskawa
  matrix, which is used to visualize how various $CP$-violating and
  $CP$-conserving quantities combine to probe the SM. I describe the
  emergence of precision flavor physics and the role of reliable theory
  calculations to link \kkm\ to \bbmd, which was essential to confirm the
  Kobayashi-Maskawa mechanism, and present the current status of theory
  predictions. Today, the focus of the field is on physics beyond the
  SM, because \mmm\ amplitudes are sensitive to virtual effects of heavy
  particles with masses which are several orders of magnitude above the
  reach of current particle colliders.
\end{abstract}

%% 5-10 words that embody the key topics in the chapter. What terms would someone put into a search engine if they were looking for a chapter like this?
\begin{keywords}
 	%please enter 5 keywords as follows:
 	weak interaction\sep 
        flavor physics\sep
        meson-antimeson mixing\sep
        CP violation\sep
        BSM physics
\end{keywords}

%%%%%%%%%%%%%%%%%%%%%%%%%%%%%%%%%%%%%%%%%%%%%%%%%%%%%%%%%%%%%%%%
%% the following item is optonal: 
%% - Single figure visually illustrating the key topic/method/outcome described in the chapter
% \begin{figure}[h]
% 	\centering
% 	\includegraphics[width=7cm,height=4cm]{blankfig}
% 	\caption{Optional: Single figure visually illustrating the key topic/method/outcome described in the chapter. 
% 		     Please add here some text explaining the pic...}
% 	\label{fig:titlepage}
% \end{figure}
%
%%%%%%%%%%%%%%%%%%%%%%%%%%%%%%%%%%%%%%%%%%%%%%%%%%%%%%%%%%%%%%%%
%% the following item is optional: 
%% - System of abbreviations/terms/symbols used in the specific field of study/community. List and define
\begin{glossary}[Nomenclature]
  \begin{tabular}{@{}lp{34pc}@{}}
    $\alpha$, $\beta$, $\gamma$ & angles of the unitarity triangle\\
    $a_{\rm fs}$ & $CP$ asymmetry in flavor-specific decays\\ 
    $A_{CP}^{\rm dir} (M\to f)$ & direct $CP$ asymmetry in $M\to f$ \\
    $A_{CP}^{\rm mix} (M\to f)$ & mixing-induced $CP$ asymmetry in $M\to f$ \\
    $B$ & neutral $b$-flavored meson $B_d$ or $B_s$, \\
          & beauty (a.k.a.\ bottom)
                                                 quantum number \\
                BSM & beyond Standard Model \\
                C & charm quantum number, \\ & charge conjugation\\ 
    		CKM & Cabibbo-Kobayashi-Maskawa \\
    $CP$ & charge-parity conjugation \\
    $D$ & neutral $D$ meson\\
    $\delta_{KM}$ & Kobayashi-Maskawa phase\\
    \dg\ & width difference between the two mass eigenstates \\
    \dm\ & mass difference between the two mass eigenstates \\
    $F$ & flavor quantum number, $F=B,C,S,U,D$ \\
    $G_F$ & Fermi constant \\
    $\Gamma_{12}$ & off-diagonal matrix element of the \mm\ decay matrix\\
    $g_w$ & weak coupling constant \\
    HQE & Heavy Quark Expansion \\
    K & neutral Kaon \\
    $\lambda$, $A$, $\rho$, $\eta$ & Wolfenstein parameters \\
    M& any of $K$, $D$, $B_{d}$, or $B_{s}$\\
    $M_{12}$ & off-diagonal matrix element of the \mm\ mass matrix\\
    OPE &Operator Product Expansion \\
    P & parity  \\
    QCD & quantum chromodynamics \\
    $(\bar\rho,\bar\eta)$ & apex of the standard unitarity triangle \\
    S & strangeness quantum number \\
    SM & Standard Model \\
    T & time reversal  \\
    $\tau$& lifetime\\
    UT & unitarity triangle\\ 
    $V$ & CKM matrix \\
    $V_C$ & Cabibbo matrix \\
    QFT        &  quantum field theory 
\end{tabular}
\end{glossary}

%%%%%%%%%%%%%%%%%%%%%%%%%%%%%%%%%%%%%%%%%%%%%%%%%%%%%%%%%%%%%%%%
%% the following item is mandatory: 
%% List of the key points and topics a reader can expect to learn from this chapter 
\section*{Objectives}
\begin{enumerate}
\item The text intends to be a comprehensive introduction into \kk,
  \bbd, \bbs, and \ddm\ for
    people studying any of these topics in experiment or theory. It
    shall convey the special knowledge needed to 
    interpret an experimental analysis or to understand the
    concepts of a theoretical calculation. 
\item Furthermore, the text comprises an 
  overview from a larger perspective and is self-contained, so that it may serve as a basis for 
  a topical course or as material for the preparation for a PhD exam. 
\item In addition, the text aims at giving a detailed and accurate
  presentation of the historical developments  of the field, from the
  understanding of \kkm\ in the 1950s to the study of \bb\ and \ddm\ in
  modern high-statistics flavor experiments. I show how the
  interplay of excellent experimental achievements and innovative
  theoretical ideas lead to landmark results which shaped the
  Standard Model of Elementary Particle Physics. 
\item  Finally, I elucidate the importance of the precision calculations
  which were needed to link \kkm\ to \bbmd\ to confirm the
  Kobayashi-Maskawa interpretation of $CP$ violation and, today,
  allow us to precisely determine fundamental parameters  
  of the Standard Model's Yukawa sector and to constrain the parameter spaces of
  new-physics models.
\end{enumerate}

%%%%%%%%%%%%%%%%%%%%%%%%%%%%%%%%%%%%%%%%%%%%%%%%%%%%%%%%%%%%%%%%
%% the following items are mandatory: 
%% - Section: Introduction 
%% - further sections
%% - Section: Conclusion
%\section{Introduction}\label{intro}
% Please provide a very general and easy to understand introduction to
% your chapter.
\section{Introduction}\label{intro}
In this introductory section the basic notation and the fundamental concepts of \mmm\ are
introduced, mostly in a qualitative way, with quantitative details
relegated to later sections.  

\begin{figure}[tbp]  
\begin{center}
\begin{tabular}{c@{\hspace{1cm}}c}
\includegraphics[width=0.4\textwidth]{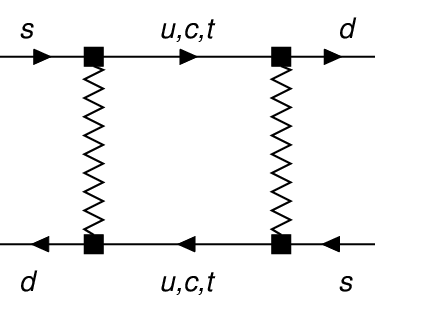} &
\includegraphics[width=0.4\textwidth]{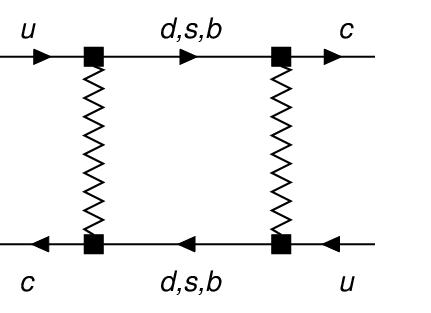} \\[3mm]
\includegraphics[width=0.4\textwidth]{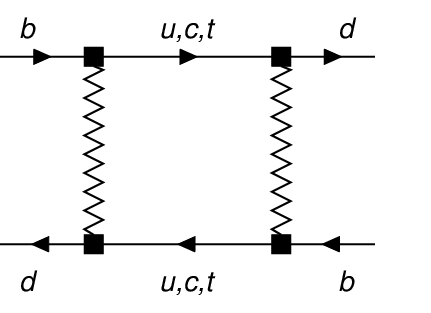} &
\includegraphics[width=0.4\textwidth]{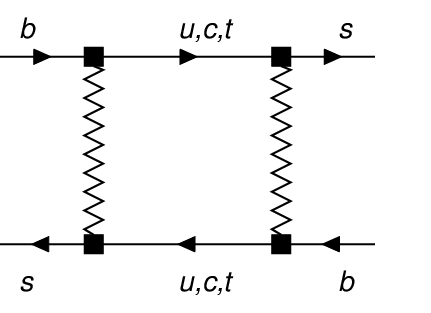} 
\end{tabular}
\end{center}
\caption{Box diagrams for \kk, \dd, \bbd\ and \bbms\ with zigzag lines
  representing W bosons. The diagrams show the transition from  antimeson $\Mbar$
  meson entering the diagram from the left into meson $M$ leaving the
  diagram to the right.  
  For each process there is also a second box diagram, 
  obtained by a 90$^\circ$ rotation.\label{fig:boxes}}
\end{figure}

Mesons can be labeled by flavor quantum numbers, which characterize the
quark-antiquark pair from which they are formed. For example, a $D_s^+$
meson has the flavor quantum numbers $C=1$ and $S=1$, which denote charm
and strangeness, respectively. One shortly writes $D_s^+ \sim c \bar s$
to indicate that $D_s^+$ has the same flavor quantum numbers as the
indicated quark-antiquark pair. We further need the beauty quantum number $B$ and,
for completeness, also introduce $D$ and $U$ for $\bar d$ and $u$ quark.
The flavor quantum numbers are $+1$ for up-type quarks and $-1$ for
down-type quarks, with opposite signs for the antiquarks. When referring
to a generic flavor quantum number we write $F$, \textit{i.e.}\
$F=B,C,S,U$, or $D$. While the strong interaction, described by quantum chromodynamics
(QCD), respects the flavor quantum numbers, the weak interaction can
change them. The most prominent examples for flavor-changing
transitions are weak decays like $D_s^+ \to K^+ \pi^0$; in this example the $C$
quantum number changes from $C=1$ to $C=0$ while $U$ increases by one
unit. Weak decays are $|\Delta F|=1$ processes mediated by the exchange
of one $W$ boson. But the Standard Model (SM) of Elementary Particle
Physics also permits $|\Delta F|=2$ transitions, through Feynman diagrams with 
two $W$ bosons. This feature  makes any of the following four
neutral mesons,
\begin{align}
  K \sim \bar s d, \qquad 
  D \sim c \bar u, \qquad 
  B_d \sim \bar b d, \qquad 
  B_s \sim \bar b s,
  \label{mesons}
\end{align}
mix with its respective antimeson,  
\begin{align}
  \bar K \sim s \bar d, \qquad 
  \bar D \sim  \bar c u, \qquad 
  \bar B_d \sim b \bar d, \qquad 
  \bar B_s \sim b \bar s.
  \label{antimesons}
\end{align}
The Feynman diagrams for the four possible \mmm\ amplitudes are shown in \fig{fig:boxes}.
Meson-antimeson mixing are examples of \emph{flavor-changing neutral
  current (FCNC)}\ processes, in which a quark morphs into another quark
with the same electric charge but different flavor. In the SM FCNC
processes are rare, because they are forbidden at tree-level.
Meson-antimeson mixing has two important implications:
\renewcommand{\theenumi}{\roman{enumi}}
\begin{enumerate}\addtolength{\labelsep}{-2pt} 
\item %[(i)]
  The flavor eigenstates $\ket{M}$ and $\ket{\Mbar}$
  corresponding to the mesons in \eqsand{mesons}{antimesons} are
  not the physical mass eigenstates and do not obey exponential decay laws. Instead 
  the mass eigenstates are linear
  combinations of $\ket{M}$ and $\ket{\Mbar}$. For $K$, $\bar K$ the
  mass eigenstates are $\ket{K_{\rm short}}$ and $\ket{K_{\rm long}}$, with the
  subscript referring to their lifetime $\tau$. Since
  $\tau_{K_{\rm    long}} \gg \tau_{K_{\rm short}}$ it is natural to use
  the mass eigenstates to describe observables in Kaon physics: 
  For sufficiently large times the $K_{\rm short}$ component
of a neutral Kaon has decayed away and one can study
$K_{\rm long}$ decays, while for times $t\sim \tau_{K_{\rm short}}$
decays of the   $K_{\rm short}$ component of the Kaon are dominant.
  % The degeneracy is lifted and 
  %         we can denote the two mass eigenstates by $M_H$ and $M_L$, where 
  %         ``$H$'' and ``$L$'' stand for ``heavy'' and ``light'',
  %         respectively. $M_H$ and $M_L$ not only differ in their masses, 
  %         but also in their lifetimes.
\item %[(ii)]
  If we produce a meson $M$ at some time $t=0$, 
           the corresponding state will 
           evolve into a superposition of $M$ and $\Mbar$ at later times
           $t>0$, leading to meson-antimeson oscillations. This property
           is used in $D$, $B_d$, and $B_s$ physics, where the lifetime
           differences of the mass eigenstates is small.  
\end{enumerate}
As an important  consequence, \mmm\ permits the study of a
\emph{quantum-mechanical superposition of  a particle with its
  antiparticle}. This feature gives access to the relative complex phase
between the $M\to f$ and $\Mbar \to f$ decay amplitudes for final states $f$
into which both $M$ and $\Mbar$ can decay.

To calculate the two mass eigenstates in terms of the flavor eigenstates
$M$ and $\bar M$ one must solve a quantum-mechanical two-state system
by diagonalising a $2\times 2$ matrix. The off-diagonal elements of this
matrix are calculated from the box diagrams in \fig{fig:boxes}; this
calculation will be explained in detail in this chapter. As a
consequence of \mmm, the eigenstates differ in their masses and decay width.
One may label the eigenstates by their lifetimes, as we did above in (i)
for the neutral Kaons. While $\Gamma(K_{\rm short}) =1/\tau(K_{\rm
  short}) > \Gamma(K_{\rm long})=1/\tau(K_{\rm long}) $ by definition,
the sign of the $K_{\rm long}$--$K_{\rm short}$ mass difference is not
fixed and must be determined by measurement.  In $B_{d,s}$ and $D$
physics one commonly labels the eigenstates by their masses as
$\ket{M_H}$ and $\ket{M_L}$ with the labels referring to ``heavy'' and
``light''. With this definition  
\begin{align}
  \dm &\equiv\; M_H- M_L >0, \qquad\qquad \mbox{while }\quad
   \dg \equiv\;  \Gamma_L - \Gamma_H   \label{defdmdg}
\end{align}
can have either sign. Here $ M_{H,L}$ and $ \Gamma_{H,L}$ denote masses and
width of the eigenstates.\footnote{Using the same notation for a generic
meson $M$ and its mass should not lead to confusion.} 
For neutral Kaons the sign of $\dg/\dm$ is firmly
established, so that today we know that
\begin{align}
  \ket{K_H} &=\;  \ket{K_{\rm long}}, \qquad\qquad
              \ket{K_{L}} =\; \ket{K_{\rm short}}. 
\end{align}
(I refrain from employing the usual notation $K_{L,S}$ for $K_{\rm long, short}$,
because I use ``L'' for ``light'' and the lighter Kaon happens to be
$K_L=K_{\rm short}$.) Note the
choice of the sign of $\dg$ in \eq{defdmdg}. This choice is motivated
by aiming at positive numbers for both $\dm$ and $\dg$ for neutral
Kaons. With this definition also the SM expectation for $\dg$ for the
$\bb$ systems is positive and this is experimentally confirmed for
$B=B_s$, while no data are yet available for $B=B_d$.

We will also need the average mass and average width, 
\begin{align}
  M &=\; \frac{M_H+M_L}2, \qquad \qquad
        \Gamma \;=\; \frac{\Gamma_H + \Gamma_L}2   \label{avgmg}
\end{align}
and note that the \emph{average lifetime} is defined as $\tau\equiv 1/
\Gamma$, that is, it is \emph{not}\ the average of $\tau(M_H)$ and
$\tau(M_L)$.  The Particle Data Table lists the such defined average masses 
and lifetimes for the neutral mesons except for $K_{\rm long}$ and
$K_{\rm short}$, for which the individual lifetimes are quoted
\cite{ParticleDataGroup:2024cfk}. 

The mass eigenstates follow exponential decay laws,
\begin{align}
  \ket{M_{L,H}(t)} &=\; e^{-i M_{L;H}t -\Gamma_{L,H} t/2} \ket{M_{L,H}} \qquad\qquad
   \mbox{with }\quad \ket{M_{L,H}(0)}=\ket{M_{L,H}}.
\end{align}
Here the oscillatory term in the exponent with the meson mass $M_{L,H}$
is the usual time-evolution factor $\exp(-i E t)$ with
$E=M_{L;H}$ in the meson rest frame; here and throughout this chapter I
use natural units with $\hbar=c=1$.  We can use
$ \ket{M_{H}(t)},\ket{M_{L}(t)}$ as a basis to express any neutral meson
state (i.e.\ any chosen superposition of $\ket{M}$ and $\ket{\Mbar}$),
\begin{align}
   \ket{M_{\rm any}} &=\; \alpha  \ket{M_{L}(t)}+ \beta   \ket{M_{H}(t)}
                      \;=\; e^{-i M t -\Gamma t/2}
                      \lt[  \alpha  e^{i \dm t/2 -\dg t/4} \ket{M_{L}}+
                              \beta e^{-i \dm t/2 +\dg t/4}
                      \ket{M_{H}} \rt] ,  \label{many}
\end{align}
where I have used \eqsand{defdmdg}{avgmg} as $M_{H,L}=M\pm \dm/2$ and
$\Gamma_{H,L}=\Gamma \mp \dg/2$. 
The first factor in \eq{many} is just the time evolution of a particle state which
does not mix, such as a charged-meson state. The term in square bracktes
shows that $\dm \neq 0$ introduces oscillatory terms and that further
$\dg\neq 0$ changes the familiar exponential decay law to a
two-exponential formula. For the cases that $\ket{M_{\rm any}} =\ket{M}$
or $\ket{M_{\rm any}} =\ket{\bar M}$ we will learn that $|\alpha|\simeq|\beta|\simeq 1/\sqrt2$.
Now the probability to observe $M_{\rm any}$ as  $M$ at time $t$ is given
by $ \lt| \langle M \ket{M_{\rm any}} \rt|^2 $ which involves oscillatory
terms like $\sin^2 (\dm t/2) = (1-\cos (\dm t))/2$  and 
$\sin (\dm t/2) \cos  (\dm t/2) =  (\sin (\dm t))/2 $, so that
the oscillation frequency in observable quantities is $\dm $ and not $\dm/2$. Likewise
$ \lt| \langle M \ket{M_{\rm any}} \rt|^2 $ and other observables involve
$\sinh (\dg t/2) $ and $\cosh (\dg t/2) $. The detailed expressions for
the time evolution of states and observables will be derived later in
Sec.~\ref{sec:time}. 

This chapter is organized as follows:  In
Secs.~\ref{sec:kkm}-\ref{sec:ddm} I will discuss \kk, \bb, and \ddm\ with
emphasis on the phenomenology and the historical evolution of the
field. This includes the presentation of theoretical and experimental
landmark results. In the context of these discussions I will derive the
necessary theoretical formulae avoiding lengthy derivations as much as
possible. Technical details are relegated to Secs.~\ref{sec:time} and
\ref{sec:sm}. In Sec.~\ref{sec:time} I derive the formulae for the time evolution  
of the neutral mesons, the relation between flavor and mass eigenstates,
and expressions linking these to physical observables.  
Sec.~\ref{sec:sm} presents the origin of flavor mixing in the SM and
beyond, and discusses how \mmm\ contributes to constrain ---or eventually
discover--- new physics. Finally, Sec.~\ref{sec:con} contains the Conclusions.

\boldmath
\section[{\kkm}, discrete symmetries, and the Cabibbo-Kobayashi-Maskawa
matrix]{\kkm, discrete symmetries, and the
  Cabibbo-Kobayashi-Maskawa matrix\label{sec:kkm}}
\unboldmath%
In this section I describe \kkm\ and the historical role which this
process played to shape the SM, with emphasis on the discrete symmetries
parity, charge conjugation, and time reversal. I use \kkm\ to exemplify
general concepts of \mmm\ and to introduce basic concepts of flavor
violation in the SM.

Historically, for three decades \kkm\ was the only known \mmm\ process.
\kkm\ was predicted in 1955 by Gell-Mann and Pais from the following
observations \cite{Gell-Mann:1955ipe}:
\begin{enumerate}\addtolength{\labelsep}{-2pt}
\item% [(i)]
  The decays $K \to \pi^- e^+\nu_e $ and
$\bar K \to \pi^+ e^-\bar\nu_e $ have shown that there are two
neutral Kaons. There was confidence in a $\Delta S =\Delta Q$ rule
linking the changes in strangeness $S$ and electric charge $Q$ of the
hadron in semileptonic decays. 
(In modern language: It was (correctly) assumed that the lepton
charge \emph{tags}\ $S$.) Thus it was clear that $K$ and $\Kbar$ are
distinct particles characterized by $S=1$ and $S=-1$, respectively.
\item %[(ii)]
  The observation that neutral Kaons decay
to $\pi\pi$ states implies the possibility of $K \leftrightarrow \pi\pi
\leftrightarrow \bar K$ transitions, and the principles of quantum physics
dictate that $\ket{K}$ and $\ket{\Kbar}$ must mix. (In modern language:
A virtual $\pi\pi$ loop permits a non-zero $K\to \Kbar$ transition
amplitude.)
\end{enumerate}
With a symmetry argument  Gell-Mann and Pais concluded that
the mass eigenstates are not close to $\ket{K}$ or $\ket{\Kbar}$,
but instead coincide with maximally mixed states
$(\ket{K}\pm\ket{\Kbar})/\sqrt{2}$. I discuss their arguments a few
paragraphs below in the context of discrete symmetries. An important
prediction of Ref.~\cite{Gell-Mann:1955ipe} was the existence of
$K_{\rm long}$, for which soon after evidence \cite{Fry:1956pg} and observation
\cite{Lande:1956pf} were reported.  To facilitate the study of the
original literature I remark that an early notation was  $\theta^0\equiv
K$; later  $K_1^0$ and $K_2^0$ were introduced to denote  $K_{\rm short}$ and  $K_{\rm
    long}$, respectively.

Both \mmm\ and the $M$ decays involve the weak interaction mediated by
the $W$ boson, which is the only SM particle with flavor-violating couplings.  
The corresponding piece of the SM
Lagrangian for quarks reads%
\beq %
  L_W = \frac{g_w}{\sqrt{2}} \sum_{j,k=1,2,3} \lt[ V_{jk} \,
\bar{u}_{jL} \, \gamma^{\mu} d_{kL} \, W^{+}_{\mu} + V_{jk}^* \,
\bar{d}_{kL} \, \gamma^{\mu} u_{jL} \, W^{-}_{\mu} \rt] .
 \label{wex}%
 \eeq%
 Here $g_w$ is the weak coupling constant and I have used the
 notations $(d_1,d_2,d_3)=(d,s,b)$ and $(u_1,u_2,u_3)=(u,c,t)$.
$V$ is a unitary $3\times 3$ matrix, %
\bea%
V &=& \left( \begin{array}{ccc}
                V_{ud} & V_{us} & V_{ub} \\
                V_{cd} & V_{cs} & V_{cb} \\
                V_{td} & V_{ts} & V_{tb}
        \end{array} \right), \label{defv}   %
\eea%
the \emph{Cabibbo-Kobayashi-Maskawa (CKM) matrix}. 
We can decompose any four-component Dirac spinor field
$\psi(x)$ as $\psi(x)=\psi_L(x)+\psi_R(x)$ with  the subscripts
``L'' and ``R'' referring to left and right chirality, respectively.
The parity
transform $P$ flips the signs of the spatial components of the four-vector $x$ as
$\vec x \to -\vec x$ and maps Nature onto a fictitious mirror-world.
$P$ also exchanges $\psi_L(x) \leftrightarrow \psi_R(x)$ and the $W$
interactions in \eq{wex} do not involve the right-handed components of
the quark spinor fields $u_{jL}(x)$ and $d_{kL}(x)$ at all. This feature
is called \emph{maximal parity violation}. Until 1956 it was believed
that $P$ is a good symmetry of Nature, but the observation that a Kaon
can decay into two-pion states with parity quantum number $P=+1$
as well as three-pion states with parity quantum number $P=-1$ lead
%Tsung-Dao
Lee and %Chen Ning
Yang to the conclusion that the weak interaction violates parity
\cite{Lee:1956qn}. By contrast, the strong and electromagnetic
interactions obey parity symmetry.

There are three fundamental discrete symmetries which are useful to
characterize interactions within and beyond the SM; apart from $P$ these
are the \emph{charge conjugation ($C$)}\ and \emph{time reversal $T$}\
symmetries. $C$ maps particles onto antiparticles and vice versa or
Nature onto a fictitious antiworld.  More precisely, $C$ maps a spinor
field $\psi(x)$ which destroys a fermion and creates an antifermion
onto the field $\psi^c(x)$ which instead destroys an antifermion and
creates a fermion. $\psi^c(x)$ is calculated from the adjoint spinor
field $\bar\psi=\psi^\dagger \gamma^0$ (with the Dirac matrix
$\gamma^0$), but we do not need the explicit form of $\psi^c(x)$ in this
chapter.  As an important feature, $C$ also flips the chirality, e.g.\
if $\psi=\psi_L$ is left-handed, then $\psi^c=\psi_L^c=(\psi^c)_R$ is
right-handed. Thus the maximal $P$ violation in \eq{wex} implies also
maximal $C$ violation. In 1956 nothing was known about quarks and the
role of quark currents in meson decays, but it was clear that in
(semi-)leptonic decays $P$ violation implies $C$ violation
\cite{Lee:1957qq}.\footnote{The seminal paper by Wu et al.\
  \cite{Wu:1957my} mentions the observation of both $P$ and $C$
  violation in an angular asymmetry in $\beta$ decay
  and Ref.~\cite{Wu:1957my} does so for
  $\pi^+\to\,  \mu^+[\to e^+ 2\nu]\, \nu$.} Still no conclusion was drawn on
$C$ violation in hadronic weak decays or \kkm\ and there was no
consensus on the question for a long time.

Since more than five decades studies of \mmm\ are instrumental to
explore $CP$ violation. 
The $CP$ transformation is a consecutive application of $C$ and $P$.
The order in which the operations are carried out does not matter,
i.e.\ $C$ and $P$  commute. $CP$ is intimately related to 
the time reversal operation $T$ which maps Nature onto a fictitious world in which time goes
backwards.  $T$ is better described as a reversal of particle motion,
to study $T$ one could compare a scattering process $A+B \to C+D$ with
$C+D \to A+B$, but in practice one studies the violation of the
associated quantum number $T= \pm 1$ in a suitable process or identifies
$T$-odd observables, just as one does in the study of $P$
violation.
The famous CPT theorem, states that any local Poincar\'e-invariant
quantum field theory (QFT) is invariant under the successive
application of $C$, $P$, and $T$ (in any order)
\cite{Luders:1954zz,pauli:1955,Luders:1957bpq}.  That is, if we take a
video of some physical process, it will be indistinguishable from the
video of the corresponding process with all particles exchanged by their
antiparticles shown backwards in a mirror.  Thus under the very wide
prerequisites of the CPT theorem $CP$ violation is identical to
$T$ violation. It is easier to work with $CP$ rather than $T$, because
$CP$ is a unitary operation on quantum fields and states, while $T$ is
anti-unitary, meaning that it combines a unitary transformation with a
complex conjugation.

In their prediction of \kkm\ and the existence of $K_{\rm long}$ in
Ref.~\cite{Gell-Mann:1955ipe} Gell-Mann and Pais assumed that $C$ is a
good symmetry and concluded that $\ket{K_{\rm short}}$ and
$\ket{K_{\rm long}}$ must be eigenstates of $C$ with opposite quantum
numbers, because the hamiltonian $H$ and the $C$ operator must have
common eigenstates if $[H,C]=0$.  Ironically, today we know that $C$ is
maximally broken, but their argument applies as well for $CP$ which is
almost a good symmetry for the \kk\ system: If we assume that
$K_{\rm long}$ is CP-odd and that further the weak decay process obeys
the $CP$ symmetry, the decay $K_{\rm long}\to \pi\pi$ (with a pair of
neutral or charged pions) into a CP-even final state is forbidden.
Instead the dominant $K_{\rm long}$ decay modes involve three pions and
the small phase space suppresses the decay rate to a level that
$\tau_{K_{\rm long}}\sim 500\times \tau_{K_{\rm short}}$.

The width difference $\dg_K$ in the \kk\ system is an unspectacular quantity,
it is essentially equal to $\Gamma_{\rm short}$, which in turn is
completely dominated by $K_{\rm short}\to \pi\pi$ decays. The decay
rates of $K_{\rm short}\to \pi^+\pi^-$ and $K_{\rm short}\to \pi^0\pi^0$
cannot be reliably calculated from first principles. Experimentally we
have \cite{ParticleDataGroup:2024cfk}
\begin{align}
%  \dg_K^{\rm exp}
  \dg_K^{\rm exp}
  &=\;  (7.338 \pm 0.003 ) \, \mbox{$\mu$eV} 
          \; =\; (11.149 \pm 0.005) \cdot 10^{-3} \, \mbox{ps}^{-1} .
\label{dgkexp}
\end{align}
Already in 1958 the time evolution of neutral kaons was used for a
measurement of $|\dm_K|$ \cite{Boldt:1958zz} to find
$|\dm_K/\Gamma_{\rm short}| \sim 1$ \cite{Boldt:1958zz}. This reference
plots a likelihood function, but does not quote an error on the
measurement.  A later measurement employing the idea to regenerate
$K_{\rm short}$'s from a $K_{\rm long}$ beam passing through matter
\cite{Muller:1960ph} found $|\dm_K/\Gamma_{K_{\rm short}}|< 1.1$
at 95\% C.L., preferring values smaller than 1. Today we know
\cite{ParticleDataGroup:2024cfk}
\begin{align}
  \dm_K & =\; (3.476 \pm 0.006)\, \, \mbox{$\mu$eV} 
         \;=\;  (5.281 \pm 0.009)  \cdot 10^{-3}  \,    \mbox{ps}^{-1}.
\label{dmkexp}
\end{align}
The regenerator method also permitted to determine the sign of
$\dm_K/\dg_K$.  

Already in 1958 S.~Weinberg estimated that $CP$ violation in
$K_{\rm long}$ decays must be smaller than 1\% and concluded ``Probably
it will be some time before experiments are performed which are capable
of detecting such small charge asymmetries.'' \cite{Weinberg:1958zz}.
The discovery of $CP$ violation had to wait until 1964, when
Christenson, Cronin, Fitch, and Turlay discovered the decay
$K_{\rm long}\to \pi^+\pi^-$ and concluded that $CP$ is violated at the
permille level in \kkm\ \cite{Christenson:1964fg}. Their grant proposal
was primarily aiming at a better measurement of
% the coherent
% regeneration of $K_{\rm short}$ from a $K_{\rm long}$ beam passing
% through matter
$K_{\rm short}$ regeneration, but further mentions ``Other results to be obtained will
be a new and much better limit for the partial rate of
$K_2^0 \to \pi^+\pi^-$'' and the authors expect their apparatus to ``set
a limit of about one in a thousand for the partial rate of
$K_2\to\pi\pi$ in one hour of operation.''  The experimental result ---a
discovery rather than a limit--- was interpreted as the discovery of
$CP$ violation in the \kkm\ amplitude, i.e.\ in a $|\Delta S|=2$ process,
and subsequent theory papers shared that view, because
models attempting to explain the measurement with \emph{direct CP
  violation}, i.e.\ $K\to \pi^+\pi^-$ and $\bar K\to \pi^+\pi^-$
amplitudes of different magnitude, were considered disfavoured by other
measurements \cite{Sachs:1964zz}.
Nevertheless, Refs.~\cite{Cabibbo:1964zza,Truong:1964} ascribed the $CP$ violation to
the weak $|\Delta S|=1$ amplitudes with Ref.~\cite{Truong:1964}
exploiting the $K \leftrightarrow \pi\pi
\leftrightarrow \bar K$ mechanism of
Ref.~\cite{Gell-Mann:1955ipe} to generate $CP$ violation in the
$|\Delta S|=2$ \kkm\ amplitude.  Yet the connection to the weak
interaction was not obvious at all at the time, recall that there was no
SM yet (and clearly nothing was known about box diagrams).  In
Ref.~\cite{Bernstein:1965hj} it was speculated that instead the electromagnetic
interaction of hadrons violates $C$ and $T$. Furthermore, the authors of
Ref.~\cite{Lee:1965hi} decomposed the hamiltonian as $H_G+H_F$ with $H_G$
comprising ``the usual weak interaction which \ldots is invariant under
CP'' and further describing $H_F$ as ``a new interaction which does not
conserve CP''. The paper discusses the three cases that $H_F$ contains
$\Delta S=0$, $|\Delta S|=1$, and $|\Delta S|=2$ interactions, and the third
possibility, called \emph{superweak}\ model has been a benchmark model
which the SM was compared to for a long time.\footnote{The superweak
  model constrains $CP$ violation to \kkm\ and was disproven, when
  $|\Delta S|=1$ $CP$ violation was discovered by the CERN NA31 and NA48
  as well as the Fermilab KTeV collaborations
  \cite{NA31:1993tha,NA48:2002tmj,KTeV:2002qqy}.} The 1965 status of the
field is well-described in the talk by Prentki in
Ref.~\cite{Moorhouse:1966rea}. This talk and the summary talk by Salam
also show that the situation with $C$ violation was not clear at the time.  

The landmark result of Ref.~\cite{Christenson:1964fg} was the branching ratio
\begin{align}
B(K_{\rm long}\to \pi^+\pi^-) &=\; (2.0\pm 0.4)  \cdot 10^{-3}. 
\end{align}
Today's world average  is \cite{ParticleDataGroup:2024cfk}
\begin{align}
B(K_{\rm long}\to \pi^+\pi^-) &=\; (1.967 \pm 0.010)  \cdot 10^{-3}. \label{brcpv}
\end{align}
The further interpretation of this result needs the theoretical machinery of
Sec.~\ref{sec:time} and is relegated to later sections of this chapter.

In the SM FCNC processes can only be studied in a meaningful way since
the introduction of the charm quark field by Glashow, Iliopoulos, and
Maiani in 1970 \cite{Glashow:1970gm}. The old three-quark version of the
SM involved the FCNC coupling $\bar s_L \gamma^\mu d_LZ_\mu$ of the
$Z$ boson, while the new four-quark
model, treating two SU(2) doublets $(u_L,d_L)^T$ and $(c_L,s_L)^T$
equally, lead to flavor-conserving $Z$ couplings. This feature is called
\emph{tree-level GIM mechanism}. The authors were guided by the three-flavor-SM
prediction of an unduly large \kkm\ as well as
$K_{\rm long}\to \mu^+\mu^-$ and $K^+\to \pi^+ e^+e^-$ decay amplitudes,
in contradiction to experiment. In the four-quark model \kkm\ involves
the box diagram of \fig{fig:boxes} with all four combinations of $u$ and
$c$ quarks on the two internal quark lines. In the four-quark model, $V$
in \eq{defv} reduces to the $2\times 2$ Cabibbo matrix $V_C$ which is
the upper left sub-matrix of $V$. The unitarity of $V_C$ makes the \kkm\
amplitude vanish exactly in the limit $m_c=m_u$ of equal up and charm
quark masses. In this limit the box diagram is identical for all four
combinations of $u$ and $c$ quarks on the internal lines and the CKM
elements combine to $( V_{us} V_{ud}^* + V_{cs} V_{cd}^*)^2 $, which
vanishes because a unitary $2\times 2$ matrix satisfies
$ V_{us} V_{ud}^* = - V_{cs} V_{cd}^*$. Setting $m_u=0$ and keeping
$m_c\neq 0$ one finds that the \kkm\ amplitude is \emph{GIM-suppressed}\
by a factor of $m_c^2/M_W^2\sim \; 10^{-4} $, where $M_W$ is the $W$
boson mass. Details on the calculation will be presented in
Sec.~\ref{sec:sm}.  The suppression by factors $m_c^2/M_W^2$ or
$m_c^2/M_W^2\, \ln(m_c^2/M_W^2)$ is a common feature of Kaon FCNC
processes called \emph{loop-level GIM mechanism}. In summary, the GIM
mechanisms have a dramatic effect on the \kkm, reducing the prediction
of the tree amplitude of the three-quark model (involving $Z$ exchange)
% (proportional to $1/M_Z^2$ with the $Z$
% boson mass $M_Z$) 
by a factor of roughly $m_c^2/(4\pi^2 M_W^2)$.
Gaillard and Lee \cite{Gaillard:1974hs} 
have estimated $m_c\approx 1.5\gev$ 
and Vainshtein and Khriplovich \cite{vainshtein:1973}
have found $m_c\approx 1 \gev$
from \kkm, which are numbers surprisingly close to the value
inferred from the $c$-$\bar c$ bound state $J/\psi$  after its discovery 
\cite{E598:1974sol,SLAC-SP-017:1974ind}. 

What did the four-quark model say about $CP$ violation?  $CP$ maps $W^+$
onto $W^-$ and exchanges quark and antiquark fields in \eq{wex}. Picking
out the $CP$ transformation of the $s$ decay vertex one finds
\begin{align}
  L_W \supset &  \; \frac{g_w}{\sqrt{2}} \;\lt[ V_{us} \,\bar u \gamma^\mu s
       \, W^{+}_{\mu} \;+ \;  V_{us}^* \,\bar s \gamma^\mu u
             \, W^{-}_{\mu} \rt]   \label{start} \\ %[2mm]
  & \; \qquad\qquad\quad     \big \downarrow  \; CP \qquad\qquad   \big\downarrow \;
      CP \no \\ % [2mm]
  & \; \phantom{\frac{g_w}{\sqrt{2}}} \;\;\;     
      V_{us} \,\bar s \gamma^\mu u
             \, W^{-}_{\mu}   \;+ \;  V_{us}^* \,\bar u \gamma^\mu s W^{+}_{\mu}  .   \label{end}
\end{align}
While owing to $ V_{us}\neq V_{us}^*$ \eq{end} looks different form
\eq{start} these two expressions are nevertheless physically equivalent,
because we can rephase any quark field as%
\bea%
d_j \to e^{i \phi^d_j} d_j,&& \qquad\qquad u_k \to e^{i \phi^u_k} u_k
. \label{fph}%
\eea%
without changing the physics. This rephasing affects $V_{jk}$ as
$V_{jk} \to V_{jk} e^{i (\phi^d_j-\phi^u_k)}$ entailing that we are free
to multiply any row or any column of $V$ by a common phase factor.  We
can do such a rephasing in \eq{end} and arrange the phase
$\phi_s-\phi_u\equiv \phi_2^d-\phi_1^u $ to bring the result into
agreement with \eq{start}.  Of course, in the four-quark model, \eq{fph}
has been conventionally applied already in \eq{wex} to obtain a real
$V_C$ with $V_{ud}=V_{cs}=\cos\theta_C$ and
$V_{us}=-V_{cd}=\sin\theta_C$ in terms of the Cabibbo angle $\theta_c$
\cite{Cabibbo:1963yz}.  For this standard choice of the $V_C$ phase
convention the $CP$ invariance of \eq{start} and all other $W$ couplings
to quarks is manifest. Thus in 1970 the origin of CP violation,
discovered six years before, was not clear. In 1973 Kobayashi and
Maskawa proposed three possibilities to accomodate $CP$ violation
\cite{Kobayashi:1973fv}. One of these was based on the observation that
the $n^2$ parameters characterising a unitary $n\times n$ matrix involve
$n(n-1)/2$ angles (which would suffice for a real orthogonal matrix) and
$n(n+1)/2$ complex phases.  With \eq{fph} we can rephase the $n$ rows
and $n$ columns with $2n-1$ phase differences $\phi^d_j-\phi^u_k$ at our
disposal to render $2n-1$ elements of $V$ real. This leaves
$n(n+1)/2-2n+1= (n-1)(n-2)/2$ complex phases as physical parameters and
for the case of $n=3$ there is exactly one physical phase in
$V$. Applying our $CP$ transformation in \eq{end} to \eq{wex} will not
leave $L_W$ invariant. The mentioned physical phase is the only
CP-violating parameter appearing in the weak interaction of quarks and
is called \emph{Kobayashi-Maskawa (KM) phase}\ $\delta_{KM}$. Our
exercise further tells us that one cannot locate $CP$ violation in a
particular term in \eq{wex}, because by rephasing our quark fields we
can render several chosen CKM elements real and transfer $\delta_{KM}$
to other elements. CP-violating observables always involve CKM elements
in combinations which are independent of phase conventions such as
$V_{us}V_{ud}^* V_{ts} V_{td}^*$, which originate from two interfering
amplitudes governed by different combinations of CKM elements.

The standard phase convention of the CKM matrix \cite{Chau:1984fp} 
adopted by the Particle Data Group chooses $V_{ud}$, $V_{us}$, $V_{cb}$,
and $V_{tb}$ real and positive. The KM phase appears in $V_{ub}$ as
$|V_{ub}| e^{-i\delta_{KM}}$ and apart from $V_{td}$ all remaining
CKM elements have phases close to 0 or $\pi$.  The structure of this
matrix is best seen in the approximate \emph{Wolfenstein
  parametrization} \cite{Wolfenstein:1983yz},
\begin{align}
  V&=\; \begin{pmatrix}
       1-\frac{1}{2}\lambda^2 & \lambda & A\lambda^3(\rho-i\eta) \\
                -\lambda & 1-\frac{1}{2}\lambda^2 & A\lambda^2 \\
                 A\lambda^3(1-\rho-i\eta) & -A\lambda^2 & 1
               \end{pmatrix}
     + {\cal O} (\lambda^4)\, ,        \label{wolf}                                                  
\end{align}
which employs an expansion in the small parameter
$\lambda \simeq |V_{us}|$ with three more positive parameters $A$,
$\rho$, $\eta$ targeted to be of order 1. \eq{wolf} nicely exhibits the
hierarchy of the CKM matrix, with diagonal elements close to 1 and the
smallest elements in the upper right and lower left corner. The origin
of this hierarchy is part of the ``flavor puzzle'' of the SM and not
understood. $CP$ violation is implemented through $\eta\neq 0$.  Today we
know that $\lambda= 0.225$, $A = 0.82$, $\rho=0.16\epm{0.01}{0.00}$, and
$\eta= 0.36\epm{0.01}{0.00} $ if there are no BSM contributions to
flavour-changing decays \cite{Charles:2015gya}. At this level of experimental precision,
the approximation in \eq{wolf} is too crude and one better works with
exact expressions (see Sec.~\ref{sec:sm}).

Ref.~\cite{Kobayashi:1973fv} can be viewed as the paper predicting the
third fermion generation, but did not receive much attention at first,
with only six citations by the end of 1975.  At the time the alternative
explanation in terms of spontaneous $CP$ violation with a second Higgs
doublet was more popular \cite{Lee:1973iz}. In the six-quark model the
prediction of the $K_{\rm long}\to \pi^+\pi^-$ decay rate involves
$\imag \lt( V_{ts}V_{td}^*/(V_{us}V_{ud}^*)\rt)$, which apart from
$\delta_{KM}$ depends on other ---at the time--- poorly known CKM
parameters (namely $A$ and $\rho$ in \eq{wolf}) as well as the unknown
top mass. Moreover, the hadronic, non-perturbative piece of the
prediction could only be roughly estimated. Thus the confirmation of the
KM mechanism had to wait for more data on flavor-changing processes and better theory
predictions.

While the third fermion generation is essential for \eq{brcpv}, the
impact of tops in the loop is negligible in  $\dm_K$, which scales like
$G_F^2/(4 \pi^2) m_c^2$, where the dependence on $M_W$  is contained in
the Fermi constant $G_F\propto 1/M_W^2$ and $1/(4\pi^2)$ is the loop
suppression factor found from calculating the box diagram. 

In retrospective, the 1964 discovery of $CP$ violation
\cite{Christenson:1964fg} revealed the virtual effect of a very heavy
particle, the top quark, with mass $m_t\sim 350 M_K$. The box diagram
with two top quarks does not suffer from GIM suppression, which partly
offsets the smallness of the CKM factor
$\lt( V_{ts}V_{td}^* \rt)^2$. Thus the absence of GIM
suppression enhances the
sensitivity to top effects.  As a general feature, FCNC processes, and
especially \mmm\ observables, probe mass scales far above the energy of
the experiment at which they are carried out. Today, FCNC processes
serve as efficient probes of physics beyond the SM (BSM physics),
Ref.~\cite{Altmannshofer:2025rxc} finds a reach of \kkm\ to BSM particle
masses up to  9000\tev, if they contribute to \kkm\ at tree level with
${\cal O} (1)$ couplings. 

We close this section by mentioning two important consequences of the
CPT theorem: Any particle and its antiparticle have the same mass and
lifetime. In the context of \mmm\ the equalities $M_K=M_{\bar K}$ and
$\Gamma_K=\Gamma_{\bar K}$ are essential to correctly relate the tiny
mass and width differences of the mass eigenstates as well as the size
of CP-violation to the box diagrams in \fig{fig:boxes}. In \kkm\ even
CPT-violating quantities have been analyzed and measured to be
consistent with zero. One may speculate that CPT symmetry is violated by
the dynamics of quantum gravity associated with the energy scale of the
Planck mass $M_{\rm Planck}\sim 10^{18}\gev$.  The current experimental
accuracy is $|M_K-M_{\bar K}|/M_K < 8 \cdot 10^{-19}$
\cite{ParticleDataGroup:2024cfk}, which is 
of order $M_K/M_{\rm Planck}$, but CPT breaking ---if it exists at all---
needs not be linear in this parameter.

\boldmath
\section{\bbm, flavor oscillations, ${CP}$ asymmetries, and the
  unitarity triangle\label{sec:bbm}}
\unboldmath
The mass eigenstates of the $B_d$ mesons have almost identical
lifetimes, so that one needs different methods to study \bbmd\ compared
to \kkm.  The DESY laboratory had operated the DORIS collider with the
ARGUS experiment, which was used as a \emph{B factory}, which is an
$e^+$-$\,e^-$ collider with the center-of-mass energy of the
$\Upsilon(4S)$ resonance. This resonance essentially only decays to
$(B^+,B^-)$ or $(B_d,\bar B_d)$ pairs.  ARGUS had studied dilepton
events, i.e.\ decays in which both $B^{\pm}$ or $\BdorBdbar$ mesons
decay semileptonically (into final states with electron/positron $e^\mp$
or (anti-)muon $\mu^\mp$). In the case of $B^\pm$ mesons the lepton
charges have necessarily opposite signs, since the lepton charge tags
the beauty quantum number. ARGUS also observed like-sign dilepton events
and concluded that they originate from a $(B_d,\bar B_d)$ pair in which
one of the $\BdorBdbar$ mesons has oscillated into its antimeson. In
addition, ARGUS has used events with fully reconstructed kinematics such
as $B_d \to D^{*-} \pi^+$ (and also decays with more than one pion) in
which the charged pion serves as the tag: The up quark in
$\bar b \to \bar c u \bar d $ ends up in the $\pi^+$ while the
charge-conjugate mode will give a $\bar u$ hadronizing into a $\pi^-$.
Tagging modes are also called \emph{flavor-specific}, characterized by
the property that a decay $B_d\to f$ and its $CP$-conjugate decay $\bar
B_d\to \bar f$
is allowed while $\Bbar_d \to f$ and $B_d \to \bar f$ are
forbidden. Here I have used the definition
\begin{align}
  \ket{\bar f} &\equiv\; CP\ket{f} .    \label{fbar} 
\end{align}
which I use for all final multi-particle states, while the $CP$
transformation of the one-particle states of our four neutral mesons
$K$, $D$, and $B_{d,s}$ is given in \eq{defcandcp}.  It is understood
that $CP$ is applied in the rest frame of the decaying particle where
$\vec p_M=0$. Thus $CP$ reverses the momenta of the particles in
$\ket{f}$, but in two-body final states we can bring
$\ket{\bar h_1(\vec p) \bar h_2(-\vec p)}$ to
$\ket{\bar h_1(-\vec p) \Bar h_2(\vec p)}$ by a $180^\circ$
rotation. This feature is crucial for the $CP$ physics discussed below,
because otherwise we could not define two-body $CP$ eigenstates in a
useful way.

For example, $B_d\to \pi^+\pi^-$ is \emph{not}\
flavor-specific. Strictly speaking, the decay $B_d \to D^{*-} \pi^+$
used by ARGUS is not exactly flavor-specific, because
$\Bbar_d \to D^{*-} \pi^+$ is allowed via $b \to u \bar c d $, but this
amplitude is suppressed by a factor of $\lambda^2$ compared to
$b \to c \bar u d $ (see \eq{wolf}) and could be neglected in 1987.

We will see in Sec.~\ref{sec:time} that the oscillation frequency is
equal to the difference $\dm_d$ of the masses of the two eigenstates of
the \bbd\ system and is proportional to the absolute value of the box
diagram in \fig{fig:boxes}.  Recalling our use of natural units with
$\hbar=c=1$, we realize that energy, mass, and frequency have the same
dimension. One usually quotes $\dm_d$ in units of inverse picoseconds,
because it is measured as the mentioned oscillations frequency and the
relevant time scale is the $B_d$ lifetime of 1.5$\,$ps.  $\dm_d$ is
proportional to the magnitude of the \bbm\ amplitude, which one can
calculate in terms of $|V_{tb} V_{td}^*|^2$ and $m_t$. Contrary to \kkm,
box diagrams with other CKM elements are negligible. To verify this, we
observe from \eq{wolf} that all three CKM combinations
$V_{tb} V_{td}^*$, $V_{cb} V_{cd}^*$, and $V_{ub} V_{ud}^*$ are
quadratic in $\lambda$ and thus of similar size and recall the GIM
mechanism suppressing contributions with light quarks.  ARGUS could not
track the time evolution of the mesons, but did a time-integrated
measurement yielding \cite{ARGUS:1987xtv}
\begin{align}
  x_d & \equiv \; \dm_d \, \tau_{B_d} \;=\; 0.73 \epm{0.17}{0.18} 
         \qquad\qquad\mbox{ARGUS 1987.} \label{argus} 
\end{align}
Confronting this with the theory prediction, ARGUS concluded that $m_t$
must be larger than 50\gev, which was the first evidence for a heavy top
quark. Shortly before, the UA1 collaboration had reported evidence for
an excess of dilepton events stemming from $B$ mesons produced in
$p\bar p$ collisions, in which all $b$-flavored hadrons are produced,
and erroneously ascribed the effect to \bbms\ \cite{UA1:1986fuh}. 
This interpretation is compatible with a roughly five times  lighter top, because
$|V_{ts}|$ in \eq{wolf} is larger than $|V_{td}|$ and the box diagram
roughly grows as $m_t^2$.

Using the 2025 value $\tau_{B_d}=(1.517\pm 0.004) \,$ps, the ARGUS
measurement in \eq{argus} implies
$\dm_d=0.48\epm{0.11}{0.12}\,\mbox{ps}^{-1}$ which perfectly complies
with the actual number found from the world average of the oscillation frequency in
time-dependent studies,
\begin{align}
    \dm_d &=\; \lt( 0.5069\pm 0.0016_{\rm stat} \pm 0.00116_{\rm
            syst}\rt) \,\mbox{ps}^{-1}\qquad\qquad 
            \mbox{HFLAV 2025\,
            \cite{HeavyFlavorAveragingGroupHFLAV:2024ctg},}  \label{wavdmd}
\end{align}
which involves data from LEP, Tevatron, BaBar, Belle(-II), LHCb, but is
dominated by LHCb measurements \cite{LHCb:2016gsk}. 

The proximity of $x_d$ to 1 and the smallness of $|V_{cb}|$ and
$|V_{ub}|$ constitute the \emph{B physics miracle}: The latter property
makes the $B$ meson long-lived, since only $b\to c$ and $b\to u$ decay
channels are open and suppressed by the small CKM elements,
so that the sizable $B_d$ lifetime around 1.5 ps permits
the study of time-dependent observables. The former property means that
the oscillation frequency is in the right range to analyze observables
governed by $\sin(\dm_d t)$ and $\cos(\dm_d t)$, i.e.\ after two
lifetimes a meson produced as $B_d$ has oscillated into a $\Bbar_d$. To
study such time-dependent quantities one must produce the $B$ mesons
with a sufficiently large boost. This was the case at the LEP~I collider
at CERN, where $b$-flavored hadrons were produced from $Z$ decays.  The
\emph{asymmetric B factories}\ \emph{Super KEK-B}\ (KEK, Tsukuba, Japan)
and \emph{PEP-II}\ (SLAC, Menlo Park, USA) with the experiments
\emph{Belle} and \emph{BaBar}, respectively, have been built to study
time-dependent observables in decays of $(B_d,\bar B_d)$ pairs
originating from the $\Upsilon(4S)$ resonance.  The different energies
of the $e^+$ and $e^-$ beams boosted the center-of-mass of the
$(B_d,\bar B_d)$ pair in the detector permitting to measure the
difference of the times at which the mesons decay. Currently the
upgraded experiment \emph{Belle II}\ is running. Hadron colliders also
provide sufficiently energetic $B$ mesons and there was a rich $b$
physics program at the $p\bar p$ collider \emph{Tevatron}\ at Fermilab
(Batavia, USA) with the experiments CDF and D\O.  One of the four major
experiments at the $p\bar p$ collider LHC at CERN is LHCb, which is a
dedicated forward-spectrometer experiment optimized for studies in $b$
(as well as $c$) physics. Furthermore, the high-$p_T$ LHC experiments CMS and ATLAS
contribute to the field as well.

B factories produce the $(B_d,\bar B_d)$ pair in a coherent state with
the quantum numbers of the $\Upsilon(4S)$. At any given time $t$ each of
the two involved mesons is a quantum-mechanical superposition of $B_d$
and $\bar B_d$, but their correlation is such that the overall beauty
quantum number is 0, as that of $\Upsilon(4S)$.\footnote{The time
  evolution of a coherent $(B_d,\bar B_d)$ pair is discussed e.g.\ in
  Ref.~\cite{BaBar:1998yfb}.}  If we observe at some time $t_1$ a
flavor-specific decay $B_d\to f_{\rm fs}$, the coherent wave function
collapses such that the other meson is in a $\ket{\Bbar_d}$ state as
this time. This ``starts the clock'' for the time evolution of
$\ket{\Bbar_d}$; one defines the time-dependent state $\ket{\Bbar_d(t)}$
which satisfies $\ket{\Bbar_d(t_1)}=\ket{\Bbar_d}$.  For $t>t_1$ this
state $\ket{\Bbar_d(t_1)}$ is a calculable superposition of $\ket{B_d}$
and $\ket{\bar B_d}$ and the decay $\Bbar_d(t_2)\to f$ observed at time
$t_2>t_1$ provides information on \bbm\ and, if the decay is \emph{not}\
flavor-specific, on the interference of the $B_d\to f$ and
$\Bbar_d \to f$ decay amplitudes.  The latter feature is heavily used to
explore $CP$ violation. $(b,\bar b)$ pairs produced at hadron colliders 
hadronize into multi-particle states containing many light hadrons in
addition to the pair of $b$-flavored hadrons. A $B_d$ can be produced
together with a $B^-$ or $\Lambda_b$, so that we cannot expect an
entangled  $(B_d,\bar B_d)$ pair as in a $B$ factory. Still the overall
beauty quantum number is zero, thus the observation of a $B^-$ or
$\Lambda_b$ (which contain a $b$ quark) tags the $B_d$ and
studies of time-dependent $CP$ asymmetries are possible as well.
Here the ``clock starts'' at the time the $B_d$ is produced.
Hadron colliders produce substantially more $B_d$ mesons than $B$
factories, which in turn have a better tagging efficiency. Unlike hadron
colliders the entanglement at $B$ factories can also be used to do
$CP$ tagging, in which one of the $B$'s is tagged through a decay into a $CP$ eigenstate,
so that the other $B$ collapses into the orthogonal state. 

To study time-dependent effects one defines the \emph{time-dependent
  decay rate}\ of a meson tagged at $t=0$ as $M$:%
\beq%
\gtf = \frac{1}{N_M}\, \frac{d\, N(M (t) \to f)}{d\, t} \,,
\label{defgtf}
\eeq%
where $d\, N(M (t) \to f)$ denotes the number of decays into the final
state $f$ occurring within the time interval between $t$ and $t+d\, t$.
$N_M$ is the total number of $M$'s produced at time $t=0$. An analogous
definition holds for \gbtf.  We consider a decay into a $CP$ eigenstate
$f_{\rm CP}$,
\begin{align}
  CP \, \ket{f_{\rm CP}} \, &=\, \eta_{\rm CP,f} \ket{f_{\rm CP}} \label{defcpeig}
\end{align}
with the $CP$ quantum number $\eta_{\rm CP,f}=\pm 1$. For example,
$D^+ D^-$ and $\pi^+\pi^-$ are CP-even eigenstates with
$\eta_{{\rm CP}, D^+ D^-}=\eta_{{\rm CP}, \pi^+ \pi^-}=1$.

We also need the $C$ and $CP$ transformations for the state of the decaying meson,
which I choose as
\begin{align}
  C\ket{M(\vec p)} &=\; \phantom{-} \ket{\bar M(\vec p)}, %\qquad\qquad
  &
  C\ket{\bar M(\vec p)} & =\;  \phantom{-}\ket{M(\vec p)}, % \nn
   &  CP\ket{M(\vec p)}   & =\; - \ket{\bar M(-\vec p)}, %\qquad\qquad
  &
  CP\ket{\bar M(\vec p)}   &=\; - \ket{M(-\vec p)},  \label{defcandcp}
\end{align}
where it is used that $M=K,D,B_d,B_s$ are all $P$-odd. One can put
arbitrary phase factors into these definitions like
$C\ket{M(\vec p)} = \exp(i \phi_C)  \ket{\bar M(\vec p)}$ with
$C\ket{\bar M(\vec p)} = \exp(-i \phi_C)  \ket{M(\vec p)}$, because
the phase of any state vector is arbitrary: Changing from
$\ket{M}$ to $\ket{M^\prime}\equiv  \exp(i \phi_C/2) \ket{M}$ with
$\ket{\bar M^\prime}\equiv  \exp(-i \phi_C/2) \ket{\bar M}$ and applying
$C$ of \eq{defcandcp} leads to
$C\ket{M^\prime (\vec p)} =\; \exp(i \phi_C) \, \ket{\bar M^\prime (\vec p)}$.
Similarly, such a freedom exists for the definition of the $CP$ transform of
the fields in $L_W$ in \eqsand{start}{end}. One does not need to carry
these arbitrary phases through the calculations, instead one can stick
to \eq{defcandcp} and e.g.\ confirm $CP$ invariance by  checking that
the $CP$-transformed quantity can be brought into agreement with the
original expression by changing unphysical phases of fields and states,
just as we did in the discussion of \eqsand{start}{end}.
However, sometimes one checks the (non-)dependence on $\phi_C$ to
identify physical quantities and to perform a ``sanity check'' of a
calculation by confirming that some calculated observable 
is independent of the choices for unphysical phases.

Needless to say, the ambiguity of $ \phi_C$ leads to the fact that you
can find different ``standard definitions'' of $C$ and $CP$ in the
literature, corresponding to $\phi_C=0$ and $\phi_C=\pi$, so that the
signs are flipped compared to \eq{defcandcp}. There is, however, a good
reason for my choice: While there are many phase conventions involved in
a quantum field theory, several of these conventions are related to each
other. A standard convention for the definition of the light meson
states in terms of (anti-)quark fields uses the standard Gell-Mann
matrices $\lambda^a$, $a=1,\ldots 8$, in the meson octet as
$M^a= (\bar u,\bar d,\bar s) \lambda^a (u,d,s)^T$, so that
$K \sim d \bar s$ and $\bar K \sim s \bar d$ in
\eqsand{mesons}{antimesons}, without any ``$-$'' signs (or, more
generally, without phase factors). Now we can employ the SU(3)$_F$
symmetry to rotate $\ket{K(\vec p)}$ into $\ket{\bar K (\vec p}$; this
is a so-called U-spin SU(2) rotation of $(s,d)^T$, the analogue of an
isospin rotation of the isodoublet $(u,d)^T$. Thus we can get from
$\ket{K(\vec p)}$ to $\ket{\bar K (\vec p}$ in two ways, by a U-spin
rotation or by applying $C$, and these two operations must be consistent
with each other.  A rotation by $\pi$ around the $y$-axis in U-spin
space leads to $\ket{K(\vec p)}\to - \ket{\bar K (\vec p)}$ and, by
convention, the combination of this rotation and $C$ is the $G_U$
\emph{parity}\ transformation which maps the three members of the U-spin
triplet onto themselves
\cite{Karliner:2010xb,Sahoo:2015msa,Meng:2022ozq,Bolognani:2024zno}.
$G_U$ parity is the analogue of the famous $G$ parity which combines $C$
with an isospin rotation and has the property that
$G\ket{\pi^\pm}=-\ket{\pi^\pm}$ and $G\ket{\pi^0}=-\ket{\pi^0}$, with
the ``$-$'' sign fixed from the $G$ parity of $\pi^0$, for which no
choice of phase of the $C$ transformation is possible, because it is a
$C$ eigenstate \cite{Lee:1956sw}.  Thus the definitions of $G_U$ and $G$
parities require the choice for $C$ in \eq{defcandcp} for neutral Kaons
and $C\ket{\pi^+}=\ket{\pi^-}$ for charged pions. In decays of $D$ or
$B$ into final states with one or more neutral Kaons, this subtlety
indeed matters in analyses using SU(3)$_F$ symmetry and leads to
mistakes if ignored as shown in Refs.~\cite{Muller:2015lua,Bolognani:2024zno}.  For the
heavy $D$ and $B$ mesons the argument presented above does not apply,
because nobody uses flavor symmetries rotating heavy mesons into their
antimesons. Nevertheless, I use the same phase conventions for $C$ and
$CP$ for all neutral mesons.

The \emph{time-dependent CP asymmetry}\ for $M\to f_{\rm CP}$ is defined
as
\begin{align}
  a_{\rm CP} (M(t)\to f_{\rm CP})
  &\equiv\;   \frac{ \gbtfcp - \gtfcp }{ \gbtfcp + \gtfcp }
    \label{acp}.
\end{align}
Specifying to $M=B_d$ one finds  
\begin{align}
   a_{\rm CP} (B_d (t)\to f_{\rm CP})
  &= \; -A_{CP}^{\rm dir} \cos ( \dm_d  \, t ) -
   A_{CP}^{\rm mix} \sin (
                \dm_d  \, t ).
                \label{acpbd}
\end{align}
We will derive \eq{acpbd} in Sec.~\ref{sec:time} and note that in
\eq{acpbd} some sub-percent corrections are set to zero.  The first term
is non-zero already for $t=0$, when $\ket{B_d(t)}=\ket{B_d}$.
$ a_{\rm CP}(B_d(0)\to f_{\rm CP})$ simply quantifies the amount by
which the decay rates $\Gamma (\Bbar_d\to f_{\rm CP}) $ and
$\Gamma (B_d \to f_{\rm CP}) $ differ from each other. This
feature is called \emph{direct CP violation}\ and motivates the notation
$A_{CP}^{\rm dir} $ in \eq{acpbd}. As time elapses, the initially
produced $\BdorBdbar$ oscillates into a superposition of $B_d$ and
$\Bbar_d$ which makes $ a_{\rm CP} (B_d (t)\to f_{\rm CP})$ sensitive to
the interference of $B_d\to f_{\rm CP}$ and $\Bbar_d\to f_{\rm CP}$ and
the size of this effect is encoded in $A_{CP}^{\rm mix}$, the
\emph{mixing-induced CP asymmetry}. Thus
$A_{CP}^{\rm mix} \sin ( \dm \, t )$ quantifies \emph{CP violation in
  the interference of mixing and decay}.  Of course,
$A_{CP}^{\rm dir/mix}= A_{CP}^{\rm dir/mix}(B_d \to f_{\rm CP})$ depends
on the decay mode, but to keep the notation short I omit this dependence
wherever this does not lead to confusion.

\begin{figure}
%\begin{minipage}{0.3\textwidth}  
  \centering\includegraphics[width=0.2\textwidth,angle=-90]{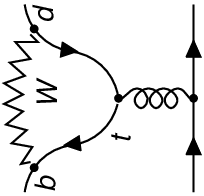} 
\caption{Penguin diagram contributing to a $\Delta S=0$ decay of a $b$-flavored
  hadron. \label{fig:tpeng}}
\end{figure}
For a non-zero $CP$ asymmetry we need two interfering amplitudes. In the
case of $A_{CP}^{\rm dir}$ these are two decay amplitudes governed by
different CKM elements. For example, $\Bbar_d \to D^+D^-$ is dominated by
the tree-level $W$-mediated $b\to c \bar c d$ amplitude but also
receives a contribution from the top penguin depicted in \fig{fig:tpeng}
and a similar diagram with internal up quark. These three contributions
are proportional to $V_{cb}V_{cd}^*$, $V_{tb}V_{td}^*$, and $V_{ub}V_{ud}^*$, 
respectively. There is further a charm penguin diagram which comes with
the same CKM structure as the tree contribution. 

Using unitarity we can eliminate one of these CKM combinations, e.g.\
$V_{tb}V_{td}^*=- V_{cb}V_{cd}^* -V_{ub}V_{ud}^*$, so we are left with
a decay amplitude of the form $A(\Bbar_d \to D^+D^-) = V_{cb}V_{cd}^* A_T
+ V_{ub}V_{ud}^* A_P$ with complex $A_{T,P}$. This (commonly used)
notation is reminiscent of ``tree'' and ``penguin'', although $A_T$ also
comprises the charm penguin and a part of the top penguin. 
The CP-conjugate mode has the amplitude $A(B_d \to D^+D^-) = -V_{cb}^*V_{cd} A_T
- V_{ub}^*V_{ud} A_P$, with the ``$-$'' sign stemming from 
$CP\ket{B_d}=-\ket{\Bbar_d}$ in \eq{defcandcp}.
The phases of the CKM elements flip their
signs, because the quarks flow in the opposite direction and the
corresponding vertex Feynman rules  in \eq{wex} involve the complex
conjugate of  the CKM element entering  $A(\Bbar \to D^+D^-)$.
That is, \emph{CP-violating}\ phases flip signs when going from a
process to its CP-conjugate one. The remainder stays the same,
 $A(\Bbar \to D^+D^-)$ and  $A(B \to D^+D^-)$ involve the same $A_T$ and
 $A_P$, because the SM has no other CP-violating parameters beyond the
 elements of $V$.\footnote{QCD could violate CP, but bounds on electric
   dipole moments constrain the corresponding
   parameter $\theta_{\rm QCD}$ to be smaller than $10^{-10}$.}  
 While the hadronic dynamics in $A_{T,P}$ is complicated and
 uncalculable, the $CP$ invariance of QCD  ensures that these quantities
 are equal in $A(\Bbar \to D^+D^-)$ and  $A(B \to D^+D^-)$. The phases
 of $A_{T,P}$ are dubbed \emph{CP-conserving}\ or \emph{strong phases}.
 In decays in which  the strong phases of $A_T$ and $A_P$ are the same, one readily
 finds $|A(\Bbar \to D^+D^-)|= |A(B \to D^+D^-)|$, so that a non-zero 
 $ A_{CP}^{\rm dir} $ needs $\arg A_T \neq \arg A_P$. It is
 impossible to calculate these phases from first principles, making
 predictions for $ A_{CP}^{\rm dir} $ impossible and rendering
 essentially all direct $CP$ asymmetries useless for the determination of
 $\delta_{KM}$ or potential BSM $CP$ phases.\footnote{In exceptional cases
 one can relate different decays to each other and
 eliminate uncalculable amplitudes \cite{Gronau:1990ka,Gronau:1990ra,Gronau:1991dp}.}  

Whenever two amplitudes $A_T$ and $A_P$ contribute to the decay, also
 $ A_{CP}^{\rm mix} $ cannot be calculated. However, there are cases in
 which one of the two amplitudes is highly suppressed or even absent.
 Such decays are called \emph{gold-plated modes}.   Thus gold-plated
 modes have necessarily $ A_{CP}^{\rm dir} =0$. (The converse is
 not true, $A_T$ and $A_P$ could have the same strong phase leading to
 $ A_{CP}^{\rm dir} =0$. Therefore by measuring  $ A_{CP}^{\rm dir} =0$
 one cannot conclude that the mode is gold-plated.) The prime example of
 a gold-plated mode is $B_d\to J/\psi K_{\rm short}$, proposed by Bigi and Sanda
 \cite{Bigi:1981qs}, in which  the tree amplitude $A_T$ is multiplied by
 $V_{cs} V_{cb}^*$ which is proportional to  two powers of the
 Wolfenstein parameter $\lambda$, while  $A_P$ instead involves $ V_{us}
 V_{ub}^* \propto \lambda^4$.  In this context $A_P$ is dubbed ``penguin
 pollution'', as it inflicts an uncertainty of a few percent on the
 value of the $CP$ phase extracted from a measurement of
 $A_{CP}^{\rm mix}  (B_d\to J/\psi K_{\rm short})$. Experimentally one detects
 the lepton pair from the $J/\psi$ decay and a $\pi^+\pi^-$ pair with
 the invariant mass of the neutral Kaon. Thus really
 $ a_{\rm CP}(B_d(0)\to J/\psi [\pi^+\pi^-]_{M_K}) $ is measured. This
 feature is important, because the $\bar b \to \bar c c \bar s$ transition in
 the $B_d$ decay produces a $K$ meson while $b \to c \bar c s$ triggering
 the $\Bbar_d$ decay produces a $\bar K$. The interference of the
 $K\to\pi^+\pi^-$ and $\bar K\to\pi^+\pi^-$ decays is needed to obtain a
 meaningful $A_{CP}^{\rm mix} $.

Interestingly, we can deduce which $CP$ phase is measured from
$A_{CP}^{\rm mix}$ without performing the 
detailed calculations of Secs.~\ref{sec:time} and \ref{sec:sm}.
$A_{CP}^{\rm mix}$ must involve the phase of the \bbd\ box diagram,
which is $\pm (V_{tb}V_{td}^*)^2$ with the sign to be determined by a
calculation of the box diagram. 
That is, we only need the sign, not the
full analytical result of this diagram. Neglecting the penguin
pollution, the decay amplitude can be written as 
\begin{align}
  % \bar A_{f_{\rm CP}} & \equiv \; A (\Bbar_d \to J/\psi \bar K [\to
  %                       \pi^+\pi^-] )\; =\; 
  %                       V_{cb} V_{cs}^* \, V_{us} V_{ud}^* \,    A_T
  %                       \nn
  A_{f_{\rm CP}} & \equiv \; A \lt( B_d \to J/\psi K [\to
                        \pi^+\pi^-] \rt)\; =\;  
                        V_{cb}^* V_{cs} \, V_{us}^* V_{ud} \,    A_T
                        \label{abjps} 
\end{align}
where the second CKM factor $V_{us}^* V_{ud}$ originates from the
$\bar s\to \bar u u\bar d$ amplitude in the $K\to \pi^+\pi^-$ decay. The
charge conjugate mode with decay amplitude 
$\bar A_{f_{\rm CP}} \equiv \; A (\Bbar_d \to J/\psi \bar K [\to
\pi^+\pi^-])$ involves the complex-conjugate CKM elements instead.

Apart from the phase of the box diagram, the desired physical $CP$ phase
$\phi_{CP,B_d}^{\rm  mix}$ can only depend on the relative phase of
$ A_{f_{\rm CP}}$ and $ \bar A_{f_{\rm CP}}$, that is, the phase of
$ \bar A_{f_{\rm CP}}/A_{f_{\rm CP}}$.
If we adopt the standard CKM phase convention explained before
\eq{wolf}, we find the CKM elements in  \eq{abjps} real, except for
$V_{cs}$, whose phase is far below $0.01^\circ$ and negligible. Thus, the
only contribution to $\phi_{CP,B_d}^{\rm  mix}$ stems from the box
diagram and (for a positive sign) one deduces
\begin{align}
  \phi_{CP,B_d}^{\rm  mix} &=\; \arg\lt( (V_{tb} V_{td}^*)^2\rt)
                             \;=\; 2 \arg  V_{td}^*,  \qquad
                             \mbox{valid for the standard CKM phase
                             convention}.
                             \label{phmvtd}
\end{align}
From \eq{wolf} one realizes that this is not expected to be a small
number, there is no suppression by powers of $\lambda$ in
\eq{phmvtd}. The observation that ---contrary to what people were used to
from Kaon physics--- $B_d$ decays can exhibit large $CP$ violation is
due to Carter and Sanda \cite{Carter:1980hr,Carter:1980tk}.  Thus,
loosely speaking, $A_{CP}^{\rm mix}(B_d\to J/\psi K_{\rm short})$ measures the
phase of the \bbmd\ box diagram. But one must keep in mind that the
latter is convention-dependent and unphysical, and the physical $CP$ phase
in any $B_d(t)\to f_{\rm CP}$ decay is the relative phase between the
box diagram and the phase of
$ \bar A_{f_{\rm CP}}/A_{f_{\rm CP}}$.

In Sec.~\ref{sec:time} we will derive 
\begin{align}
       A_{CP}^{\rm mix}(B_d\to J/\psi K_{\rm short}) &=\; -\sin
                                              \phi_{CP,B_d}^{\rm  mix}
                                             .  \label{amixres1}
\end{align}
The overall sign of $ A_{CP}^{\rm mix}(B_d\to J/\psi K_{\rm short})$ is
related to the $CP$ quantum number $\eta_{{\rm CP}, J/\psi K_{\rm short}}$
of the final state. To determine this we recall that $K_{\rm short}$ is
a placeholder for $\pi^+\pi^-$ which is $CP$ even. $J/\psi$ is a
$J^{PC}=1^{--}$ resonance, thus it is also CP-even. Now the quantum
number of the total angular momentum of the $ J/\psi K_{\rm short}$
state is $j=0$, because $B_d$ has zero spin. That implies that the final
state has angular momentum quantum number $l=1$, meaning that the wave
function is proportional to $ Y_{m=0}^{l=1}(\theta,\phi)$ if the
$z$-axis points in the flight direction of $K_{\rm short}$ or
$J/\psi$. The parity transformation of $CP$ maps
$ Y_{m=0}^{l=1}(\theta,\phi)$ onto
$ Y_{m=0}^{l=1}(\pi-\theta,\phi+\pi)= - Y_{m=0}^{l=1}(\theta,\phi)$, so
that we arrive at the $CP$ quantum number
$\eta_{{\rm CP}, J/\psi K_{\rm short}}=-1$.  If we considered a
$b\to c\bar c s$ decay into a CP-even final state, we would find
$A_{CP}^{\rm mix}=\sin \phi_{CP,B_d}^{\rm mix}$ instead of
\eq{amixres1}.

To describe the impact of  $CP$ asymmetries on CKM metrology one
introduces \emph{unitarity triangles (UTs)}. The unitarity of $V$
implies  for $j\neq k$:
\begin{eqnarray}
   V_{1j}^* V_{1k} + V_{2j}^* V_{2k} + V_{3j}^* V_{3k} &=& 0
   \qquad \mbox{columns}\label{un1} \\ 
\mbox{and}\qquad\qquad
    V_{j1}^* V_{k1} + V_{j2}^* V_{k2} + V_{j3}^* V_{k3} &=& 0
  \qquad \mbox{rows}\label{un2} .
\end{eqnarray}
The first equation expresses that any two columns of $V$ are orthogonal to
each other, the second one does this for rows. We have already used the
first relation for $j=1$ and $k=3$ as
$V_{tb}V_{td}^*=- V_{cb}V_{cd}^* -V_{ub}V_{ud}^*$ above. 
Each of the 
relations in \eqsand{un1}{un2} defines a triangle in the complex plane, e.g.\
for \eq{un1} the three corners are located at $0$, $ V_{1j}^* V_{1k}$
and $-V_{2j}^* V_{2k}$. The three sides of this triangle are
$|  V_{1j}^* V_{1k} |$, $|  V_{2j}^* V_{2k} |$, and $|  V_{3j}^* V_{3k}
|$. The UTs have the important feature that
the phase transformations of
\eq{fph} rotate the unitarity triangles in
the complex plane, but leaves their shape fixed. That is, both sides and
angles of the UTs are independent of phase conventions and, indeed, we
can associate physical observables with them. The angles are related to
$CP$ asymmetries.  

\begin{figure}
\centering\includegraphics[width=0.4\textwidth]{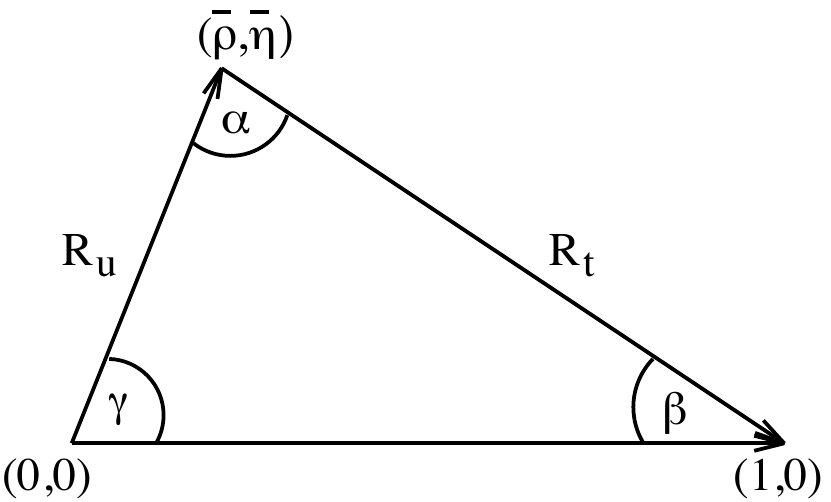}
\caption{The (standard) unitarity triangle. The non-trivial sides are
  $R_u=\sqrt{\bar\rho^2+\bar\eta^2}$ and
  $R_t=\sqrt{(1-\bar\rho)^2+\bar\eta^2}$. \label{fig:ut}}
\end{figure}

The area of all six triangles is the same and
given by $J/2$, where $J$ is the \emph{Jarlskog invariant} \cite{Jarlskog:1985ht}
\begin{eqnarray} 
 J &\equiv & \imag \lt[ V_{td}^* V_{tb} V_{ub}^* V_{ud} \rt]  
% \; =\, c_{12} c_{23} c_{13}^2 s_{12} s_{23} s_{13} \sin\delta_{13} 
 \; \simeq \; A^2\lambda^6\eta .
\end{eqnarray}
Here last result uses the Wolfenstein approximation of \eq{wolf}.  Four
of the six \emph{unitarity triangles} are squashed, the three sides are
similar only for the choice $(j,k)=(3,1)$. Moreover, within the
Wolfenstein approximation the shapes of the ``column'' and ``row''
of  \eqsand{un1}{un2} are equal for $(j,k)=(3,1)$.   Seeking a definition
of a completely rephasing-invariant unitarity triangle (which does not
rotate under rephasings) we
divide \eq{un1} (for $(j,k)=(3,1)$) by $V_{23}^* V_{21}=V_{cb}^*
V_{cd}$ to arrive at
\begin{eqnarray} 
  \frac{V_{ub}^* V_{ud}}{V_{cb}^* V_{cd}} +     
  \frac{V_{tb}^* V_{td}}{V_{cb}^* V_{cd}} +  1 &=& 0
\label{sut}
\end{eqnarray}
When people speak of ``the'' unitarity triangle they mean
the rescaled triangle defined by \eq{sut}. Since its baseline coincides with
the interval $[0,1]$ of the real axis, the unitarity triangle is
completely determined by the location of its apex $(\bar \rho,\bar \eta)$,
where 
\begin{eqnarray}
\bar \rho + i \bar \eta \equiv - \frac{V_{ub}^* V_{ud}}{V_{cb}^* V_{cd}} 
 \label{defre} .
\end{eqnarray}
This is an exact expression; comparing it with the Wolfenstein
approximation in \eq{wolf} one finds that $(\bar \rho,\bar \eta)$ agrees
with $(\rho,\eta)$ to an accuracy of 3\% \cite{Buras:1994ec}.
The UT was used in the Wolfenstein approximation since the late 1980s
\cite{Bigi:1987in}, the notation $\bar \rho$, $\bar \eta$ was
introduced in Ref.~\cite{Buras:1994ec} in which the Wolfenstein
approximation was refined by expanding $V$ to order $\lambda^5$.
The UT is depicted in \fig{fig:ut}.
The two non-trivial sides of the triangle 
are 
\begin{eqnarray}
  R_u &\equiv& \sqrt{\bar \rho^2 + \bar \eta^2 }, \qquad\qquad 
  R_t \;\equiv \; \sqrt{(1-\bar \rho)^2 + \bar \eta^2 }
\label{defsd} .
\end{eqnarray}
$CP$-violating quantities are associated with the triangle's three angles 
\begin{equation}
    \alpha = \arg\left[ - \frac{V_{td}V_{tb}^*}{V_{ud}V_{ub}^*}\right],
    \qquad
    \beta  = \arg\left[ - \frac{V_{cd}V_{cb}^*}{V_{td}V_{tb}^*}\right],
    \qquad
    \gamma = \arg\left[ - \frac{V_{ud}V_{ub}^*}{V_{cd}V_{cb}^*}\right].
    \label{eq:beta}
\end{equation}
These angles were used since the late 1980s \cite{Bigi:1987in} in the Wolfenstein
approximation and then in the improved version \cite{Buras:1994ec};   
\eq{eq:beta} is the exact definition, which does not employ any expansion in
$\lambda$ \cite{Lenz:1998qp}. The Belle (-II) collaborations use the
notation
\begin{align*}
  \phi_1&\equiv\,  \beta, &
  \phi_2&\equiv\,  \alpha, &
  \phi_3&\equiv\,  \gamma,
\end{align*}
with the concept that in the (un-rescaled) UT the angle $\phi_j$ is
opposite to the side involving $V_{j1} V_{j3}^*$.
  
The angle $\gamma$ coincides with the Kobayashi-Maskawa phase $\delta_{KM}$ at the
sub-permille level.  With \eqsto{defre}{eq:beta} one obtains
\begin{eqnarray}
 \ov{\rho} +  i \ov{\eta} &=&
              R_u  e^{i \gamma},\qquad\qquad 
1-\ov{\rho} - i \ov{\eta} \; = \; R_t  e^{-i \beta}
\label{defrb} .
\end{eqnarray}
The unitarity relation of \eq{sut} now simply reads 
\begin{eqnarray}
R_u  e^{i \gamma} +R_t  e^{-i \beta} &=& 1 \label{utrt}
\end{eqnarray}
Taking real and imaginary parts of \eq{utrt}  allows us to express any two of
the four quantities $R_u,R_t,\gamma,\beta$ in terms of the remaining two
ones. By multiplying \eq{utrt} with either $\exp(-i\gamma)$ or
$\exp(i\beta)$ one finds analogous relations involving
$\alpha=\pi-\beta-\gamma$. 

In the standard CKM phase convention three of the six elements entering
the UT are real.  Now with the improved Wolfenstein expansion of Ref.~\cite{Buras:1994ec}
one verifies
\begin{align}
  \arg (-V_{cd}) &=\; A^2 \ov\eta \lambda^4\,+ \, {\cal O} (\lambda^8)
                   \;=\; 6\cdot 10^{-4} \;=\; 0.03^\circ , \label{vcd}
\end{align}
so that one can safely neglect this phase and take $V_{cd}$ as a negative number. 
Then
\begin{align}
  V_{ub} &=\; |V_{ub}| e^{-i\gamma} ,\qquad\qquad
           V_{td} \;=\; |V_{td}| e^{-i\beta} ,
           \label{vubvtd} \\
  \phi_{CP,B_d}^{\rm  mix} &\stackrel{\mathrm{SM}}{=}\; 2\beta   \label{phmixsmd}      
\end{align}
%in the standard phase convention and
\eqsand{phmvtd}{amixres1} boil down to
the famous result
\begin{align}
         A_{CP}^{\rm mix}(B_d\to J/\psi K_{\rm short}) &\stackrel{\mathrm{SM}}{=} \; -\sin (2\beta)
                                             ,  \label{amixres2}
\end{align}
which is independent of any phase conventions if $\beta$ is defined as
in \eq{eq:beta}.

We can associate with each of the four \mmm\ processes one of the six
unitarity triangles, which we consider rescaled as in \eq{sut}. Then
\bbd\ probes the standard UT, while \kkm\ discussed in
Sec.~\ref{sec:kkm}, involving $V_{qs}V_{qd}$ with $q=u,c,t$, is related
to the UT expressing the orthogonality of the first two columns. The
height of this  rescaled  ``squashed'' triangle is of order
$\bar\eta A^2\lambda^4\simeq 6\cdot 10^{-4}$, which is tiny. $CP$ violation is thus small in
\kkm\ and the enhancement from the large top mass is instrumental to get
to the ---still small, but detectable--- branching ratio in \eq{brcpv}. 

CP violation in the $B_d$ system was discovered through measurements 
of $ A_{CP}^{\rm mix}(B_d\to J/\psi K_{\rm short}) $ by BaBar and Belle in
2001. These experiments and the B factories hosting them were designed
to measure this quantity and to thereby establish $CP$ violation beyond
Kaon decays. Since the measured value complied with the expectation from
the SM, this measurement was viewed as a confirmation of the KM
mechanism \cite{Kobayashi:1973fv}, prompting the Nobel Prize for
Kobayashi and Maskawa in 2008. 

Today's experimental world average is
\begin{align}
    A_{CP}^{\rm mix, exp}(B_d\to J/\psi K_{\rm short}) &=\;  - 0.710\pm 0.011
                                          \qquad\qquad
                                \mbox{HFLAV 2025\, \cite{HeavyFlavorAveragingGroupHFLAV:2024ctg},}
\label{acpmexp}
\end{align}
which also involves other $b\to c \bar c s$ decay modes and uses data
from BaBar \cite{BaBar:2009byl}, Belle \cite{Belle:2012paq}, Belle II
\cite{Belle-II:2024lwr}, and LHCb \cite{LHCb:2017mpa,LHCb:2023zcp}.  
\eq{acpmexp} implies
\begin{align}
 \beta &=\; 22.62^\circ \pm 0.45^\circ  .  \label{betaex}
\end{align}
To clarify this point, I first mention that in 2004 BaBar had determined
$\sgn\cos(2\beta)$ to be positive by scanning over the invariant mass of
the $K_{\rm short}\pi^0$ pair in $B_d\to J/\psi K_{\rm short} \pi^0$ around the $K^{*(0)}$
resonance and used interference effects of $S$ and $P$ waves
\cite{BaBar:2004xhu}. Thus $2\beta =\phi_{CP,B_d}^{\rm mix}$ is
determined to lie in the first quadrant with
$\sin \phi_{CP,B_d}^{\rm mix}>0$, $\cos \phi_{CP,B_d}^{\rm mix} >0$.
One could still add $\pi$ to $\beta$, but
$22.62^\circ+180^\circ= 202.62^\circ$ (implying $\bar \eta<0$ and
$\bar \rho>1$ from \eq{defrb}) is incompatible with other
measurements by far, even if these had contributions from BSM physics.

At the level of precision in \eq{betaex} one must worry about the
penguin pollution of order
$\imag(V_{ub}^*V_{us}/(V_{cb}^* V_{cs})) \simeq \bar \eta \lambda^2 \sim
0.02 $, which is neglected in \eq{betaex}.  One can estimate the size by
measuring the mixing-induced $CP$ asymmetry in the ``control channel''
$B_s\to J/\psi K_{\rm short}$ which is a $\bar b\to \bar c c \bar d $
decay in which both $A_T$ and $A_P$ come with a CKM factor of order
$\lambda^3$, so that one can determine the here large penguin
contribution from data and relate it to the decay of interest by using
the approximate U-spin symmetry of QCD \cite{Barel:2020jvf}. This
amounts to exchanging $d$ and $s$ quark lines, which relates $A_{T,P}$
in $B_d\to J/\psi K_{\rm short}$ to $A_{T,P}$ in
$B_s\to J/\psi K_{\rm short}$ and would be exact if down and strange
quark had the same mass. Currently this is not feasible, because
$ A_{CP}^{\rm mix}(B_s\to J/\psi K_{\rm short})$ is measured consistent
with zero with an error around 0.41 \cite{LHCb:2015brj}. Alternatively
one can calculate the penguin pollution using soft-collinear
factorization, which can be applied to both $b\to c \bar c d$ and
$b\to c \bar c s$ decay modes. The result, however, involves an unknown
phase and varying this phase between 0 and $2\pi$ only results in an
upper bound on the penguin pollution, which reads
$|\delta^{\rm peng} \, A_{CP}^{\rm mix}(B_d\to J/\psi K_{\rm short})|
\leq 0.0086$ inflicting an uncertainty of
$|\delta^{\rm peng} \, \beta|\leq 0.34^\circ$ on $\beta $ in \eq{betaex}
\cite{Frings:2015eva}.  For the control channel this reference finds
$|\delta^{\rm peng} \, A_{CP}^{\rm mix}(B_s\to J/\psi K_{\rm short})|
\leq 0.26$.

\bbmd\ is very sensitive to BSM physics,
Ref.~\cite{Altmannshofer:2025rxc} finds a reach of \bbmd\ to virtual BSM
particle effects with masses up to 350\tev. In general, BSM physics will
affect both $\dm_d$ and $\phi_{CP,B_d}^{\rm mix}$, and $V_{td}$ can no
more be found from these quantities. A BSM analysis of \bbmd\ requires
the determination of this CKM element from other observables.

The width difference $\dg_d$ stems from all decays into final states $f$
which are common to $B_d$ and $\Bbar_d$.  That is, only
non-flavor-specific decays contribute to  $\dg_d$. To understand this
feature, note that the mass eigenstates $\ket{B_H}$ and $\ket{B_L}$ are linear
combinations of $\ket{B_d}$ and $\ket{\Bbar_d}$,
$\ket{B_{L,H}}=\alpha_{L,H} \ket{B_d} + \beta_{L,H} \ket{\Bbar_d}$ with
$|\alpha_{L,H}|^2+|\beta_{L,H}|^2=1$.
Introducing the decay amplitudes $A_f= \langle f\ket{B_d}$ and
$\bar A_f= \langle f\ket{\Bbar_d}$, normalized such that $\Gamma(B_d\to f)=|A_f|^2$.
one finds
\begin{align}
  \Gamma (B_{L,H}\to f)  
  &= \; | \langle f \ket{B_{L,H}}|^2 \;=\;
    (\alpha_{L;H}^* A_f^* +  \beta_{L;H}^* \bar A_f^* )
    (\alpha_{L;H}  A_f +  \beta_{L;H} \bar A_f)   
  \; = \;
                            |\alpha_{L,H} |^2 |A_f|^2 +
                          |\beta_{L,H} |^2 |\bar A_f|^2 \,+\, 2 \,\real
                          \lt( \alpha_{L;H}^* \beta_{L,H} \, A_f^* \bar
    A_f  \rt),
    \label{blhf}
\end{align}
so that the total widths read
\begin{align}
  \Gamma_{L,H} &= \; \sum_f  \Gamma (B_{L,H}\to f) \;= \; 
                          |\alpha_{L,H} |^2 \Gamma_{\rm tot} (B_d)  +
                 |\beta_{L,H} |^2 \Gamma_{\rm tot} (\Bbar_d) 
                 \,+\, 2\, \real
                 \lt( \alpha_{L,H}^* \beta_{L,H} \, \sum_f A_f^*
                 \bar A_f  \rt) \nn 
                 &=\; \Gamma_{\rm tot} (B_d) \,+\, 2\, \real
                   \lt( \alpha_{L,H}^* \beta_{L,H} \,   \sum_f A_f^*
                   \bar A_f  \rt),  \label{dgexc}
\end{align}
where I have used that the CPT theorem dictates equal total decay rates
for $B_d$ and $\bar B_d$,
$\Gamma_{\rm tot} (B_d)=\Gamma_{\rm tot} (\Bbar_d)$.  The first term in
\eq{dgexc} is the same for $\Gamma_L$ and $\Gamma_H$, thus
$\dg_d=\Gamma_L-\Gamma_H$ only receives contributions from the last term
in \eq{dgexc} to which only decays into final states with
$A_f\neq 0 \neq \bar A_f$ contribute. Now $ \sum_f A_f^* \bar A_f $
corresponds to the $\bbd$ box diagram with light quarks $u,c$ on the
internal lines. It can be read such that a $\Bbar_d$ meson enters the
diagram from the left and a $B_d$ enters from the right to decay into
the same final state with the quark content of the internal lines. The
optical theorem links inclusive quantities like $\sum_f A_f^* \bar A_f $
to the so-called \emph{absorptive part}\ of a loop diagram, which can be
cut into two pieces such that $f$ corresponds to the cut internal lines
of this diagram. The absorptive part is calculated by taking the
imaginary part of the loop integral while retaining the couplings (i.e.\
the CKM elements) with the full complex phases.  One can view the last
term in \eq{dgexc} as a $\Bbar_d \leftrightarrow f \leftrightarrow B_d$
``rescattering'' transition. If we take a box diagram with internal
charm and anti-up lines as an example, we can cut it through these lines
and the diagrams to the left and right of the cut correspond to
$\bar A_f$ and $A_f^*$ with the sum over $f$ carried out over all $C=1$
final states such as $D^+\pi^-, D^{*+}\pi^0 K^- K_{\rm short},\ldots$.  The
crucial point is that only states which are kinematically accessible in
a $B_d$ decay can contribute to $\dg_d$, so that the box diagrams with
one or two internal top quarks do \emph{not}\ contribute to
$\dg_d$. While $\dm_d \propto m_t^2$, one instead finds
$\dg_d \propto m_b^2$ and $\dg_d/\dm_d \sim 10^{-3} $ from the
calculation presented in Sec.~\ref{sec:sm}.

There is yet no measurement of  $\dg_d$ in the
\bbd\ system. The experimental situation is as follows:
\begin{align}
  \lt| \frac{\dg_d}{\Gamma_d} \rt|^{\rm exp} &=\; 0.001\pm 0.010  \qquad\qquad
    \mbox{HFLAV \cite{HeavyFlavorAveragingGroupHFLAV:2024ctg}}
\label{dgdexp}
\end{align}
with the average width $\Gamma_d=1/\tau_{B_d}$ entailing
$|\dg_d|^{\rm exp} = (0.7 \pm 6.6)\cdot 10^{-3} \; \mbox{ps}^{-1}$. The value in
\eq{dgdexp} is dominated by measurements by ATLAS \cite{ATLAS:2016mln},
CMS \cite{CMS:2017ygm}, and LHCb \cite{LHCb:2014qsd}. One measures
$\dg_d$ by measuring the lifetime in a $B_d$ decay into a $CP$
eigenstate like $J/\psi K_{\rm short}$ and uses the knowledge on $\Gamma_d$.
HFLAV quotes \eq{dgdexp} without the
``$|$'' for the absolute value, but the cited measurements are not
sensitive to the sign of $\dg_d$, i.e.\ whether the lighter mass
eigenstate has the shorter or longer lifetime. However, they give
information on whether the longer-lived or the shorter-lived eigenstates
is closer to the (appropriately defined) $CP$-even eigenstate, as
discussed in Sec.~\ref{sec:time}.

I next discuss \bbms, which is very similar to \bbmd\ discussed
above. The corresponding box diagram involves $V_{ts}^2$ instead of
$V_{td}^2$. We have seen above that \bbd\ physics probes the standard
unitarity triangle, as we have encountered CKM elements of the first and
third columns of $V$.  Now \bbms\ involves the CKM elements of the second
and third columns and thus probes a ``squashed'' unitarity triangle with height of
order $\bar\eta \lambda^2$, so that we expect much smaller CP
asymmetries than in \bbd.  It is not useful to depict results in terms of 
``squashed'' triangles; recall that $V$ only involves four parameters, so quoting
values for $V_{us}\simeq \lambda$ and $V_{cb}=A \lambda^2$ and drawing the
allowed range for the apex $(\bar\rho,\bar\eta)$ of the standard UT are
sufficient to describe $V$.

We can estimate the expected value for $\dm_s$ from $\dm_d$ by rescaling
$\dm_d$ in \eq{wavdmd} by $|V_{ts}^2/V_{td}^2|\sim 23 $. Since
$\tau_{B_s}\simeq \tau_{B_d}$, one realizes that \bbs\ oscillations are
very rapid, which prohibited their detection until 2006, when the CDF collaboration
at the Fermilab Tevatron collider observed these oscillations with
\cite{CDF:2006hcy,CDF:2006imy}   
\begin{align}
  \dm_s &=\; \lt( 17.77\pm 0.10_{\rm stat} \pm 0.07_{\rm syst}\rt) \,
          \mbox{ps}^{-1}
          \qquad\qquad \mbox{CDF 2006.}
\end{align}
This number  excellently agrees with today's world average from CDF, LHCb and CMS,
\begin{align}
 \dm_s &=\; \lt( 17.766\pm 0.004_{\rm stat} \pm 0.004_{\rm syst}\rt) \,
          \mbox{ps}^{-1},
         \qquad\qquad
         \mbox{HFLAV 2025\, \cite{HeavyFlavorAveragingGroupHFLAV:2024ctg},}
\label{wavdms}
\end{align}
which is also dominated by LHCb data \cite{LHCb:2020qag,LHCb:2021moh}.
The hadronic physics in \bbmd\ and \bbms\ would be the same in the
U-spin symmetry limit of $m_s=m_d$, but in practice $\dm_s/\dm_d=35$ is
larger than $|V_{ts}^2/V_{td}^2|\sim 23 $  because of sizable (but
calculable) U-spin breaking. In fact,  $\dm_d/\dm_s$ from
\eqsand{wavdmd}{wavdms} is the most accurate way to determine
 $|V_{td}/V_{ts}|$ which gives the side $R_t$ of the UT in
 \fig{fig:ut} because of $|V_{td}/V_{ts}|=\lambda R_t \lt (1+\lambda^2
 (1/2- \bar \rho) +{\cal O} (\lambda^4)\rt)$.
 But as a caveat, \bbm\ is also most sensitive to new physics 
 and whenever we want to \emph{test}\ a BSM  hypothesis in mixing
 observables we must not  use numerical values for $V_{td}$ or $V_{ts}$ 
 obtained by assuming that the SM is correct. The state-of-the-art is to
 perform a combined fit to CKM elements and BSM parameters to all
 available data and to  study the likelihood ratio of the BSM best-fit
 point and the SM hypothesis \cite{Lenz:2010gu,Lenz:2012az,UTfit:2022hsi}.

In the SM one finds that to an excellent approximation $\dg_s/\dm_s=\dg_d/\dm_d$.
Since $\dm_s \gg \dm_d$, one finds $\dg_s$ sizable,
$\dg_s/\Gamma_s=\dg_s \tau_{B_s} \simeq 0.12$. The current world average
\begin{align}
  \dg_s^{\rm exp} &=\; (0.0781 \pm 0.0035)\; \mbox{ps}^{-1} \qquad\qquad
          \mbox{HFLAV 2025\,
          \cite{HeavyFlavorAveragingGroupHFLAV:2024ctg},}
          \label{dgsexp}
\end{align}
agrees well with the SM prediction discussed in Sec.~\ref{sec:sm}.  The
quoted number uses data from CDF, D\O, ATLAS \cite{ATLAS:2020lbz}, CMS
\cite{CMS:2015asi,CMS:2024ihd}, and LHCb
\cite{LHCb:2014iah,LHCb:2017hbp,LHCb:2016tuh,LHCb:2016tuh,LHCb:2021wte,LHCb:2023sim,LHCb:2023xtc}. Most
of these measurements use $B_s\to J/\psi \phi [\to K^+K^-]$ decays or
the corresponding decay with $\psi(2S)$. These decays also provide
information on the $CP$ phase of mixing-induced $CP$ violation, i.e.\ the
analogue of $2\beta$ in the \bbs\ system. Since $\phi$ is a vector
meson, there are three final states $(J/\psi \phi)_l$ characterized by
the angular orbital momentum quantum number $l=0,1,$ or 2. The $l=0$
S-wave state is dominant and CP-even and  the $l=2$ state is CP-even as
well. The P-wave
state, with $l=1$ and $J/\psi$ and $\phi$ polarizations perpendicular to
each other, is CP-odd, so that one needs an angular analysis to separate
the corresponding decay modes
\cite{Rosner:1990xx,Dighe:1995pd,Fleischer:1996aj,Dighe:1998vk,Dunietz:2000cr}. One
observes two exponential decay distributions following
$\exp(-\Gamma_L t)$ and $\exp(-\Gamma_H t)$, which determines
$|\dg_s|$. One further observes that the larger of $\Gamma_L$ and
$\Gamma_H$ occurs with decays into CP-even final states, so that the
mass eigenstates are, to a good approximation, $CP$ eigenstates, with the
shorter-lived state being dominantly CP-even.  Interestingly, all three
LHC experiments find errors for $|\dg_s|$ of comparable size, namely $0.004$. While
the quoted measurements determine $|\dg_s|$, it is important to note
that $\sgn \dg_s$ is firmly established to be positive: LHCb had scanned
the $K^+K^-$ invariant mass in $B_s\to J/\psi K^+K^-$ though the $\phi$
resonance and, since the variation of strong phases around a resonance
is known (from e.g.\ the Breit-Wigner formula), $\sgn \dg_s$ could be
determined \cite{LHCb:2012fgm} in 2012.
  The result complies with the SM
expectation, that $\ket{B_{s,L}}$ is the shorter-lived eigenstate (and
almost exactly CP-even).  Thus in this respect there is no difference
between \bbs\ and \kk\ systems.

The sizable width difference in the \bbs\ system opens the door to new
observables. Dunietz has pointed out that one can use \emph{untagged}\
$B_s$ samples to study $CP$ violation, essentially mimicking the situation
with \kkm, because the longer-lived $B_{s,H}$ state is mostly decaying
to CP-odd final states, just as $K_{\rm long}$, and  a $\BsorBsbar $
beam enriches itself with $B_{s,H}$ in time, because the $B_{s,L}$
component decays faster \cite{Dunietz:1995cp}. Thus one can access CP
properties from lifetime information. I will come back to this point  at
the end of Sec.~\ref{sec:tde}.

The analogue of the $\sin(2\beta)$ measurement of \bbmd\ in the \bbs\
system involves the time-dependent $CP$ asymmetry in the decay
$B_s\to J/\psi \phi$, which is driven by $\bar b \to \bar c c \bar s$
like $B_d\to J/\psi K_{\rm short}$. Unlike the $\dg_s$ measurements described in
the previous paragraph  this analysis
requires flavor tagging; in the papers
\cite{LHCb:2014iah,LHCb:2017hbp,LHCb:2016tuh,LHCb:2016tuh,LHCb:2021wte,LHCb:2023sim} 
cited above the $CP$ phase $\phi_{CP,B_s}^{\rm  mix}$ is determined
together with $\dg_s$. We can take the formulae used for the phase of
the \bbd\ box diagram in \eq{phmvtd} with the
substitution $2\beta \to -2\beta_s$ with the definition of $\beta_s$ through
\begin{align}
  \frac{V_{ts}V_{tb}^*}{V_{cs}V_{cb}^*} &\equiv \; \lt|
                                          \frac{V_{ts}V_{tb}^*}{V_{cs}V_{cb}^*}
                                          \rt| e^{i\beta_s} .
\end{align}
Thus in the standard CKM phase convention we have
\begin{align}
   V_{td} & \stackrel{\,\eq{vubvtd}\,}{=}\; |V_{td}| e^{-i\beta}, \qquad\qquad
            V_{ts} = \; |V_{ts}| e^{i\beta_s}, \qquad\qquad
   \phi_{CP,B_s}^{\rm  mix}  \;  \stackrel{\mathrm{SM}}{=}\; -2\beta_s             
\label{phmixsms} 
\end{align}
where $\arg(-V_{cd})=0.03^\circ$ (see \eq{vcd}) and
$\arg(V_{cs})\simeq -A^2 \bar\eta \lambda^6 = -0.002^\circ$ are
neglected. \eq{phmixsms} is the analogue of \eq{phmixsmd} in \bbmd.  The
different signs in the definitions of the phases of $V_{td}$ and $V_{ts}$, first
introduced in \cite{Proceedings:2001rdi}, were motivated by the goal
to have both $\beta$ and $\beta_s$ positive:
\begin{align}
  \beta_s &= \; \bar\eta \lambda^2
            \lt(1 + \lambda^2 (1-\bar\rho) \rt) +{\cal O}(\lambda^6)
            \; = \; 0.019 \;=\; 1.1^\circ.  
\end{align}
In $ A_{CP}^{\rm mix}(B_s\to (J/\psi \phi)_l) $ we must take care of the
CP quantum number  $\eta_{{\rm CP}, (J/\psi \phi)_l}=(-1)^l$  
and furthermore include the effect from a non-negligible
$\dg_s$:
\begin{align}
   a_{\rm CP} (B_s (t)\to  (J/\psi \phi)_l)
  &= \; -\frac{A_{CP}^{\rm dir} (B_s (t)\to  (J/\psi \phi)_l) \,
    \cos(\dm_s  \, t )-A_{CP}^{\rm mix} (B_s (t)\to  (J/\psi \phi)_l) \,
    \sin (\dm_s  \, t )}{\cosh (\dg_s \, t/ 2) -
       (-1)^l \cos(\phi_{CP,B_s}^{\rm  mix}) \sinh  (\dg_s \, t / 2) }
                                        \label{acpbs} \\
 A_{CP}^{\rm mix}(B_s\to  (J/\psi \phi)_l)
  &=\; 
    (-1)^l\,  \sin  \phi_{CP,B_s}^{\rm  mix}
    \;=\; \stackrel{\mathrm{SM}}{=}\; -(-1)^l\,  \sin  (2\beta_s) \, 
                                             .  \label{amixres3}
\end{align}
This  formula, which generalizes \eq{amixres1} to $\dg_s\neq 0$, will be derived in
Sec.~\ref{sec:time}.
We see that $ A_{CP}^{\rm mix}(B_s\to  (J/\psi
\phi)_l)$ is expected to be small; the quantity will not significantly contribute to
CKM metrology (though it could provide us with the height $\bar\eta$ of
the UT), but instead is a superb testing ground for BSM physics.
Ref.~\cite{Altmannshofer:2025rxc} finds a reach of \bbms\ to virtual BSM
particle effects with masses up to  70\tev.

The experimental effort 
\cite{LHCb:2014iah,LHCb:2017hbp,LHCb:2016tuh,LHCb:2016tuh,LHCb:2021wte,LHCb:2023sim}
lead to the 2025 world average 
\begin{align}
  \phi_{CP,B_s}^{\rm  mix, exp} 
  &=\;  - 0.052 \pm 0.013 \;=\; 2.98 ^\circ \pm 0.74^\circ
                                          \qquad\qquad
                                \mbox{HFLAV 2025\, \cite{HeavyFlavorAveragingGroupHFLAV:2024ctg},}
\label{acpmexps}
\end{align}
which complies with the SM expectation of $-2\beta_s=-0.038\pm 0.001$
\cite{Charles:2015gya} at 1$\sigma$.  

I further mention that $\sgn\dg_s$ is correlated with
the \bbs\ mixing phase as
$\sgn \cos \phi_{CP,B_s}^{\rm  mix}=\sgn\dg_s$ \cite{Dunietz:2000cr} in
the SM and 
any BSM theory,  in which the BSM contribution to the $b\to c\bar c s$
decay amplitudes constituting $\dg_s$ in \eq{dgexc} is smaller than the
SM contribution. This is plausible, because $b\to c\bar c s$ are large tree
amplitudes and furthermore the excellent agreement of \eq{dgsexp} with the SM
prediction do not suggest BSM dominance in  \eq{dgexc}.

In Ref.~\cite{Nandi:2008rg} it was proposed to determine 
  $\sgn \dg_s=\sgn \phi_{CP,B_s}^{\rm  mix}$ with $B_s\to D_s^\pm K^\mp$
  which involves the CKM angle $\gamma$ in the combination 
  $\phi_{CP,B_s}^{\rm  mix}+\gamma$ which breaks the degeneracy of the
  two solutions for $\phi_{CP,B_s}^{\rm  mix}$ found from $B_s\to J/\psi
  \phi$.
  Conversely, once the  information on $\sgn \dg_s$ 
  from Ref.~\cite{LHCb:2012fgm} was available, one could exploit the large
  width difference to remove discrete ambiguities in
  studies of $B_s\to D_s^\pm K^\mp$ aiming at the determination of
  $\phi_{CP,B_s}^{\rm  mix}+\gamma$ \cite{DeBruyn:2012jp}.

In \eq{amixres3} and \eq{acpmexps} the penguin pollution has been set to
0. The estimate of the size of penguin pollution using SU(3)$_F$
symmetry is difficult, because $\phi$ has a large SU(3)$_F$ singlet
component and one would need precise data on $CP$ asymmetries is
$B_{d,s}\to J/\psi \omega $ decays. The dynamical calculation of
Ref.~\cite{Frings:2015eva} finds $|\delta^{\rm peng}
\, \phi_{CP,B_s}^{\rm  mix} | \leq 0.97^\circ$, $\leq 1.22^\circ$, and $\leq 0.99^\circ$
for the longitudinal, parallel, and perpendicular polarizations of $J/\psi
\phi$, respectively, from which the $(J/\psi)_l$ amplitudes are
constructed. (Only the amplitude with perpendicular polarizations has $l=1$,
i.e.\ is CP-odd.)
We see that the penguin pollution might matter in view of the experimental
error in \eq{acpmexps}. For the decay mode $B_s\to J/\psi f_0[\to
\pi^+\pi^-]$ employed to determine $\dg_s$ in Ref.~\cite{LHCb:2023xtc} 
no estimates of the penguin pollution are available and future
determinations of $ \phi_{CP,B_s}^{\rm  mix}$ from $B_s\to J/\psi f_0$ should not be averaged
with other measurements.

\boldmath
\section{\ddm \label{sec:ddm}}
\unboldmath
\ddm\ is an example of a FCNC process of \emph{up-type
  quarks}, since it involves $c\to u$ transitions. Charm physics is the
only way to study such processes with the needed precision because
branching ratios of $t\to c$ or $t\to u$ decays (suffering from the
large total width of the top quark) or $u\to t$ or $c\to t$ FCNC
single-top production (i.e.\ without a $\bar b$ quark in the final
state) are not competitive yet.

The \ddm\ amplitude, described by the \dd\ box diagram in
\fig{fig:boxes}, involves the down-type quarks $d$, $s$, and $b$ on the
internal lines.  The CKM elements are hierarchical, with a situation
similar to \kkm; if we neglect the small CKM factor $V_{ub} V_{cb}^*$
with magnitude
$|V_{ub} V_{cb}^*| \simeq R_u A^2\lambda^5 \simeq 10^{-4} $, we are back
to the 2-flavor SM with the $2\times 2$ submatrix $V_C$.  In this
2-flavor theory we  find the \ddm\ amplitude far more GIM
suppressed than its \kkm\ counterpart: Firstly, we observe that the CKM
factor in both cases is $\sin^2\theta_C \cos^2\theta_C$. Secondly, we
see that the GIM suppression factor which was $m_c^2/M_W^2$ in \kkm\
will (in the limit
$m_d=0$) involve the much smaller $m_s$ instead of $m_c$ in \ddm. Yet the suppression is even
stronger, because $m_c\sim 1.3\gev$ is the largest mass scale in the
\kkm\ diagram and we can set external momenta (of order $m_s$ or the QCD
scale $\lqcd\sim 400\mev$) to zero. In \ddm\ the largest mass scale is
the momentum of the external charm quark, thus it is of order $m_c$ as
well and we must keep both $m_s$ and $m_c$ in the calculation. The
result is a GIM suppression of order $m_s^4/(m_c^2 M_W^2)$, which is
smaller by a factor of $m_s^4/m_c^4$ compared to \kkm. 
%%% dm_K propto m_K f_K^2  mc^2/mw^4   lambda^2 
%%% dm_D propto m_D f_D^2 ms^4/mc^2 mw^4   lambda^2 
%%% Thus dm_k/dm_D = m_K f_K^2/(m_D f_D^2) ms^4/mc^4
%%% f_K =0.16 GeV, f_D=0.212 GeV,  
Accounting for the different masses and hadronic parameters this results
in a naive estimate of  $\dm_D\sim 2\cdot 10^{-4} \cdot \dm_K =
 10^{-6}\, \mbox{ps}^{-1}$ from \eq{dmkexp}. 

If we include the contributions from the internal $b$ quarks, we
encounter six combinations of $d,s,b$ quarks on the internal lines
of the box diagram. As usual, CKM unitarity permits to eliminate one of
these, commonly this is done for $V_{ud}V_{cd}^*=-V_{us}V_{cs}^*-V_{ub}V_{cb}^*$.
% Writing the \dd\ mixing amplitude as
% \begin{align}
%   {\cal M}_{\dd} &=\; \lt(V_{us}V_{cs}^*\rt)^2 {\cal M}_{ss} \, +\,
%                    \lt(V_{us}V_{cs}^*  V_{ub}V_{cb}^*   \rt) {\cal M}_{sb} \, +\,
%                    \lt(V_{ub}V_{cb}^*\rt)^2 {\cal M}_{bb}  
% \end{align}
% Recalling $m_d=0$,  
This results in three contributions to the \ddm\ amplitude, proportional
to $\lt(V_{us}V_{cs}^*\rt)^2$, $\lt(V_{us}V_{cs}^*  V_{ub}V_{cb}^*
\rt) $, and  $\lt(V_{ub}V_{cb}^*\rt)^2 $, respectively. While the latter
two CKM elements are smaller than the first one, this suppression is
partially offset by  the weaker GIM suppression  of the loop diagrams
with one or two $b$ quarks. So while numerically sub-leading, these terms
are not negligible compared to the $\lt(V_{us}V_{cs}^*\rt)^2\,
m_s^4/(m_c^2 M_W^2)$ term.\footnote{For an analysis of this feature for
  $\dg_D$ see Ref.~\cite{Bobrowski:2010xg}.}

The experimental situation of $\dm_D$ and $\dg_D$ is summarized in
Ref.~\cite{Friday:2025gpj}. In charm physics it is common standard to
quote results in terms of
\begin{align}
  x_D^{\rm exp} &=\; \frac{\dm_D}{\Gamma_D}, \qquad\qquad
        y_D^{\rm exp} =\; - \frac{\dg_D}{2\Gamma_D}, \label{defxy}
\end{align}
where the ``$-$'' sign in $y_D$ stems from the definition in
\eq{defdmdg} which differs from the one in Ref.~\cite{Friday:2025gpj}.
The first discovery of \ddm, $(x_D,y_D)\neq (0,0)$, from a combination of
several measurements was in 2007
\cite{HeavyFlavorAveragingGroupHFLAV:2024ctg},
in 2012  LHCb achieved this in a single experiment \cite{LHCb:2012zll} and in 
2021 LHCb \cite{LHCb:2021ykz} has established $x_d\neq 0$ with a significance of
more than 5$\sigma$, see Ref.~\cite{Friday:2025gpj} for details. Apart
from LHCb also BaBar, Belle, and CDF made important contributions to the
discovery and quantification of \ddm. The 2025 world average is
\begin{align}
  x_D &=\;  (0.407 \pm 0.044) \cdot 10^{-2}, \qquad\qquad
      y_D =\;  \lt( 0.645\epm{0.024}{0.023} \rt) \cdot 10^{-2}
                                          \qquad\qquad
                                \mbox{HFLAV 2025\, \cite{HeavyFlavorAveragingGroupHFLAV:2024ctg}}
\label{xyexp} 
\end{align}
which is found from a global fit in which the CP-violating quantities
are fitted together with $x_D$ and $y_D$. No evidence for $CP$ violation
is found until today. With $1/\Gamma_D=\tau_D = (0.4103 \pm 0.0010) \, \mbox{ps}$
the numbers in \eq{xyexp} imply
\begin{align}
  \dm_D &=\;  (0.0099 \pm 0.0011) \, \mbox{ps}^{-1}, \qquad\qquad
      \dg_D =\;  - \lt( 0.01572 \epm{0.00058}{0.00056} \rt) \, \mbox{ps}^{-1}
\label{dmixexp} 
\end{align}
We notice two important points here: Firstly, $\dm_D$ is larger than our
naive estimate by 4 orders of magnitude. For $|\dg_D|$ one finds the
same order-of-magnitude discrepancy between box diagram estimate and
data \cite{Bobrowski:2010xg}. Secondly, $\sgn (\dm_D/\dg_D)$ is
negative, opposite to what is observed in \kkm\ and \bbms\ and what the
SM predicts for \bbmd!

To understand the failure of the box diagram calculation I discuss the
contribution with the large CKM factor $\lt(V_{us}V_{cs}^*\rt)^2$.
The smallness of the result of the box diagram Results from four
contributions which are of order 1 and combine to a result of order
$m_s^4/(m_c^2 M_W^2)$. % Now, by symmetry consideration,
The box diagrams
vanish in the U-spin symmetry limit $m_d=m_s$. The
$\lt(V_{us}V_{cs}^*\rt)^2$ piece of the \ddm\ is second order in the U-spin
breaking  parameter $m_s-m_d$. Indeed, if we calculate the box diagram 
with $m_s\neq 0\neq m_d$, we find a result proportional to 
\begin{align}
   \lt( m_s^2-m_d^2 \rt)^2 &=\; \qquad 
                             \underbrace{(m_s -m_d )^2} \quad \cdot
                             \quad  \underbrace{(m_s + m_d )^2}\nn
  &\qquad\quad\;\;  \parbox[t]{1.3cm}{\centering U-spin\\  \centering breaking} 
      \parbox[t]{3.0cm}{\centering{artifact of}\\ 
                            \centering{perturbation theory}}  
\end{align}  
Thus indeed, the U-spin breaking is correctly reproduced, but the
additional suppression by $(m_s + m_d )^2/m_c^2$ is an artifact of the
calculational method, originating from the fact that the $W$ boson only couples
to left-chiral fields and one needs an even number of chirality flips on
each quark line.
% In Nature, left-right flips can also come from a
% non-perturbative object of QCD called quark condensate.
There is no theoretical reason why the $\Delta C=2$ transition
requires any chirality flip at all. 
So we are lead to consider alternatives to the box diagram.  In
Refs.~\cite{Georgi:1992as,Ohl:1992sr} it was pointed out that
contributions in which one or both of the internal $s$ or $d$ lines of
the box diagram are cut, so that they become external lines, (and
eventually a gluon is added to get a connected diagram), the artificial
suppression factor $(m_s + m_d )^2/m_c^2$ can be avoided and the
SU(3)-breaking could reside in a hadronic matrix element. To date, the
proposed contributions cannot be reliable calculated.  In summary, the
theory community was unable to predict the results in \eq{xyexp} and
further did not come up with convincing postdictions.

The result in \eq{xyexp} nevertheless teaches us about possible BSM
explanations. BSM physics which contributes to the \ddm\ amplitude 
through new box or even tree-level diagrams with heavy particles
typically contribute much more to $x_D$ than to $y_D$ and it is unlikely
that $x_d \sim y_D$ originates from BSM physics in $x_D$ and
long-distance QCD dynamics in $y_D$. More convincingly, BSM physics
comes with new complex couplings. Recall that the standard form of $V$
with real $V_{ud}$, $V_{us}$ is the result of the rephasings in \eq{fph} 
and these rephasing will also appear in the couplings of hypothetical BSM
particles. Thus, by default, a dominant BSM contribution to $x_D$ would
give ${\cal O}(1)$ mixing-induced $CP$ asymmetries in e.g.\ $\DorDbar \to
K^{\pm}\pi^{\mp}$ decays, which are not observed. This means that BSM
explanations of \eq{xyexp} must have a mechanism which suppresses the
effect in $x_D$ to the level of $y_D$ and further aligns new complex phases
such that  $CP$ violating observables are suppressed.
The eventual discovery of BSM physics 
will clearly come from $CP$ asymmetries and not from better
measurements of $x_D$ and $y_D$.

As discussed in the previous sections, we use \kkm\ and \bbm\ to study
the interference effects in $M \to f$ and $\bar M \to f$ decays. Since
\dd\ oscillations are very slow, most of the $\DorDbar$ sample has
decayed once a sizable superposition of $\ket{D}$ and $\ket{\Dbar}$ has
evolved. But for such studies one can also use CP-tagged mesons, the CP
eigenstates are $\ket{D_{CP\pm}}=(\ket{D} \mp \ket{\Dbar})/\sqrt2$ and
are thus maximal superpositions of $\ket{D}$ and $\ket{\Dbar}$.
CP-tagged states are used at BES III to determine the strong phase in
$D\to K^\pm \pi^\mp$. There are ways to prepare such states even at
hadron colliders \cite{Naik:2021rnv}. $D_{CP\pm}\to f$ decays are not
sensitive to the \dd\ box diagram, they probe $CP$ eigenstates rather
than mass eigenstates. Similarly to mixing-induced $CP$ asymmetries,
$D_{CP\pm}\to f$ decays involve $CP$-violating observables
which do not need a non-vanishing strong phase.

\section{Time evolution of neutral mesons and associated
  CP-violating quantities}\label{sec:time}
In this section I will derive the formulae needed to describe \mmm\ and the
CP-violating observables related to it,  with emphasis on the time
evolution of neutral meson states. Compared to the previous sections the
presentation is more technical but is nevertheless of interest to both
the theoretical and experimental community. Especially, presented  aspects of the
CP asymmetry in flavor-specific decays,
$a_{\rm fs}$, which is still to be discovered in $B_{d,s}$  and $D$
decays, might be helpful to devise future measurements. The formalism of
the section is general and applies to the SM as well as any BSM
theory. SM predictions will be presented in Sec.~\ref{sec:sm}.

\subsection{Time evolution}
In order two understand the QFT formalism of the weak interaction of hadrons in
general and of \mmm\ in particular, we first seek a description of fields
and states which permits the application of perturbation theory to the
electroweak interaction, which involves small coupling constants in
which our amplitudes of interest  can be expanded.  In fact, for FCNC
transitions the lowest non-vanishing order in the weak coupling $g_w$ is
sufficient, e.g.\ \mmm\ occurs first at order  $g_w^4$ (box diagrams)
and corrections from diagrams with an additional electroweak boson are
negligible in view of QCD uncertainties.  In a first step the lagrangian
of any QFT is expressed in terms of fields and states in the Heisenberg
picture, meaning that the quantum fields depend on the space-time variable
$x^\mu$, while the states (describing the particles) are time-independent.
One then splits the hamiltonian density $H$ as
\begin{align}
H&=\; H_0 + H_{\rm int} \label{hint}
\end{align}
with $H_{\rm int} = -L_{\rm int}$ containing the interaction terms with
the small couplings in which we want to expand in our calculations. In
the second step one applies a unitary tranformation to fields and states
to arrive at the interaction picture (a.k.a.\ Dirac picture). As a result the states
$\ket{\psi(t)}$ become time-dependent and one encounters  a time evolution operator
${\cal U}(t_2,t_1)$,
\begin{align}
   \ket{\psi(t_2)} &=\; {\cal U}(t_2,t_1)   \, \ket{\psi(t_1)},
                     \qquad\qquad
                     \mbox{where }\quad
                      {\cal U}(t_2,t_1) = \mathrm{T}\!\exp \lt[-i \!
                     \int_{t_1}^{t_2} dt \! \int \, d^3 \vec x \, H_{\rm
                     int}  \rt] \label{defu}   .
\end{align}
Here $\mathrm{T}\!\exp$ is the time-ordered exponential and $H_{\rm int}$
depends on $x^\mu=(t,\vec x)$ through the fields.  In the calculation of
a $2\times 2$ scattering process, where $\ket{\psi}$ is a two-particle state,
one encounters the S-matrix $S= {\cal U}(\infty,-\infty) $ and
conveniently applies covariant perturbation theory in momentum space to
calculate the cross section (or other observables of interest) from the
S-matrix elements in terms of the momentum four-vectors $p_j^\mu$ of
the in-going and out-going particles. For the treatment of \mmm\ and the
time-dependent decay rates in \eq{defgtf} we need to depart from the
described standard path in two aspects: Firstly, we must keep the time
in the formalism, since we need ${\cal U}(t_2,t_1)$ for
$t_2\neq \infty \neq -t_1$. This means that we cannot trade $t$ for the
energy $E=p^0$ with the usual Fourier transform w.r.t.\ to time. Starting from
the formalism of covariant perturbation theory we can Fourier-transform back to a world
in which states are labeled as $\ket{\vec p, t}$ which gives us the
``old-fashioned time-dependent perturbation theory'' employed in early
days when QFT emerged from quantum mechanics. Secondly, in perturbation
theory applied to processes of leptons and bosons (such as QED) one chooses
$H_0$ in \eq{hint} as the free hamiltonian. In perturbation theory the
initial and final states of a scattering or decay process are
constructed from $H_0$, accounting for the fact that sufficiently widely
separated particles (forming so-called \emph{asymptotic states}) behave
like free particles. In our cases of interest, we encounter hadrons as
asymptotic states, which are bound states of the strong interaction.
Not surprisingly, since QCD is non-perturbative at large distances, we
cannot apply perturbation theory in terms of the strong coupling
$\alpha_s$ from the start. Thus $H_0$ in \eq{hint} is $H_0=H_{\rm QCD}$,
which contains both the quadratic pieces of the free hamiltonian and all
interaction terms with the gluon field. That is, our fields and states
are still in the Heisenberg picture w.r.t.\ QCD. In particular, the
state vectors are eigenstates of $\widehat H_0=\widehat H_{\rm QCD}=
 \int d^3\vec x \, H_{\rm QCD}$. For example, in a
weak two-body meson  decay $M \to h_1 h_2$ into ground-state hadrons $h_{1,2}$
the asymptotic states for initial and final state satisfy
\begin{align}
  \widehat H_{\rm QCD} \ket{M(\vec{p}_M)} &= \;  E_M \ket{M(\vec{p}_M)} \;=\;
                                  \sqrt{M_M^2+ \vec{p}{}_M^2}\,
                                   \ket{M(\vec{p}_M)},\nn
 \widehat H_{\rm QCD} \ket{h_1(\vec{p}_1)\, h_2(\vec{p}{}_2) }
                                   &=\; (E_1+E_2) \,
    \ket{h_1(\vec{p}_1)\, h_2(\vec{p}_2)} \;=\;
                                   \lt(\sqrt{M_1^2+ \vec{p}_1^{\,2}}+\sqrt{M_2^2+\vec{p}_2^{\,2}}\rt)
                                   \,
    \ket{h_1(\vec{p}_1)\, h_2(\vec{p}_2)} \label{hqcdhei}
\end{align}
and, in our decay, $E_M=E_1+E_2$.  Since the time dependence stems from
$H_1= H_{\rm int}$ and asymptotic states (describing $M$ before the
decay and $h_{1,2}$ when they hit the detector (or, in a cascade decay,
at the time when they decay to other particles) are obtained from
$H_0=H_{\rm QCD}$, there is no time-dependence yet in \eq{hqcdhei}.
Ground state hadrons are those which do not decay through the strong
interaction, this characterization applies to our decaying mesons $K$,
$B_{d,s}$, and $D$, as well as to charged or neutral $\pi$, $K$, and
$\eta^{(\prime)}$ in the final states. A vector meson like $K^*$ is not
in this category, the strong decay $K^*\to K \pi$ is instantaneous and
the asymptotic state is $K\pi$ in this case.  In fact, there are
observables in which one exploits the interference of
$K^{*0}\to K_{\rm short} \pi^0$ and $\bar K^{*0}\to K_{\rm short} \pi^0$ in a decay chain,
thus the neutral vector meson is clearly not an asymptotic state
observed in a detector. The most prominent example of a quantity using
the shown interference is
$ A_{CP}^{\rm mix}(B_d\to J/\psi \KorKbar{}^{*0} [\to K_{\rm short} \pi^0])$.

To derive the desired formula for the time-evolution of a weakly
decaying meson, we start with the case of a charged meson $M^+$, for
which no flavor oscillations occur. In the interaction picture, the 
ket $\ket{M^+}$ of a meson produced at time $t=0$ evolves in time as
\begin{eqnarray} 
  {\cal U}(t,0) \ket{M^+} &=&  \ket{M^+ (t)} + 
            \sum_f \ket{f} \bra{f}  {\cal U}(t,0) \ket{M^+} ,
                              \label{chevol}
\end{eqnarray}
with 
\begin{eqnarray} 
   \ket{M^+ (t)} &\equiv & \ket{M^+} \bra{M^+} {\cal U}(t,0) \ket{M^+}
\label{defmt}
\end{eqnarray}
describing the situation that $M^+$ has not decayed at the considered
time $t>0$. The second term in \eq{chevol} involves the sum over all
final states $\ket{f}$ into which $M^+$ can decay.  In the description
of time-dependent decay processes it is further common to switch from the
interaction picture kets $\ket{\ldots}_{I}$ to the Schr\"odinger picture
$\ket{\ldots}_{S}$ defined as
\begin{align}
  \ket{\psi}_{S} &\equiv \; e^{-i \widehat H_0 t} \ket{\psi }_{I}
                     \;=\; e^{-i E_\psi t} \ket{\psi }_{I}. 
\end{align}
so that the one-particle state $ {\cal U}(t,0) \ket{M^+}$ in \eq{chevol}
picks up an extra factor of $\exp (-i E_{M^+} t)$ in the Schr\"odinger
picture. We do not transform the fields, e.g.\  $ {\cal U}(t_2,t_1)$
is not changed in the $I\to S$ transformation. 
By energy conservation these factors of $\exp (-i E t)$ drop out between bras and
kets, so that ${}_S\langle \ldots \rangle_S = {}_I\langle \ldots
\rangle_I $. We can determine $\ket{M^+ (t)}$ in \eq{chevol} 
by  employing the exponential decay law to deduce 
\begin{eqnarray}
   \ket{M^+ (t)} &=& e^{-i M_M t} e^{-\Gamma t/2} \ket{M^+}
  \label{explaw}  
\end{eqnarray}
in the meson rest frame.  The first term is the familiar time evolution
factor of a stable state with energy $E=M_M$. We understand the second factor
involving the total width $\Gamma$ by considering the
probability to find an undecayed meson at time $t$:
\begin{eqnarray}
\lt| \langle M^+ \ket{M^+ (t)}\rt|^2
   &=& e^{-\Gamma t} \no
\end{eqnarray}
Here I have normalized the states as $\langle M^+\ket{M^+}=1 $. 

\begin{figure}[t]
\centering 
\includegraphics[scale=0.35]{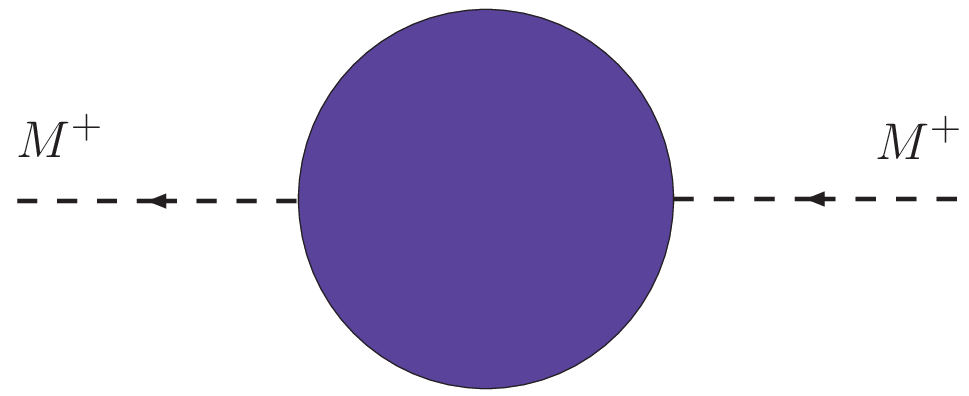} 
\hspace{2cm}
\includegraphics[scale=0.35]{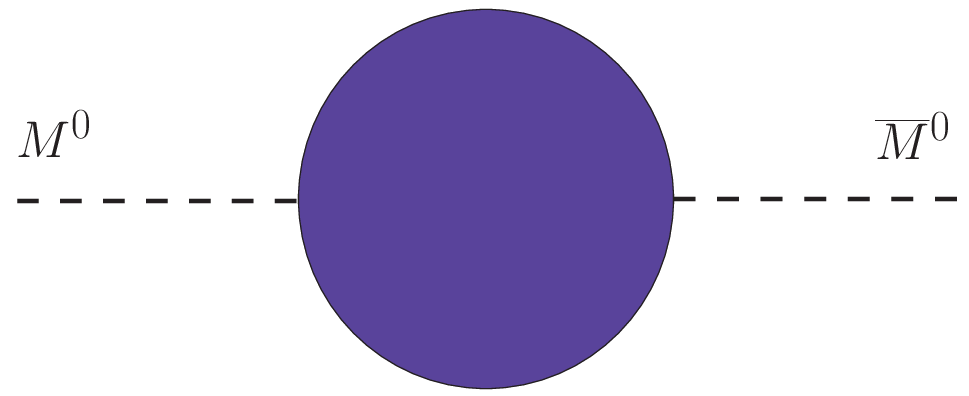} 
\caption{Left: generic self energy $\Sigma$ of a charged meson. Right: 
$M^0\!-\bar{M}{}^0$ mixing amplitude $\Sigma_{12}$.  
\label{fig:self}}
\end{figure}

Since $M_M-i \Gamma/2$ is independent of $t$, we can compute it using
the familiar covariant formulation of quantum field theory and in the
following calculations  I comply with the
standard relativistic normalization of the meson states,
\begin{align}
\bra{M(\vec
  p\,{}^\prime)} M(\vec p )\rangle &=\; 2E \, (2\pi)^3 \delta^{(3)} (\vec
  p\,{}^\prime-\vec p ). \label{norm} 
\end{align}
The optical
theorem tells us that $M_M$ and $-\Gamma/2$ are given by the real and
imaginary parts of the self-energy $\Sigma$ (depicted in the left
diagram of \fig{fig:self}), where%
\bea%
-i (2\pi)^4 \delta^{(4)}(\vec p\,{}^\prime-\vec p ) \Sigma &=&
\frac{\bra{M^+(\vec p\,{}^\prime)} S \ket{M^+ (\vec p)} }{2 M_M}%
\label{defsic}
\eea%
This defines $\Sigma$ in the interaction picture, in the Schr\"odinger
picture we must add the mass $M_M$, i.e.\ the ``tree-level self-energy'',
to $\Sigma$. 
The factor of $1/(2 M_M)$ in \eq{defsic} originates from the
normalization in \eq{norm}. The truncated self-energy diagram of a boson
has mass dimension 2 and by dividing by $2 M_M$ we arrive at the correct
mass dimension 1 for  $\Sigma=M_M-i\Gamma/2$. 
From \eq{explaw} we find
\begin{eqnarray}
 i \frac{d}{d\, t}  \ket{M^+ (t)} &=& 
    \Big( M_M-i \frac{\Gamma}{2}\, \Big) \ket{M^+ (t)} . \label{tevc} 
\end{eqnarray}
Now this equation can be generalized to a two-state system describing
neutral meson mixing. We may view $(\ket{K},\ket{\Kbar})$ as a two
component object with strangeness (or U-spin quantum number $U_3=\pm 1$) 
as an inner degree of freedom distinguishing these components. Then
\begin{eqnarray}
 i \frac{d}{d\, t} \left( \!
\begin{array}{c}
\ds \ket{M (t)} \\[1mm]
\ds \ket{\,\bar{M} (t)}
\end{array}\! \right) &  = & 
\Sigma 
\left(\!
\begin{array}{c}
\ds \ket{M(t)} \\[1mm]
\ds \ket{\,\bar{M}(t)}
\end{array}\!\right) \label{schr}
\end{eqnarray}
where now $\Sigma$ is the $2\times 2$ matrix defined as
\begin{eqnarray}
  -i (2\pi)^4 \delta^{(4)}(p_i^\prime- p_j)  \Sigma_{ij} 
     &=& \frac{\bra{i, \vec p_i{}^\prime} 
               S^{\rm SM}
               \ket{j, \vec p_j} 
              }{2 M_M} \label{defaij} 
\end{eqnarray}
in the interaction picture with $\ket{1, \vec p_1}= \ket{ M (\vec p_1)}$ and $\ket{2, \vec p_2}=
\ket{\,\bar{M} (\vec p_2)}$. In the Schr\"odinger picture used in \eq{schr}
$\Sigma$ gets an extra additive term and is to be read as
$\Sigma_S=\Sigma_I + M \cdot 1$ with the $2\times 2$ unit matrix $1$. 
Since the shift between ``$I$'' and ``$S$'' is trivial and only affects
equations describing time evolutions, I omit the corresponding
index. 

Recalling that any matrix can be written 
as the sum of a hermitian and an antihermitian matrix, we write
\begin{eqnarray}
%\Sigma &=& \left( M - i\, \frac{\Gamma}{2}  \right) \label{defmg}
\Sigma &=& M - i\, \frac{\Gamma}{2} \label{defmg}
\end{eqnarray}
with the \emph{mass matrix}\ $M=M^\dagger$ and the \emph{decay matrix}\
$\Gamma=\Gamma^\dagger$. Then
\begin{eqnarray}
  M_{12} &=& \frac{\Sigma_{12}+\Sigma_{21}^*}{2}, \qquad  \qquad  \qquad  
  \frac{\Gamma_{12}}{2} \;=\; i \, \frac{\Sigma_{12}-\Sigma_{21}^*}{2}. 
\label{absdisp}
\end{eqnarray}
Again, $\Sigma_{12}$ and the quantities derived from it here and below
are different for the four cases $M=K$, $B_{d,s}$, and $D$ and I drop the
corresponding index. 
 
The expressions on the RHS of \eq{absdisp} are called \emph{dispersive}\
and \emph{absorptive}\ parts of $\Sigma_{12}$, respectively.  The right
diagram in \fig{fig:self} generically represents all contributions to
$\Sigma_{12}$. The operational definitions of dispersive and absorptive
parts, mentioned already after \eq{dgexc}, amount to factoring out all
CKM elements (or other complex couplings when BSM theories are
considered) and taking the real and imaginary part of the remainder for
dispersive and absorptive part, respectively. To shed light on this we
discuss the \bbd\ box diagram in \fig{fig:boxes} and write the \bbd\
amplitude as $\Sigma_{12}=(V_{tb}V_{td}^*)^2 A_{\rm rest}$. This
quantity describes a
$\ket{2}\equiv\ket{\Bbar} \to \ket{1}\equiv \ket{B}$ transition in which
a $b$ quark enters the diagram and a $\bar b$ quark leaves it.  Now
$\Sigma_{21}$ in \eq{absdisp} instead describes the
$\ket{B} \to \ket{\Bbar}$ transition which involves the box diagram in
which the direction of all quark lines are reversed, so that the CKM
factor is complex-conjugated w.r.t.\ $\Sigma_{12}$. The remainder of
the amplitude is the same, because  $\Sigma_{12}$ and $\Sigma_{21}$ 
are related by $CP$ conjugation and there are no sources of $CP$ violation
beyond the CKM
factors (or, in the case of a BSM theory, the complex coupling constants
factored out). Thus $\Sigma_{21}=(V_{tb}^*V_{td})^2 A_{\rm rest}$ and we
realize that  $\Sigma_{21}^*$ appearing in the definitions of
dispersive and absorptive parts in \eq{absdisp} are both proportional to
$(V_{tb}V_{td}^*)^2$. Thus 
\begin{eqnarray}
  M_{12} &=& (V_{tb}V_{td}^*)^2\,  \real A_{\rm rest}
             \qquad  \qquad  \qquad  
             \frac{\Gamma_{12}}{2} \;=\;
             - (V_{tb}V_{td}^*)^2\,  \imag A_{\rm rest}
\label{mga12} .
\end{eqnarray}
From the optical theorem we know that $\imag A_{\rm rest} \neq 0$
requires contributions from on-shell intermediate states, so that only
box diagrams with internal $u,c$ quarks contribute to $\Gamma_{12}$ for
\bbmd.

% To compute $\Sigma_{12}$ we can certainly use
% perturbation theory for the weak interaction (which to lowest order
% amounts to the calculation of the box diagram in \fig{fig:boxes}), but
% we must take into account the non-perturbative nature of the strong
% binding forces.
The diagonal elements $M_{11}$ and $M_{22}$ are the (equal) masses of
$M$ and $\bar M$ and are generated from the quark mass terms in the
lagrangian $L$ and from the binding energy of the strong
interaction. Had we stayed in the interaction picture, these diagonal
elements would only contain $\Sigma_{11}=\Sigma_{22}$, i.e.\ the
electroweak contributions to the meson self-energy, which are tiny
electroweak corrections to the meson masses and completely negligible.
Thus the extra term $M_M \delta_{jk}$ in the Schr\"odinger picture puts
the full meson mass into the diagonal elements of the mass matrix.

By contrast, the off-diagonal elements $M_{12}= M_{21}^*$ and all
elements of $\Gamma$ stem from the weak interaction and are therefore
tiny in comparison with $M_{11}$ and $M_{22}$. The only reason why we
can experimentally access $M_{12}$ roots in the $CPT$ theorem
\cite{Luders:1954zz,pauli:1955,Luders:1957bpq}: 
Applying $CPT$ to \eq{defaij}, which maps $\Sigma_{11}\leftrightarrow
\Sigma_{22}$, one finds
\begin{align}
  M_{11} &\;=\; M_{22}, \qquad\qquad   
  \Gamma_{11} \;=\;  \Gamma_{22}, 
\label{cptmg}
\end{align}
so that the eigenvalues of $\Sigma$ are exactly degenerate for 
$\Sigma_{12}=\Sigma_{21}=0$. Even the smallest $\Sigma_{12}$ can lift
the degeneracy and can lead to large \mmm. 

The presented derivations of \eqsand{explaw}{schr} were developed in
\cite{Nierste:2009wg} and use a shortcut: I have avoided to prove that \eq{tevc}
holds with time-independent $M$ and $\Gamma$ and instead used the
phenomenological input that we know the time-evolution of a decaying
particle.  However, \eq{tevc} and the equivalent equation \eq{explaw}
with the exponential decay law are not valid exactly, but receive tiny
(and phenomenologically irrelevant) corrections \cite{Khalfin:1957}. The same
statement is true for \eqsand{schr}{defmg}, a proper derivation of
\eq{schr} using time-dependent perturbation theory for the weak
interaction employs the so-called \emph{Wigner-Weisskopf approximation}\
\cite{Weisskopf:1930au,Lee:1957qq}. Corrections to this approximation have been addressed in
Ref.~\cite{Chiu:1990cm} and are below the $10^{-10}$ level.

\eq{schr} is sometimes referred to as a ``Schr\"odinger equation'',
while the correct phrasing is ``time evolution equation in the
Schr\"odinger picture''. A Schr\"odinger equation involves a
hamiltonian, but $M-i\Gamma/2$ is not (the matrix representation of)
a hamiltonian, because it is not hermitian. It is also not some piece of
a hamiltonian acting in a 2-dimensional subspace of the full
infinite-dimensional Fock space of particle physics,
because the exponential decay described by $\Gamma$ is an
effective description of the $M\to f$ transitions in \eq{chevol}.

I will now present the solution of \eq{schr}.
% \eq{defpq} means that the
% eigenvectors of $\Sigma$ in \eq{defaij} are $(p,q)^T$ and $(p,-q)^T$.
We can diagonalize $\Sigma$ as
\begin{eqnarray}
  Q^{-1} \Sigma \,Q &=& \lt( 
\begin{array}{cc} 
 M_L - i \Gamma_L/2 & 0 \\ 
 0 & M_H - i \Gamma_H/2 
\end{array} \rt)  \label{qsq}
\end{eqnarray}
with 
\begin{eqnarray}
  Q&=& \lt( 
\begin{array}{rr} 
p & p \\ q & - q \\
\end{array} \rt) \qquad \mbox{and} \qquad 
  Q^{-1} \;=\; \frac{1}{2 pq}  \lt( 
\begin{array}{rr} 
q &  p \\ q & - p \\
\end{array} \rt)   \label{defq}
\end{eqnarray}
and $\lt|p\rt|^2+\lt|q\rt|^2 = 1$.  The ansatz in \eq{defq} (with just
two coefficients $p$, $q$) works because of $\Sigma_{11}=\Sigma_{22}$.
Thus the eigenvectors of $\Sigma$ in \eq{defaij} are the columns of $Q$,
$(p,q)^T$ and $(p,-q)^T$, which leads to  the mass eigenstates \cite{Lee:1957qq}
\begin{align}
\ket{M_L} &=\;
    p \ket{M} + q \ket{\Mbar} \,,  \qquad\qquad
\ket{M_H} = \;
    p \ket{M} - q \ket{\Mbar}   .
\label{defpq}
\end{align}
$\ket{M_{L,H}(t)}$ obey an exponential decay law like
$\ket{M^+ (t)}$ in \eq{explaw} with $(M_M,\Gamma)$ replaced by
$(M_{L,H},\Gamma_{L,H})$.  Transforming back to the flavour basis gives%
\bea%
\lt( \!\begin{array}{c} 
\ket{M (t)} \\[2mm] 
\ket{\,\bar{M} (t)} 
\end{array}\! \rt) 
&=& Q \, \lt(
\begin{array}{cc} 
\ds e^{-i M_Lt -  \Gamma_L t/2} & \ds 0 \\[2mm] 
\ds  0 & e^{-i M_H t - \Gamma_H t/2} 
\end{array} \rt)\, Q^{-1} 
\lt( \! \begin{array}{c} 
\ket{M } \\[2mm] 
\ket{\,\bar{M} } 
\end{array} \!\rt)
\label{tes}
\eea%
The average mass $m$ and average width $\Gamma$ have been defined in
\eq{avgmg} and the definitions of the mass and width difference can be
found in \eq{defdmdg}. 
The matrix appearing in \eq{tes} can be compactly written as 
\begin{equation}
 Q \, \lt( 
\begin{array}{cc} 
\ds e^{-i M_L t- \Gamma_L t/2} & \ds 0 \\[2mm] 
\ds  0 & e^{-i M_H t- \Gamma_H t/2} 
\end{array} \rt) \, Q^{-1} \;=\; 
\lt( 
\begin{array}{rr}
\ds g_+(t)              &\ds \frac{q}{p} g_-(t) \\[2mm]  
\ds \frac{p}{q} g_-(t)  &\ds g_+ (t)  
\end{array} \rt) \label{qlhq}
\end{equation}
with%
\bea%
 g_+ (t) &=& e^{-i m t} \, e^{-\Gamma t/2} \lt[ \phantom{-} \cosh\frac{\dg \,
  t}{4}\, \cos\frac{\dm\, t}{2} - i \sinh\frac{\dg \, t}{4}\, \sin \frac{\dm
  \, t}{2} \, \rt] , \nn g_- (t) &=& e^{-i m t} \, e^{-\Gamma t/2} \lt[-
\sinh\frac{\dg \, t}{4}\, \cos \frac{\dm \, t}{2} + i \cosh\frac{\dg \,
  t}{4}\, \sin \frac{\dm \, t}{2} \, \rt] .
\label{gpgm}
\eea%
Inserting \eq{qlhq} into \eq{tes} gives us the desired expression for 
\mm\ oscillations: 
\bea%
\ket{M (t)} &=& \phantom{\frac{p}{q}\,}
  g_+ (t)\, \ket{M} + \frac{q}{p}\, g_- (t)\, \ket{\,\bar{M}} \,, \nn
\ket{\,\bar{M} (t)} &=& \frac{p}{q}\, g_- (t)\, \ket{M}
  + \phantom{\frac{q}{p}\,} g_+(t)\, \ket{\,\bar{M}} \,,
  \label{tgg}
\eea%
  We verify $g_+(0)=1$ and $g_-(0)=0$ and find that $g_\pm (t)$ has no
  zeros for $t>0$ if $\dg\neq 0$. Hence the two lifetimes $1/\Gamma_L$
  and $1/\Gamma_H$ lead to a dispersion, so that an initially produced
  $M$ will never turn into a pure $\bar M$ or back into a pure $M$.

We will frequently encounter the combinations%
  \bea%
  | g_\pm (t) |^2 & = & \frac{e^{- \Gamma t}}{2} \lt[ \phantom{-} \cosh
  \frac{\Delta \Gamma \, t}{2} \pm \cos\lt( \dm\, t \rt) \rt] , \nn
  g_+^* (t)\, g_- (t) & = & \frac{e^{- \Gamma t}}{2} \lt[ - \sinh
  \frac{\Delta \Gamma \, t}{2} + i \sin \lt( \dm\, t \rt) \rt] .
\label{gpgms}
\eea%
% The coefficients are, of course, different for the four considered
% neutral meson systems, and the notation $p_M$, $q_M$ with $M=K$,
% $M=B_{d,s}$, or $M=D$, will be used where needed to avoid
% confusion. $p$ and $q$ will be calculated in terms of the 
% box diagrams in \fig{fig:boxes} below. 
Note that $M-i\Gamma/2$ is not a hermitian matrix, so that we cannot
expect $Q$  linking $(\ket{M},\ket{\Mbar})$ to $(\ket{M_L},\ket{M_H})$
to be unitary. As a consequence, the mass eigenstates need not be
orthogonal to each other, that is 
\begin{align}
  \langle M_L \ket{M_H} &=\;    |p|^2-|q|^2 \label{mlmh}
\end{align}
need not vanish. We can use this observation to find a criterion for
$CP$ violation. If $CP$ is conserved in  \mmm, then $\ket{M_{L,H}}$ are
eigenstates of the $CP$ operator with different eigenvalues $\pm 1$.
Since the $CP$ operator is unitary, its eigenvectors are  orthogonal
and in this case therefore \eq{mlmh} must vanish. Thus we conclude 
that $|q/p| \neq 1$ implies that $CP$ is violated in $|\Delta F|=2$
transitions. This phenomenon is called \emph{$CP$ violation in mixing}\
and should not be confused with the mixing-induced $CP$ violation
defined by $A_{CP}^{\rm mix}\neq 0 $ in \eq{acpbd}. The latter quantity
is specific to the studied $M\to f$ decay mode, while  $CP$ violation in
mixing is a universal effect, which only depends on $|q/p|$ and
therefore affects \emph{all}\ decays of $M$.

$CPT$ transforms the state $\alpha \ket{M(\vec p)}+\beta \ket{\Mbar
(\vec p)} $ to $\alpha^* \ket{\Mbar (\vec p)}+\beta^* \ket{M
(\vec p)} $ times an arbitrary phase factor. Thus $\ket{M_{L;H}}$ are not
$CPT$ eigenstates for $|p| \neq |q|$. Since $CPT$ is a good symmetry of
the SM, this looks strange. But there is no contradiction here, because
$\ket{M_{L;H}}$ are eigenstates of $M-i \Gamma t/2$, which is \emph{not}\
a hamiltonian. 

\boldmath
\subsection{\dm, \dg, and $CP$ violation in mixing}
\unboldmath
I will next solve the eigenvalue problem in \eq{qsq} to express $\dm$,
$\dg$, $p$ and $q$ in terms of $M_{12}$ and $\Gamma_{12}$. 

The secular equation for the two eigenvalues
$\sigma_{L,H}=M_{L,H}-i\Gamma_{L,H}/2$ of $\Sigma$ is
$(\Sigma_{11}-\sigma_{L,H})^2-\Sigma_{12}\Sigma_{21} =0$. The two
solutions of this equation therefore satisfy %
\bea%
\lt(\sigma_H -\sigma_L\rt)^2 &=& 4\,\Sigma_{12}\Sigma_{21} \no \eea%
or \bea%
( \dm + i \frac{\dg }{2} )^2 \;= \; 4\, \lt( M_{12} - i
\frac{\Gamma_{12}}{2} \rt) \lt( M_{12}^* - i \frac{\Gamma_{12}^*}{2}
\rt).  \eea%
Taking real and imaginary part of this equation gives%
\bea%
\lt( \dm \rt)^2 - \frac{1}{4} \lt( \dg \rt)^2 &=& 4 \lt| M_{12} \rt|^2 -
\lt| \Gamma_{12} \rt|^2 \,,
  \label{mgqp:a} \\[2mm]
\dm\, \dg &=& -4\, \real ( M_{12} \Gamma_{12}^* ) \,,
  \label{mgqp:b} 
\eea%
From \eq{qsq} we further infer $[Q^{-1} \Sigma Q]_{12}=[Q^{-1} \Sigma
Q]_{21}=0$, which determines 
\bea%
\frac{q}{p} &=& - \frac{\dm + i \, \dg/2}{2 M_{12} -i\, \Gamma_{12} }
  = - \frac{2 M_{12}^* -i\, \Gamma_{12}^*}{\dm + i \, \dg/2} \,.
\label{mgqp:c}
\eea%
(The second solution, with opposite sign, is discarded by imposing
$\dm >0$.)  For the simplification of \eqsto{mgqp:a}{mgqp:c} it is
useful to identify the physical quantities of the mixing problem in
\eqsand{schr}{defmg}. Rephasing 
$\ket{M}$ or $\ket{\,\bar{M}}$ cannot change 
the physics, but it changes the phases of $M_{12}$,
$\Gamma_{12}$ and $q/p$, none of which can therefore have any physical
meaning. Thus only% 
\bea%
&& |M_{12}|, \qquad\qquad |\Gamma_{12}|, \qquad\quad \mbox{and} \quad
\phi \equiv \arg \left( -\frac{M_{12}}{\Gamma_{12}} \right).
        \label{defphi} %
\eea%
can be  physical quantities of \mmm.
\eq{mgqp:b} then reads 
\bea%
\dm\, \dg &=& 4\, |M_{12}|  |\Gamma_{12}| \cos\phi .
\label{mgqp:d}
\eea%
We can easily solve \eqsand{mgqp:a}{mgqp:d} to express $\dm$ and $\dg$,
which we want to confront with the measurements in
Eqs.~{(\ref{dgkexp})}, {(\ref{dmkexp})}, {(\ref{wavdmd})},
(\ref{dgdexp}), {(\ref{wavdms})}, {(\ref{dgsexp})}, and
{(\ref{dmixexp})}, in terms of the theoretical quantities $|M_{12}|$,
$|\Gamma_{12}|$ and $\phi$.

Before doing so, we recognize that a non-vanishing phase $\phi$ is
responsible for $|q/p| \neq 0$ identified after \eq{mlmh} as a criterion
for $CP$ violation: 
By multiplying the two expression for $q/p$ in \eq{mgqp:c}
with each other we find
\bea%
\lt( \frac{q}{p}\rt)^2 
 &=& \frac{2 M_{12}^* -i\, \Gamma_{12}^*}{2 M_{12} -i\, \Gamma_{12} }
 \; =\;  \frac{M_{12}^*}{M_{12}} \, 
      \frac{\ds 1 + i \lt|\frac{\Gamma_{12}}{2 M_{12}}\rt| e^{ i  \phi\;\,}}
           {\ds 1 + i \lt|\frac{\Gamma_{12}}{2 M_{12}}\rt| e^{ -i \phi}}
\label{mgqp:e} .
\eea%
and 
this expression shows that $\phi\neq 0,\pi$ indeed
implies $|q/p|\neq 1$, i.e.\ $CP$ violation in mixing.

Interestingly, $CP$ violation in mixing is found small (i.e.\
$\lt||q/p|-1\rt|\ll 1$) for all of the four $K$, $B_{d,s}$ and $D$
systems, to date it is only measured non-zero in \kkm, while otherwise
only upper bounds have been experimentally determined. I will discuss
the experimental situation below in more detail. 
% In the case of \kkm\ we have established this
% phenomenon in \eq{ekqp} from the measured value of $\real \e_K$ in
% \eq{ekexp}.
The smallness is expected in the SM: \kkm\ and \ddm\ are dominated by
the $2\times 2$ Cabibbo matrix $V_C$ and $CP$ violation requires
contributions which are sensitive to all three fermion generations.
(With the real $V_C$ of the two-generation SM, one immediately finds
$M_{12}$ and $\Gamma_{12}$ real in \eq{mga12}.)  The ``leakage'' to the
third generation from $V_{ts}V_{td}^*\neq 0$ and $V_{ub}V_{cb}^*\neq 0$
is small, suppressing $\sin \phi$.  In the \bb\ systems the line of
arguments is as follows: Since $\Gamma_{12}$ gets no contributions from
the box diagram with top quarks, which in turn dominates $M_{12}$, we
find $|\Gamma_{12}/M_{12}|= {\cal O}(m_b^2/m_t^2)$. Even BSM physics
cannot change this, because the decays contributing to $|\Gamma_{12}|$
are experimentally studied well enough to exclude order-of-magnitude
enhancements.  Thus the second term in the numerator and denominator of
\eq{mgqp:e} is small, irrespective of the value of $\phi$, and
$|q/p|\simeq 1$ for $B_d$ and $B_s$ mesons.

It is useful to define the quantity $a_{\rm fs}$ through
\begin{eqnarray}
\lt| \frac{q}{p} \rt|^2 &=& 1 -a_{\rm fs} . \label{defa} 
\end{eqnarray}
From the discussion above we understand that  $a_{\rm fs}$ quantifies
$CP$ violation in \mmm. 
For all neutral meson complexes we know that $a_{\rm fs}$ is very small from the
experimental numbers quoted in Secs.~\ref{sec:tde} and \ref{sec:sm}. 
By expanding $(q/p)^2$ in \eq{mgqp:e} in terms of $\phi$ or 
$\Gamma_{12}/M_{12}$ we find
\begin{eqnarray}
 a_{\rm fs} &=&   \frac{4 |\Gamma_{12}|\, |M_{12}|}{
               4 |M_{12}|^2 + |\Gamma_{12}|^2} \, \phi 
          + {\cal O} (\phi^2), 
     \qquad\qquad\qquad\qquad\quad\mbox{for \kkm\ and \ddm} \label{agmk} \\ 
 a_{\rm fs} &=& \imag \frac{\Gamma_{12}}{M_{12}} + 
         {\cal O} \lt(\Big(  \imag \frac{\Gamma_{12}}{M_{12}} \Big)^2   \rt)
        \; = \;  \lt| \frac{\Gamma_{12}}{M_{12}} \rt| \sin \phi \,, 
      \qquad\quad\mbox{for \bbm}.   
    \label{agmb} 
\end{eqnarray}
With this result it is straightforward to solve \eqsand{mgqp:a}{mgqp:d} 
for $\dm$ and $\dg$. Incidentally, in both cases we have 
\bea%
\dm &\simeq & 2\, |M_{12}| ,
  \label{mgsol:a} \\
\dg &\simeq & 2\, |\Gamma_{12}| \cos \phi  .
  \label{mgsol:b}
  \eea% 
  which holds up to corrections of order $\phi^2$ for Kaons and $D$
  mesons and corrections of order $|\Gamma_{12}/M_{12}|^2$ for $B$
  mesons.  In the SM we can replace $\cos \phi$ by 1 not only in \kkm\
  but also in \bbm, but there  the smallness of
  $\lt||q/p|-1\rt|$ is already implied by $|\Gamma_{12}/M_{12}|\ll 1$
  so that the experimental information on the smallness of $\lt||q/p|-1\rt|$
  leaves some space for non-negligible BSM contributions to $\phi$. Note that in \ddm\
  we have $\cos \phi \approx -1$, because $\dg_D/\dm_D <0$ is measured.

Importantly, in all neutral meson complexes one deduces from \eq{mgqp:e} that
\bea%
  \frac{q}{p} &=& - \frac{M_{12}^*}{|M_{12}|} 
     \lt[ 1 + {\cal O} (a_{\rm fs}) \rt] \label{phqp} .%
     \eea%
     That is, the phase of $-q/p$ is essentially given by the phase of
     the box diagram in \fig{fig:boxes}. $q/p$ depends on phase
     conventions and is specific to the choice for the $CP$
     transformation in \eq{defcandcp}. In documents with opposite signs
     compared to \eq{defcandcp}
     you find $q/p$ in \eq{phqp} without the ``$-$'' sign.  Since \bbm\
     is dominated by the box diagram with internal tops we readily
     infer%
\bea%
     \frac{q}{p} &=& -\frac{V_{tb}^* V_{tq}}{V_{tb} V_{tq}^*}
      \; =\; - \exp[ i \arg \lt( V_{tb}^* V_{tq} \rt)^2] \qquad\qquad 
   \mbox{for \bbmq\ with $q=d,s$} \label{qpb}   
\eea%  
up to tiny corrections of order $a_{\rm fs}$. \eqsand{mgsol:a}{mgsol:b}
further show that $\dm$ is trivially related to $|M_{12}|$ and
$\dg$ is essentially determined by $|\Gamma_{12}|$. Since $-\Gamma_{12}/2$ is
the absorptive part of $\Sigma_{12}$, which is the $\Mbar \to M$
transition amplitude, we verify that $\Gamma_{12}$ is composed of
all decays into final states $f$
which are common to $M$ and $\Mbar$ from the optical theorem:
\begin{align}
  \Gamma_{12} &= \sum_f \langle M \ket{f}    \langle f \ket{\Mbar}  
                \;=\;\sum_f A_f^* \bar A_f \label{ga12af}
\end{align}
We have found this feature
already in the discussion of $\dg$ around \eq{dgexc}. Indeed, for
$|q/p|=1$ one has $|q|=|p|=1/\sqrt2$ and inserting $\alpha_H=\alpha_L=p$ and
$\beta_L=-\beta_H=q$ into \eq{dgexc} gives
\begin{align}
  \dg \;=\; \Gamma_L-\Gamma_H 
  &=\; 4 \, \real
                   \lt(p^* q \,   \sum_f A_f^*
    \bar A_f  \rt)
    \;\stackrel{\eq{ga12af}}{=}\; 2\,  \real
    \lt( \frac{q}{p} \,   \Gamma_{12} \rt)
    \;\stackrel{\eq{phqp}}{=}\; - 2 \, \real  \lt(
    \frac{M_{12}^*}{|M_{12}|} \, \Gamma_{12}\rt)
    \nn
    &= \; - 2 \, \real  \lt(
    \frac{M_{12}^*}{|M_{12}|} \,
      \frac{|\Gamma_{12}|^2}{\Gamma_{12}^*}\rt)
      \; \stackrel{\eq{defphi}}{=}\; 2 \,\real  \lt( e^{-i\phi}
     \,
      |\Gamma_{12}| \rt) \;=\; 2  \, |\Gamma_{12}| \,\cos\phi
      \label{dgagree}
\end{align}
in agreement with \eq{mgsol:b}.

We can apply our formalism also to the width difference between $CP$
eigenstates of the \mm\ systems. This is unambiguously only possible in a
theory which conserves $CP$. But the hierarchy of the CKM matrix permits
the definition of such eigenstates also in the SM: In the dominant
CKM-favored tree-level decays modes we can neglect direct $CP$
violation, because the required interfering second amplitude is
CKM-suppressed. Furthermore, the amplitudes of all such CKM-favored decays
can be arranged to have weak phases essentially equal to 0 or $\pi$; the
standard CKM phase convention, which I use in the following discussion,
has this property.  I exemplify the topic with $B_s$ mesons, because it
was studied for this case \cite{Dunietz:2000cr}.
For example, the $b\to c\bar c s$ and $b\to c \bar u d$ amplitudes have
both essentially real CKM factors. Thus we can define
$\ket{B_s^{\rm even}}$ and $\ket{B_s^{\rm odd}}$ as orthogonal states
with the property $B_s^{\rm odd} \not\to f_{\rm CP+}$ and
$B_s^{\rm even} \not\to f_{\rm CP-}$ in CKM-favored decays, where
$f_{\rm CP+}$ and $f_{\rm CP-}$ denote CP-even and CP-odd states,
respectively. Thus e.g.\ $B_s^{\rm even}$ can decay to $D_s^+D_S^-$,
while $B_s^{\rm odd}$ cannot, as long as the CKM-suppressed penguin
amplitude is neglected. In the standard CKM phase convention in which
the considered decay amplitudes are real, one finds with \eq{defcandcp}:
\begin{align}
\ket{B_s^{\rm even}} &=\; \frac{}{}
 \frac{\ket{B_s} - \ket{\Bbar_s}}{\sqrt{2}} , \qquad 
                       \qquad
\ket{B_s^{\rm odd}} =\; 
 \frac{\ket{B_s} + \ket{\Bbar_s}}{\sqrt{2}} .\label{cpe} 
\end{align}
Note that our derivation of the mass eigenstates in \eq{defpq} does not
tell us anything about their relationship to the $CP$ eigenstates in
\eq{cpe}, i.e.\ whether
$|\langle D_s^+ D_S^- \ket{B_{s,L}}|^2 > |\langle D_s^+ D_S^-
\ket{B_{s,H}}|^2$ or not, that is, whether $B_{s,L}$ or $B_{s,H}$ is
closer to $B_s^{\rm even}$. Now we  repeat the derivation in
\eq{dgagree} with the $CP$ eigenstates; for this we simply replace $p$ and
$q$ by $+ 1$ and $-1$, respectively, in our derivation of \eq{dgagree} and find
\begin{align}
  \dg_{\rm CP} &\equiv\; \Gamma(B_s^{\rm even}) -\Gamma(B_s^{\rm odd})
                 \; =\; \sum_f \lt( |\langle f \ket{B_s^{\rm even}}|^2 -
                 |\langle f \ket{B_s^{\rm odd}}|^2 \rt) \; =\;
                 - 2\,  \real   \Gamma_{12} , \label{dgcp}
\end{align}
valid for the standard CKM phase convention. $\Gamma_{12}$ in the \bbms\
system is dominated by $b\to c\bar c s$ decays, so that  $\Gamma_{12}$
has the  essentially real CKM factor  $(V_{cb} V_{cs}^*)^2$ and one
can omit the ``$\real$'' in \eq{dgcp}. In Ref.~\cite{Dunietz:2000cr}
it is explained how one can measure $\Gamma(B_s^{\rm even/odd}) $.
We realize that $\dg_{\rm CP}$ is a mixing observable which probes
$\Gamma_{12}$  but is
not sensitive to $M_{12}$. The SM prediction discussed in Sec.~\ref{sec:sm}
predicts $\Gamma_{12}<0$ and BSM physics cannot be so large that this
sign is flipped.  Thus, independently of any BSM physics in $M_{12}$,
theory predicts $ \dg_{\rm CP}>0$, so that the $CP$-even eigenstate is
shorter-lived.  The measurement of $\sgn \dg_s>0$ (predicted in the SM
as well) was needed to identify $B_{s,L}$ with the shorter-lived
eigenstate and $ \dg_{\rm CP}>0$ means that it is also mostly CP-even.
In the limit $\cos\phi=1$ $CP$ is a good symmetry and the mass and $CP$ eigenstates coincide. 
$ \dg_{\rm CP}>0$ has been established experimentally as a byproduct of
the analyses determining $ \phi_{CP,B_s}^{\rm  mix}$ in \eq{acpmexps}
together with $1/\Gamma_{L,H}$, which find that the $\BsorBsbar$
decays into the CP-even final states $(J/\psi \phi)_{l=0,2}$ with a
lifetime $1/\Gamma_L$. 
The measurement of  $ \dg_{\rm CP}$ in the $B_s$ system, described in
Ref.~\cite{Dunietz:2000cr}, was originally proposed to determine $\cos
\phi$ through a comparison of $\dg_{\rm CP}$ in \eq{dgcp} and
$\dg$ in \eq{mgsol:b}. This is of little interest today, because we know
that $\cos\phi$ is close to 1 from bounds on $|\sin\phi|$ discussed below. 

Interestingly, in \ddm\ it is also experimentally firmly established
that the shorter-lived eigenstate is dominantly $CP$-even
\cite{HeavyFlavorAveragingGroupHFLAV:2024ctg}, which seems to be a
common feature of all four \mm\ systems.  So while we cannot calculate
$\Gamma_{12}$ reliably for \ddm, experiment tells us that
$\Gamma_{12}<0$ from \eq{dgcp}.  In the \bbd\ case $\dg_{\rm CP}>0$ is a
SM prediction and not yet verified experimentally. \bbd\ mixing is special,
because the $CP$ eigenstates defined through
$B_d^{\rm odd}\not\to D^+D^-$ are not close to the mass eigenstates,
because $M_{12}$ has the phase $2\beta$ in the standard phase convention,
so that in the SM one has 
$|\langle B^{\rm even} \ket{B_{s,\rm short}}|^2 =(1+\cos (2\beta))/2$ and
$|\langle B^{\rm even} \ket{B_{s,\rm long}}|^2 =(1-\cos(2\beta))/2$
\cite{Dunietz:2000cr,Gershon:2010wx}. This factor is taken into account
when $|\dg_d|$ is constrained from lifetime measurements in
$\BdorBdbar \to J/\psi K_{\rm short}$ in \eq{dgdexp}.

Next I discuss $CP$ violation in  mixing, which we have identified 
in \eq{mlmh} as a consequence of $|q/p|\neq 1$. The standard way to
define the corresponding $CP$ asymmetry employs the
flavor-specific decays $M \to f_{\rm fs}$ and $\Mbar \to \bar f_{\rm fs}$ with
$\Mbar \not\to f_{\rm fs}$ and $\Mbar \not\to f_{\rm fs}$. With our result in \eq{tgg}
for the states $\ket{M(t)}$ and $\ket{\Mbar(t)}$ and \eq{gpgms} we can
calculate the  time-dependent decay rates defined in \eq{defgtf} for the
decays of interest:
\begin{align}
  \Gamma(M(t)\to f_{\rm fs})
  =\; |\langle  f_{\rm fs} \ket{M(t)}|^2 \;=\;\phantom{ \lt| \frac{q}{p} \rt|^2}\,
    |g_+(t)|^2 |A_ { f_{\rm fs}}|^2 &\;{=}\;\phantom{\lt| \frac{p}{q} \rt|^2}\,
    |A_ { f_{\rm fs}}|^2 \, \frac{e^{- \Gamma t}}{2} \, \lt[ \cosh
    \frac{\Delta \Gamma \, t}{2} + \cos\lt( \dm\, t \rt) \rt] \label{gafs1}
  \\
  \Gamma(M(t)\to \bar f_{\rm fs})
  =\; |\langle  \bar f_{\rm fs} \ket{M(t)}|^2 \;=\;
    \lt| \frac{q}{p} \rt|^2|g_-(t)|^2 |\bar A_ {\bar f_{\rm fs}}|^2
    & \;{=}\;
     \lt| \frac{q}{p} \rt|^2
    |\bar A_ { \bar f_{\rm fs}}|^2 \, \frac{e^{- \Gamma t}}{2} \, \lt[ \cosh
    \frac{\Delta \Gamma \, t}{2} - \cos\lt( \dm\, t \rt)  \rt] \nn
    &=\;
    (1-a_{\rm fs})\, \frac{e^{- \Gamma t}}{2} \, \lt[ \cosh
    \frac{\Delta \Gamma \, t}{2} - \cos\lt( \dm\, t \rt)  \rt] 
    \label{gafs2}\\
  \Gamma(\Mbar (t)\to f_{\rm fs})
  =\; |\langle  f_{\rm fs} \ket{\Mbar(t)}|^2 \;=\;
    \lt| \frac{p}{q} \rt|^2 |g_-(t)|^2 |A_ { f_{\rm fs}}|^2 & \;{=}\;
     \lt| \frac{p}{q} \rt|^2
    |A_ { f_{\rm fs}}|^2 \, \frac{e^{- \Gamma t}}{2} \, \lt[ \cosh
    \frac{\Delta \Gamma \, t}{2} - \cos\lt( \dm\, t \rt) \rt] \nn
    &=\; (1+a_{\rm fs}) \, \frac{e^{- \Gamma t}}{2} \, \lt[ \cosh
    \frac{\Delta \Gamma \, t}{2} - \cos\lt( \dm\, t \rt)  \rt] 
    \label{gafs3}\\
\Gamma(\Mbar (t)\to \bar f_{\rm fs})
  =\; |\langle  \bar f_{\rm fs} \ket{\Mbar (t)}|^2 \;=\; \phantom{ \lt| \frac{q}{p} \rt|^2}\,
    |g_+(t)|^2 |\bar A_ {\bar f_{\rm fs}}|^2 & \;{=}\;  \phantom{\lt| \frac{p}{q} \rt|^2}\,
    |\bar A_ { \bar f_{\rm fs}}|^2 \, \frac{e^{- \Gamma t}}{2} \, \lt[ \cosh
    \frac{\Delta \Gamma \, t}{2} + \cos\lt( \dm\, t \rt)  \rt] \label{gafs4}
 \end{align}
I have neglected ${\cal O}(a_{\rm fs}^2)$ corrections in \eq{gafs3} and
will do so from now on. From these four equations  one can extract
$a_{\rm fs}$, $|\bar A_ { \bar f_{\rm fs}}|^2/|A_ {f_{\rm fs}}|^2$, and the
overall normalization. The second quantity determines the direct $CP$
asymmetry
\begin{align}
  A_{CP}^{\rm dir} (M\to  \bar f_{\rm fs})
  &= \;
  \frac{|A_ {f_{\rm fs}}|^2-|\bar A_ {\bar f_{\rm fs}}|^2}{|A_ { f_{\rm fs}}|^2+|\bar A_ {\bar f_{\rm fs}}|^2},
\end{align}
which is different for every decay mode, while $a_{\rm fs}$ is universal
and can be determined by combining many different decay channels. Since
four observables depend on three quantities, there is redundant
information in \eqsto{gafs1}{gafs4}, which can be used to eliminate
experimental uncertainties. One may worry about the charge symmetry of
the detector and add a parameter $\epsilon_c$ for some unaccounted
charge asymmetry to the formulae. Likewise one can do so with a
parameter $\epsilon_p$ for the production asymmetry between $M$ and
$\Mbar$. As shown in Ref.~\cite{Nierste:2004uz} one cannot disentangle
$ A_{CP}^{\rm dir} (M\to \bar f_{\rm fs})$ from $\epsilon_c$, as these
quantities always appear in the same combination. However, it is
possible to identify $a_{\rm fs}$ and $\epsilon_p$ uniquely from the
four measurements associated with \eqsto{gafs1}{gafs4}. Thus
insufficient knowledge of neither $\epsilon_c$ nor $\epsilon_p$ is a 
show-stopper for a measurement of $a_{\rm fs}$. The time-dependent $CP$
asymmetry associated with \eqsto{gafs1}{gafs4} reads
\begin{align}
  a_{\rm CP} (M(t)\to \bar f_{\rm fs})&\equiv \;
                                        \frac{ \gbtfs - \gtfbs }{ \gbtfs +
                                        \gtfbs }.
                                        \label{defacpfs}
\end{align}
One usually uses semileptonic decays to measure $a_{\rm fs}$, i.e.\ one
uses $f_{\rm fs}=X \ell^+\nu$ with e.g.\ $X=D^-$, $D^{*-}$, or the fully
inclusive final state. Then $a_{\rm CP} (M(t)\to \bar \ell^+\nu)$ is called
\emph{semileptonic $CP$ asymmetry}. Inserting \eqsand{gafs2}{gafs3} into
\eq{defacpfs} gives
\begin{align}
  a_{\rm CP} (M(t)\to \bar f_{\rm fs})&=\;
      a_{\rm fs} +  A_{CP}^{\rm dir} (M\to   f_{\rm fs}).
                                        \label{acpfs}
\end{align}
we note that the time-dependence cancels between numerator and
denominator in \eq{defacpfs}.
Moreover, the measurement of $a_{\rm fs}$ requires no flavor
tagging \cite{Yamamoto:1997cg}. Defining the \emph{untagged decay rate} as
\begin{align}
\guntf &=\; \gtf + \gbtf,         \label{guntf}
\end{align}
one finds from \eqsto{gafs1}{gafs4} \cite{Dunietz:2000cr,Nierste:2004uz}:
\begin{eqnarray}
a_{\rm fs, unt} (t) &=&
   \frac{\guntfs -  \guntfbs}{\guntfs + \guntfbs}
\;= \;   A_{CP}^{\rm dir} (M\to  f_{\rm fs}) 
        +\frac{a_{\rm fs}}{2} - \frac{a_{\rm fs}}{2} \,
        \frac{\cos (\dm\, t)}{\cosh (\dg t/2) }
        . \,  \label{fsun}
\end{eqnarray}
Thus one does not have to pay the price of the lower statistics of a
tagged sample. Unlike the tagged asymmetry in \eq{acpfs} the untagged
version in \eq{fsun} depends on $t$, which is a welcome feature to
separate $ A_{CP}^{\rm dir}$ from $a_{\rm fs}$ and signal from
background.  Note that at $t=0$ there is no sensitivity to $a_{\rm fs}$
as $M$ needs time to mix into $\Mbar(t)$. 
Adding charge and production asymmetries, $\epsilon_c$
will only change the first term, while $\epsilon_p$ only appears in the
time-dependent second term \cite{Nierste:2004uz}. One needs another observable to extract
$a_{\rm fs}$ then, for example the ``right-sign asymmetry'' 
$a_{\rm right} \equiv (\gtfs - \gbtfbs)/( \gtfs + \gbtfbs)$ which equals
$ A_{CP}^{\rm dir} (M\to   f_{\rm fs})$ in the absence of $\epsilon_c$
\cite{Nierste:2004uz}.

One can probe direct $CP$ violation in $M^\pm\to  f_{\rm fs}^\pm$ decays and, if one finds
a null result, one may gain confidence that $A_{CP}^{\rm dir} (M\to  f_{\rm
  fs})$ also vanishes in the neutral mode of interest and that one
further has $\epsilon_c$ under control.  If in addition there is no
$\epsilon_p$\footnote{There is no production asymmetry in $B$ factories,
but $\epsilon_p\neq 0$ could occur from different acceptances of  $B_d$
and $\Bbar_d$ related to the asymmetric beam energies},
one could use time-integrated measurements:
\begin{align}
A_{\rm fs,unt} &\equiv \; 
\frac{\int_0^\infty dt [ \guntfs - \guntfbs  ]}{
	\int_0^\infty dt [ \guntfs + \guntfbs] } 
\; =\; \frac{a_{\rm fs}}{2} \, \frac{x_M^2-y_M^2}{x_M^2-1} ,
\label{aunt}
\end{align}
with $x_M\equiv \dm/\Gamma$ and $y_M\equiv -\dg/\Gamma$ which we have
already encountered for the case $M=D$ in \eq{defxy}. For the
semileptonic decays one often studies \emph{dilepton asymmetries}\ at
$B$ factories by comparing the number $N_{++}$ of decays
$(M(t),\Mbar(t))\to (f,f)$ with the number $N_{--}$ of decays to
$(\bar{f},\bar{f})$ for $f=X\ell^+\nu_{\ell}$. Then one finds
$a_{\rm fs}=(N_{++}-N_{--})/(N_{++}+N_{--})$ in time-integrated
measurements. Also in \kkm\ one studies time-independent $CP$-violating
quantities; $a_{\rm fs}^K$ can be measured in 
semileptonic $K_{\rm long}$ decays. I will discuss $a_{\rm fs}^K$
and $a_{\rm fs}^D$ together with mixing-induced $CP$ violation in Sec.~\ref{sec:tde}.

The experimental situation for $CP$ violation in \bbm\ is as follows:
\begin{align}
  a_{\rm fs}^{d, \rm exp} &=  -0.0021 \pm 0.0017\,.
                            \label{eq:expafsd} \qquad\qquad
  \mbox{HFLAV \cite{HeavyFlavorAveragingGroupHFLAV:2024ctg}}\\
  a_{\rm fs}^{s, \rm exp} &= -0.0006 \pm 0.0028\,, 
                             \qquad\qquad
  \mbox{LHCb \cite{LHCb:2016ssr}}
                          \label{eq:expafss}
\end{align}  
The number for $a_{\rm fs}^{d, \rm exp}$ is an average of BaBar, Belle,
and LHCb measurements presented in
Refs.~\cite{LHCb:2024xyw,BaBar:2014bbb,Belle:2005cou,LHCb:2014dcj} and
earlier, less precise values from CLEO, OPAL, ALEPH, and D\O. All
measurements have used semileptonic decays. Since we have precise data
on $\dg_s$ in \eq{dgsexp}, we can combine \eqsto{agmb}{mgsol:b} to place
a bound on $\phi_s$, the $CP$ phase in \bbms:
\begin{align} 
  a_{\rm fs}^s &=\; \lt| \frac{\Gamma_{12}^s}{M_{12}^s} \rt| \sin \phi_s
                 \;=\; \frac{\dg_s}{\dm_s} \tan \phi_s \; =\;
                (4.40\pm 0.20) \cdot 10^{-3}   \tan \phi_s 
\end{align} 
so that \eq{eq:expafss} implies
\begin{align} 
  \phi_s&=\; -0.14 \epm{0.60}{0.52} \;=\; \lt( -7.8 \epm{34}{30} \rt)^\circ,
\end{align}
showing that the $CP$ phase in \bbms\ is not very well constrained. In
Sec.~\ref{sec:sm} we will see that the SM prediction for $\phi_s$ is unmeasurably
small, so that $a_{\rm fs}^s$ is a BSM topic. The measurement in \eq{acpmexps}
does not leave much space for BSM contributions to $\arg M_{12}^s$ and
BSM contributions to $\arg \Gamma_{12}^s$ are somewhat exotic and can
barely exceed a few percent, so that we need a reduction of the error in   
$a_{\rm fs}^s$ by at least a factor 10 compared to \eq{eq:expafss}.

$a_{\rm fs}^d$ can more easily be enhanced by BSM physics, because
$\Gamma_{12}^d$ is suppressed by two powers of $\lambda$. Furthermore,
$|M_{12}^d|$ in the denominator is 35 times smaller than $|M_{12}^s|$ in
$a_{\rm fs}^s$, which lifts $|a_{\rm fs}^d|$ into a region which is
better accessible by experiment. $a_{\rm fs}^d$ will be discussed in
Sec.~\ref{sec:sm}.

One can use many more decays beyond semileptonic ones, examples for
flavor-specific decays to measure $a_{\rm fs}^d$ are
$B_d\to J/\psi K^+ \pi^-$, $B_d\to D_s^+ D^-$, $B_d\to D^- K^+$, and
multi-body decays with strangeness $S=\pm 1$. When studying those decays
it is mandatory to include the $A_{CP}^{\rm dir}$ term in time evolution
formulae like \eq{fsun}. For $a_{\rm fs}^s$ the decays
$B_s\to D_s^- \pi^+$ and $B_s\to D_s^- X$ with $X=\pi^+\pi^+\pi^-$,
$\pi^+K^+K^-$, or any other $X$ with zero strangeness come to mind.

In summary, to fully determine the time evolution of any \mm\ system
governed by \eqsand{schr}{defmg} one must calculate and measure the
three quantities $\dm$, $\dg$, and $a_{\rm fs}$, the latter of which
determines the fundamental $CP$ phase $\phi=\arg (-M_{12}/\Gamma_{12} )$
which describes $CP$ violation in mixing through $|q/p|\neq 1$.  The
relations between the phenomenological quantities $\dm$, $\dg$,
$a_{\rm fs}$ and the theoretical quantities $|M_{12}|$, $|\Gamma_{12}|$,
$\phi$ are given in \eqsand{agmk}{agmb} as well as
\eqsand{mgsol:a}{mgsol:b}.  The mentioned quantities are universal in
the sense that they appear in the time-dependent decay rate of any
$M(t)\to f$ decay and do not depend on $f$. The smallness of
$a_{\rm fs}$ is understood within the SM in \kk, \dd, and \bbms\ from
the structure of the CKM matrix which aligns the phases of $M_{12}$ and
$\pm \Gamma_{12}$ to a large degree and suppresses $\sin \phi$, and in
\bbm\  from the smallness of $|\Gamma_{12}/M_{12}|$, which holds
also in BSM theories.  $\phi$ can be best measured in flavour-specific
decays and the corresponding $CP$ asymmetry
$a_{\rm fs} \equiv 1- | q/p |^2$ can be measured without flavor tagging.

\boldmath
\subsection{Mixing-induced $CP$ asymmetries, $CP$
  violation in
  $\kkm$, and
  time-dependence of exclusive decays\label{sec:tde}}
\unboldmath
We have seen in Sec.~\ref{sec:bbm} that mixing-induced $CP$ asymmetries
can provide a clean access to fundamental $CP$-violating quantities in
the gold-plated decay modes, which essentially only involve a single 
$CP$-violating phase, i.e.\ the penguin pollution is suppressed or even
absent. In the following we first study a mixing-induced $CP$ asymmetry
in Kaon physics. Subsequently, 
to derive expressions like \eqsand{amixres2}{amixres3} for the
mixing-induced $CP$ asymmetries, we will study $\Gamma(M(t)) \to f$
for a given final state $f$. I have already derived the corresponding
expression  for the case $f=f_{\rm fs}$ in \eqsto{gafs1}{gafs4}. In the
beginning I take $f$ arbitrary and will later specify to the case
$f=f_{\rm CP}$ used in \eqsand{amixres2}{amixres3}. 

Next I elaborate further on the decay amplitudes $A_f=A(M\to f) $ and
$\bar A_f =  A(\,\bar{M} \to f )$ introduced in \eq{abjps} for
$f=f_{\rm CP}$ and before \eq{blhf} for any $f$. Their precise definition is
\begin{align}
  (2\pi)^4 \delta^{(4)}(p_M-p_f) A_f
  &=\,   N_f  i \bra{f} S \ket{M} ,
  & %\qquad\qquad
    (2\pi)^4 \delta^{(4)} (p_M-p_f) \bar A_f \,=\,   N_f
             i  \bra{f} S \ket{\bar{M}}  .
\label{defaf}% 
\end{align}
with the S-matrix
\begin{align}
 S &\equiv \mathrm{T}\!\exp \lt[-i \!
     \int \! d^4 x \, H_{\rm int}  \rt] 
     \label{sma}
\end{align}
involving $H_{\rm int} =-L_{\rm int}$ which is the hamiltonian of the
electroweak interaction of the SM (possibly amended by BSM terms), in
the interaction picture. The decay rate is calculated from $|A_f|^2$,
$|\bar A_f|^2$ and multiplies these expressions with inverse powers of
$\pi$ and other numerical factors.  All these factors are absorbed into
the normalization factor $N_f$ such that $\Gamma(M \to f)=|A_f|^2$, cf.\
the calculation in \eqsand{blhf}{dgexc}.  $N_f$ is the same for $A_f$,
$\bar A_f$, $A_{\bar f}$, and $\bar A_{\bar f}$. In two-body decays the
amplitudes are just numbers, because the kinematics in fixed. 
In multi-body decays
the amplitudes depend on the kinematical variables specifying the
studied point in the 
Dalit plot. The variation of $CP$ asymmetries over the Dalitz plot is
extensively used in the experimental analyses, e.g.\ to find
regions with a large strong phase difference between the tree and penguin
amplitudes adding to $A_f$ to maximize  $|A_{CP}^{\rm dir}| $.
In decays into
polarized final states (like $J/\psi \phi$) or differential decay rates
the amplitudes are further labeled with the polarization of the final state
or, equivalently, with the angular momentum quantum number.

If we switch off QCD and replace the external hadrons by quark states,
we can simply calculate $A_f$ and $\bar A_f$ in perturbation theory. In
the case of tree-level decays, \eq{sma2} is expanded to second order
in $H_{\rm int}$ (i.e.\ to second order in
$g_w$) and we find  Feynman diagrams with one virtual $W$ boson
connecting the quark line with the decaying $b$, $c$, or $s$ quark with
a quark or lepton line.

The key quantity to describe mixing-induced $CP$ violation is the combination% 
\begin{align}
\lambda_f &=\; \frac{q}{p}\, \frac{\ov{A}_f}{A_f} . \label{deflaf}% 
\end{align}
$\lambda_f$ encodes the essential feature of the interference of the
$M\to f $ and $\bar{M} \to f$ decays: $\arg{\lambda_f}$ is the
relative phase between $\bar A_f/A_f$ (stemming from the decay) and
$-M_{12}$ (from $q/p$ in \eq{phqp}).

In a first application,
I discuss the decays of neutral Kaons into two charged or neutral pions.
Kaons are simpler than $D$ or $B_{d,s}$, because the observables
are expressed in terms of the mass eigenstates, so that no explicit time
appears in the formulae. 

A neutral $K$ or $\ov K$ meson state is a superposition of $K_H=K_{\rm long}$
and $K_L=K_{\rm short}$. At short times the decays of the $K_{\rm short}$
component of our Kaon beam will vastly dominate over the $K_{\rm long}$ decays
and one can access the decay rates $\Gamma (K_{\rm short} \to \pi \pi)$ for
$\pi\pi=\pi^+\pi^-,\pi^0\pi^0 $. At large times, say, after 20 times the
$K_{\rm short}$ lifetime, our beam is practically a pure $K_{\rm long}$ beam
and we can study the $CP$-violating $\Gamma (K_{\rm long} \to \pi \pi)$
decays. For this discussion I switch to the eigenbasis of strong
isospin $I$:% 
\begin{align}
\ket{\pi^0 \pi^0 } & = \; \sqrt{\frac{1}{3}} \, \ket{\lt(\pi \pi
  \rt)_{I=0}} - \sqrt{\frac{2}{3}} \, \ket{\lt(\pi \pi \rt)_{I=2}} \,,
\nn \ket{\pi^+ \pi^- } & = \; \sqrt{\frac{2}{3}} \, \ket{\lt(\pi \pi
  \rt)_{I=0}} + \sqrt{\frac{1}{3}} \, \ket{\lt(\pi \pi \rt)_{I=2}} \,,
                             \label{siso}
\end{align}
The strong interaction respects strong-isospin symmetry to an accuracy
of typically 2\%, so that we can neglect any rescattering between the
$I=0$ and $I=2$ states. Any \emph{direct}\ $CP$ violation requires two
interfering amplitudes which differ in their weak and strong phases.
Since we can neglect all final states beyond $\pi^+\pi^-$ and
$\pi^0\pi^0$, we encounter a two-state problem in which
$(\pi \pi)_{I=0}$ can only scatter elastically into itself and the same
statements holds for $(\pi \pi)_{I=2}$.  Consequently, neither
strong-isospin eigenstate can interfere with another state and no direct
$CP$ violation contributes to the famous $CP$-violating quantity%
\begin{align}
\e_K \equiv \frac{ \bra{(\pi\pi)_{I=0}} K_{\rm long} \rangle }{
  \bra{(\pi\pi)_{I=0}} K_{\rm short} \rangle }
\label{ek} .
\end{align}
Abbreviating $A_0\equiv A_{(\pi\pi)_{I=0}}$, $\bar A_0\equiv \ov
A_{(\pi\pi)_{I=0}}$ and (see \eq{deflaf}) $\lambda_0 \equiv
\lambda_{(\pi\pi)_{I=0}}$ I  insert \eq{defpq} into \eq{ek} to find
\begin{align}
\e_K &=\; \frac{1-\lambda_0}{1+\lambda_0} \,.
\label{ekla}
\end{align}
The experimental value \cite{ParticleDataGroup:2024cfk}
\begin{align}
%  \e_K^{\rm exp} & =\;  e^{i\, \phi_{\e}} \, (2.23 \pm 0.01) \times
  \e_K^{\rm exp} & =\;  e^{i\, \phi_{\e}} \, (2.228 \pm 0.011) \times
    10^{-3} \qquad \qquad \mbox{with } \quad
  \phi_{\e} \; =\; (43.5\pm 0.5)^\circ \;=\; (0.97 \pm 0.01 )\, \frac{\pi}{4} 
\,. \label{ekexp}
\end{align}%
therefore allows us to determine $\lambda_0$, which in our example is
apparently close to 1. The number in \eq{ekexp} is calculated 
from the measured quantities
$A(K_{\rm long}\to \pi^+\pi^-)/A(K_{\rm short}\to \pi^+\pi^-)$ and
$A(K_{\rm long}\to \pi^0\pi^0)/A(K_{\rm short}\to \pi^0\pi^0)$ by
expressing $\ket{\pi^+\pi^-}$ and $\ket{\pi^0\pi^0}$ in terms of
$\ket{(\pi\pi)_I}$ with \eq{siso}.  In our case with $|A_0|=|\bar A_0|$
(absence of direct $CP$ violation in a $K\to f_{\rm CP}$ decay) we have
$|\lambda_0|=|q/p|$. With \eq{ekla} we find
\begin{align}
\e_K \simeq \frac{1}{2} \lt[ 1 -\lambda_0  \rt] \; \simeq \;
    \frac{1}{2}
    \lt( 1 -\lt|\frac{q}{p}\rt| - i \, \imag \lambda_0 \rt) 
\label{ekqp}
\end{align}
up to corrections of order $\e_K^2$. Remarkably, from the real and imaginary
part of $\e_K$ we infer two $C\!P$-violating quantities:
\begin{align}
               a_{\rm fs}^K&=\;  4 \real \e_K\;=\; 2 \lt(1- \lt|\frac{q}{p}\rt| \rt)
\end{align}  
and the deviation of $ \imag \lambda_0$ from 0.
We have already encountered the first quantity, thus $\real \e_K$
quantifies $CP$ violation in mixing.  The second quantity,
$\imag \lambda_0$, is sensitive to the studied final state and measures
mixing-induced $CP$ violation in the decay $K\to (\pi\pi)_{I=0}$.

In the further discussion I exploit two more features of Kaon physics:
Firstly, $\dg_K= \Gamma_{\rm short}$ up to corrections of
$\Gamma_{\rm long}/\Gamma_{\rm short}=0.002$.  Secondly,
$\Gamma_{\rm short}$ is almost completely dominated by $K_{\rm short}\to (\pi\pi)_{I=0}$.
The second largest contribution is $K_{\rm short}\to (\pi\pi)_{I=2}$, whose
decay rate is smaller by a factor of 500. Thus
\begin{align}
  \Gamma_{12} &=\; \sum_f \langle{K} \ket{f}\langle{f} \ket{\bar
                  K}\;\simeq\;   A_0^* \bar A_0 \;=\; |A_0|^2 \frac{\bar A_0}{A_0}  
\label{g12k}
\end{align}
implying $ \bar A_0/A_0 = \Gamma_{12} /|\Gamma_{12} |$
and, using \eq{mgqp:c},
\begin{align}
  \lambda_0
  &\simeq \; -  \frac{2 M_{12}^{K\,*} -i\, \Gamma_{12}^{K\,*}}{\dm_K + i \,
              \dg_K/2} \, \frac{\Gamma_{12}^K }{|\Gamma_{12}^K |} \;=\;
                \frac{2 |M_{12}^K| e^{-i\phi_K}  + i\, |\Gamma_{12}^K|}{\dm_K + i \,
    \dg_K/2}, \qquad\qquad \mbox{where }\,
    \phi_K = \arg\lt( -\frac{M_{12}^K}{\Gamma_{12}^K}\rt)             .  
\end{align}  
With \eqsand{mgsol:a}{mgsol:b} we can trade $M_{12}^K$ for $\dm_K$
and $\Gamma_{12}^K$ for $\dg_K$:
\begin{align}  
  \lambda_0
  &\simeq \; 
         1 \, - \, i \phi_K \,   \frac{\dm_K}{\dm_K + i \,
              \dg_K/2} ,
                             \qquad\qquad
 \imag \lambda_0
  \; \simeq \; - \phi_K \frac{4 \dm_K^2}{4 \dm_K^2+\dg_K^2}. 
\end{align}
Here and in the following I neglect higher-order terms in $\phi_K$. 
In \eq{agmk} we can also use \eqsand{mgsol:a}{mgsol:b} to find
\begin{align}
  a_{\rm fs}^K &=\; \frac{4 \dg_K\dm_K}{4 \dm_K^2+\dg_K^2} \phi_K  
\end{align}
We can now relate $\e_K$ in \eq{ekqp} to the desired $CP$ phase $\phi_K$
and start with the phase:
\begin{align}
  \tan\phi_\epsilon &\;= \frac{\imag \e_K}{\real \e_K }\; =\;
                      \frac{-2 \,\imag \lambda_0}{a_{\rm fs}^K} \;=\; 
                      \frac{2 \dm_K}{\dg_K}  \; =\; 0.947 \pm 0.002, \label{tphi}
\end{align}
where I have used the experimental numbers in \eqsand{dgkexp}{dmkexp} in the
last step. We realize that $\phi_\epsilon$ gives redundant information,
it is determined by $\dm_K$ and $\dg_K$ and insensitive to the $CP$
phase of interest.  The result in \eq{tphi}, amounting to
$\phi_\epsilon=43.45^\circ \pm 0.051^\circ$, is in  reasonable agreement with
the experimental value in \eq{ekexp}. One may use \eq{tphi} to derive
\begin{align}
   \sin (\phi_\e) &=\; \frac{2 \dm_K}{\sqrt{4 \dm_K^2+\dg_K^2}}, \label{sinphe}
\end{align}
and to express $|q/p|$ and
$\imag\lambda_0$ in \eq{ekqp} in terms of $\phi_\epsilon$ to find the
compact formula
\begin{align}
\e_K &\simeq\,  \frac12  \sin (\phi_\e) e^{i \phi_\e} \phi_K 
         + {\cal O} (\phi_K^2) .\label{ephi}
\end{align}
In the real part of this expression only $ {\cal O} (\phi_K^2)$ terms have
been neglected, while the imaginary part is also affected by
approximating $\dg_K$ by $\Gamma(K_{\rm short}\to (\pi \pi)_{I=0})$, which might
explain the $\sim 2\sigma$ tension in $\tan \phi_\epsilon$ between
\eqsand{ekexp}{tphi}.

Next we determine $ a_{\rm fs}^K$ from a flavour-specific decay:
With \eqsand{defpq}{defa} one easily finds
\begin{align}
 A_L &\equiv \;
  {\Gamma(K_{\rm long} \to \ell^+\nu\,\pi^-) - 
          \Gamma(K_{\rm long}\to \ell^-\bar\nu\,\pi^+) \over
   \Gamma(K_{\rm long} \to \ell^+\nu\,\pi^-) + 
          \Gamma(K_{\rm long}\to \ell^-\bar\nu\,\pi^+)} 
\;=\; \frac{1 - |q/p|^2}{1 + |q/p|^2}
  \; \simeq \; \frac{a_{\rm fs}^K}{2}  
\; =\; \frac12 \sin (2\phi_\e) \, \phi_K + {\cal O} (\phi_K^2) ,
\label{aphi}
\end{align}
where I have used \eq{sinphe}; $A_L$ should be used with
$\phi_\epsilon=43.45^\circ \pm 0.051^\circ$ or
$\sin\phi_\epsilon=0.68776\pm 0.00063$ found from $\dg_K$ and $\dm_K$
in \eqsand{dgkexp}{dmkexp}; and in the prefactor $\sin\phi_\epsilon$ in \eq{ephi}
this should be done as well.

The data \cite{ParticleDataGroup:2024cfk} are 
\begin{eqnarray}
  A_L^{\rm exp} &=& \lt( 3.32 \pm 0.06 \rt) \times 10^{-3}
\end{eqnarray}
and give 
\begin{align}
  % \phi &=& (6.77 \pm 0.12) \times 10^{-3} \label{phikres} .
     \phi_K^{\rm exp} &=\; (6.64 \pm 0.12) \times 10^{-3} \label{phikres} .          
\end{align}
This number is in good  agreement with 
\begin{align}
  \phi_K^{\rm exp}&=\;(6.48 \pm 0.03) \times 10^{-3} \label{phikres2} .
  %% Re[eps_K] gives 6.46
\end{align}
found from the experimental value for $|\e_K|$ in \eq{ephi} with \eq{ephi}. The
accuracy of the various approximation used in the derivations of the
formulae above is discussed in \cite{Proceedings:2001rdi}.

Next I will generalize \eqsto{gafs1}{gafs4} to the case of any decay
$M(t)\to f$, thus $f$ is not necessary flavor-specific or $CP$
eigenstate. We seek
\begin{align}
  \gtf &=\,  \lt| \langle f \ket{M (t)}  \rt|^2 , 
  \qquad \gbtf =
         \lt| \langle f \ket{\Mbar (t)} \rt|^2
  \label{gtfaf}
\end{align}
and inserting \eq{tgg} leads to an expression involving
$A_f$ and $\bar A_f$ as well as $|g_{\pm}|^2$ and 
$g_+^* (t)\, g_- (t)$ quoted in \eq{gpgms}. The amplitudes appear in the
normalization of $\gtf$ and $\gbtf$ and otherwise combine with $q/p$ to
$\lambda_f$:
\begin{align}%
\gtf =\;& {\cal N}_f \, | A_f |^2 \, e^{-\Gamma t}\, \Bigg\{ \frac{1 +
  \lt| \lambda_f \rt|^2}2\, \cosh \frac{\dg \, t}{2} +
\frac{ 1 - \lt| \lambda_f \rt|^2}2\, \cos ( \dm \, t )  \no \\*
& \qquad \qquad \qquad - \real \lambda_f \, \sinh \frac{\dg \, t}{2} -
\imag \lambda_f \, \sin \lt( \dm \, t \rt) \Bigg\} \,,
\label{gtfres} \\
\gbtf \;=& {\cal N}_f \, | A_f |^2 \, \frac{1}{1-a_{\rm fs}} \,
  e^{-\Gamma t}\, \Bigg\{ \frac{1 + \lt| \lambda_f \rt|^2}2\,
    \cosh \frac{\dg \, t}{2}
  - \frac{1 - \lt| \lambda_f \rt|^2}2\, \cos ( \dm \, t ) \no \\*
&  \qquad \qquad \qquad\qquad
    - \real \lambda_f \, \sinh \frac{\dg \, t}{2}
    + \imag \lambda_f \, \sin ( \dm \, t ) \Bigg\} \,.
   \label{gbtfres}
\end{align}
We recall \eq{fbar} for the definition of the $CP$-transformed state.
In the $M(t) \to \bar f$ decay rates   it is advantageous to keep 
$\ov{A}_{\ov f}$ while trading $A_{\ov f}$ for $\lambda_{\ov f}$:%
\begin{align}%
\gtfb \;=&  {\cal N}_f \lt| \ov{A}_{\ov{f}} \rt|^2 e^{-\Gamma t}\,
  ( 1 - a_{\rm fs}) \, \Bigg\{ \frac{1 +
  | \lambda_{\ov{f}} |^{-2}}{2}\, \cosh \frac{\dg \, t}{2}
  - \frac{ 1 - | \lambda_{\ov{f}} |^{-2}}{2}\, \cos ( \dm \, t ) \no \\*
& \qquad \qquad \qquad\qquad
  - \real \frac{1}{\lambda_{\ov{f}}}\, \sinh \frac{\dg \, t}{2} \,
  + \imag \frac{1}{\lambda_{\ov{f}}}\, \sin (\dm \, t) \Bigg\} \,,
\label{gtfbres} \\[2pt]
\gbtfb \; =& {\cal N}_f \lt| \ov{A}_{\ov{f}} \rt|^2 e^{-\Gamma t}\,
  \Bigg\{ \frac{1 + | \lambda_{\ov{f}} |^{-2}}2\,
    \cosh \frac{\dg \, t}{2}
  + \frac{1 - | \lambda_{\ov{f}} |^{-2}}2\, \cos ( \dm \, t ) \no\\*
& \qquad \qquad \qquad
  - \real \frac{1}{\lambda_{\ov{f}}}\, \sinh \frac{\dg \, t}{2}
  - \imag \frac{1}{\lambda_{\ov{f}}}\, \sin ( \dm \, t ) \Bigg\} \,.
   \label{gbtfbres}
\end{align}%
\eqsto{gtfres}{gbtfres} and \eqsto{gtfbres}{gbtfbres} are our master
formulae to calculate any time-dependent decay rate of interest.

For $f=f_{\rm fs}$ we have $\lambda_f=1/\lambda_{\ov f}=0$ and reproduce
\eqsto{gafs1}{gafs4}.
Defining the \emph{mixing asymmetry},% 
\beq%
{\cal A}_0 (t) = \frac{\gtf - \gtfb}{\gtf + \gtfb } \,,
\label{defa0}
\eeq%
we find to order $a_{\rm fs}$:% 
\beq%
{\cal A}_0 (t) = \frac{\cos ( \dm\, t ) }{\cosh (\dg \,t/2)}
  + \frac{a_{\rm fs}}{2}
  \lt[1- \frac{\cos^2 ( \dm\, t )}{\cosh^2 (\dg \, t/2)}  \rt] .
\label{resa0}
\eeq%
Note that ${\cal A}_0 (t)$ is not a $CP$ asymmetry. Instead
$\gtf\ \propto |\bra{M} M(t)\rangle|^2$ quantifies the probability that
an ``unmixed'' $M$ decays to $f$ at time $t$, while
$\gtfb\ \propto |\bra{\ov M} M(t)\rangle|^2$ does so for the
corresponding probability for the process $M\to\ov M \to f$. The
asymmetry ${\cal A}_0 (t)$ can be employed to measure $\dm$. To
determine the expression describing the ARGUS discovery of \bbmd\ one
must integrate \eqsand{gbtfres}{gbtfbres} over $t$ to find the
probabilities $\int_0^\infty dt |\bra{B_d} B_d(t)\rangle|^2$,
$\int_0^\infty dt |\bra{\Bbar_d} B_d(t)\rangle|^2$,
and their $\Bbar_d(t)$ counterparts, to which the numbers $N_{+-}$,
$N_{++}$, and $N_{--}$ of opposite-sign and like-sign dilepton events
are related. The integrated quantities determine
$x_d\equiv x_{B_d}= \dm_d \tau_{B_d}$ (and further depend on the tiny
$y_d\equiv - \dg_d \tau_{B_d}/2$), as we have seen in \eq{argus}.

Next we apply our master formulae to decays into $CP$ eigenstates,
$M\to f_{\rm CP}$, thus $\ket{\bar f_{\rm CP}}=\eta_{\rm CP,f}
  \ket{f_{\rm CP}}$with $CP$ quantum number $\eta_{\rm CP,f}=\pm 1$.
I will set $a_{\rm fs}$ to zero, because we are interested in large $CP$ asymmetries 
like in \eq{amixres2}, compared to which  $a_{\rm fs}$ is
negligible. Thus, I use $|q/p|=1$ in the following.
The time-dependent $CP$ asymmetry reads
\begin{align}
a_{f_{\rm CP}}(t) 
  &= \frac{ \gbtfcp - \gtfcp }{ \gbtfcp + \gtfcp } \,. \label{defacp}
\end{align}
Using \eq{gtfres} and \eq{gbtfres} one finds 
\begin{align}
a_{f_{\rm CP}}(t) & = - \frac{A_{CP}^{\rm dir} \cos ( \dm  \, t ) +
       A_{CP}^{\rm mix} \sin ( \dm \, t )}{
       \cosh (\dg \, t/ 2) +
       A_{\dg} \sinh  (\dg \, t / 2) }
    + {\cal O} ( a_{\rm fs} ) \,,
    \label{acps}%
\end{align}
with (for $f=f_{\rm CP}$)% 
\begin{align}
  A_{CP}^{\rm dir} &=\, \frac{1- \lt| \lambda_f \rt|^2}{1+ \lt|
  \lambda_f \rt|^2} \,, \qquad 
A_{CP}^{\rm mix} =\, - \frac{2\, \imag
  \lambda_f}{1+ \lt| \lambda_f \rt|^2} \,, 
\qquad
A_{\dg} =\,  -
\frac{2\, \real \lambda_f}{1+ \lt| \lambda_f \rt|^2} .
    \label{dirmix}
\end{align}%
where I have used the notation of \cite{Fleischer:1999jv,Dunietz:2000cr}.
The interpretation of $ A_{CP}^{\rm dir}$ as the direct $CP$ asymmetry
is possible, because I have set $|q/p|=1$, so that $|\lambda_{f_{\rm CP}}
|=|\bar A_{f_{\rm CP}}/A_{f_{\rm CP}}|$.  
Note that $|A_{CP}^{\rm dir}|^2 + |A_{CP}^{\rm mix}|^2 + |A_{\Delta\Gamma}|^2
= 1$. Experimentally one can study the time-dependence of $a_f(t)$ and
read off the coefficients of $\cos ( \dm \, t )$ and
$\sin ( \dm \, t )$, so that one can determine $|\lambda_f|$ and
$\imag \lambda_f$.

In a
gold-plated  $M\to f_{\rm CP}$ decay we have  $|\lambda_{f_{\rm CP}}|=1$ and
thus $A_{CP}^{\rm dir}=0$ in Eqs.~(\ref{acps}), (\ref{acpbd}),
(\ref{acpbs}), and (\ref{dirmix}). Furthermore,
\begin{align}
  A_{CP}^{\rm mix} &=\; -\imag \lambda_{f_{\rm CP}} . \label{acpgold}
\end{align}%
Moreover, the phase of
\begin{align}
  \frac{\bar A_{f_{\rm CP}}}{A_{f_{\rm CP}}}
  &=\;     -\frac{V_{q_1b}V_{q_2q_3}^* A_T}{V_{q_1b}^*V_{q_2q_3} A_T}
  \;=\; - \frac{V_{q_1b}V_{q_2q_3}^*}{V_{q_1b}^*V_{q_2q_3}}
\end{align}
is trivially
read off from the phase of the CKM elements, which are here
exemplified for a $b\to q_1 \bar q_2 q_3$ decay. 
$A_T$ is the ``tree''
amplitude introduced after \eq{acpbd}. 
In $B$ physics, where we
also know the phase of $q/p$ from \eq{qpb}, we can therefore directly
relate the measured $ \imag \lambda_{f_{\rm CP}}$ to phases of CKM
elements, if  $M\to f_{\rm CP}$ is gold-plated.

In $B_d\to J/\psi K[\to \pi^+\pi^-]$ one finds (with $\eta_{ J/\psi
  K[\to \pi^+\pi^-]}=-1$)
\begin{align}
  \lambda_{J/\psi K_{\rm short}}
  &=\;  \underbrace{-\frac{V_{tb}^*V_{td}}{V_{tb}^*V_{td}} } \quad
    \underbrace{\frac{V_{cb}V_{cs}^* V_{us} V_{ud}^*}{V_{cb}^*V_{cs} V_{us}^* V_{ud}}}\label{lafjpsi}\\
  & \quad \mbox{from }-\frac{q}{p}\;\;
    \qquad \mbox{from }-
    \frac{\bar A_{\Bbar_d\to J/\psi \Kbar}}{A_{B_d\to J/\psi K}}\,
    \frac{\bar A_{\Kbar\to\pi\pi}}{A_{K\to \pi\pi}} \nn
    &\simeq\; - e^{-i 2\beta} \label{lafjpsi2}
\end{align}
Thus
\begin{align}
  A_{CP}^{\rm mix} (B_d\to J/\psi K_{\rm short})
  &=\; -\imag \lambda_{J/\psi K_{\rm short}}  \;=\; -\sin(2\beta) \label{amixres4}
\end{align}  
We had motivated this result earlier from
considerations of the box diagram and using the standard
CKM phase convention for $V$, for which the second factor in  \eq{lafjpsi}
is real.
In \eq{phmixsmd}  the result of \eq{lafjpsi} was quoted as 
$\phi_{CP,B_d}^{\rm  mix} =2\beta$.
The same derivation for $A_{CP}^{\rm mix}(B_s\to  (J/\psi \phi)_l) $ reads
\begin{align}
  \lambda_{(J/\psi \phi)_l}
  &=\;
    (-1)^l \frac{V_{tb}^*V_{ts} V_{cb} V_{cs}^*}{V_{tb}^*V_{ts} V_{cb}^* V_{cs}}                           
    \;=\; (-1)^l e^{i 2\beta_s} \label{lajpsph} \\
    A_{CP}^{\rm mix}(B_s\to  (J/\psi \phi)_l)
  &=\;  -\imag \lambda_{(J/\psi \phi)_l}
    \;=\; -(-1)^l \sin(2\beta_s)  \label{amixres5}\\
  A_{\dg} &=\;   - \real \lambda_{(J/\psi \phi)_l} \;=\;
            - (-1)^l \cos(2\beta_s) \label{amixres6}
\end{align}
which was quoted as $ \phi_{CP,B_s}^{\rm  mix}=-2\beta_{s}$ 
in \eq{phmixsms}. We further recognize $ A_{\dg}$ in
\eq{acpbs}.

One can also identify gold-plated modes in $B_{d,s}$ decays to $CP$
non-eigenstates \cite{Aleksan:1990ts}. For example, $B_s \to D_s^- K^+$
is a $\bar b\to \bar c u \bar s$ decay interfering with
$\Bbar_s \to D_s^- K^+$, which is a $b\to u \bar c s$ mode
\cite{Aleksan:1991nh,Fleischer:2003yb,Nandi:2008rg,DeBruyn:2012jp}.
(The valence quark $s$, $\bar s$ makes the final state a
$ u \bar c s \bar s$ state, with the $s$ and $\bar s$ ending up in
$D_s^-$ and $K^+$, permitting the mentioned interference.) The mentioned
decay mode is gold-plated, because there is no penguin
contribution. Since $f=D_s^- K^+$ is not a $CP$ eigenstate, we cannot
expect $|\bar A_f/A_f|=1$ and
$(|\bar A_f|^2 - |A_f|^2)/((|\bar A_f|^2 + |A_f|^2)$ is not a $CP$
asymmetry. In studies of decays to $CP$ non-eigenstates one fits the
four expressions in \eqsto{gtfres}{gbtfbres} to the four decay modes, in
the exemplied decay these are $B_s(t) \to D_s^- K^+$,
$\Bbar_s(t) \to D_s^- K^+$, $B_s(t) \to D_s^+ K^-$, and
$\Bbar_s(t) \to D_s^+ K^-$. The ratio $\bar A_f/A_f$ not only involves
the $CP$ phase of interest, but also a strong phase $\delta$. While
$\delta$ cancels from $\lambda_{f_{\rm CP}}$ in decays to $CP$ eigenstates, this is not the case for
$CP$ non-eigenstates, where $\bar A_f$ and $A_f$ are completely
unrelated.  Studying the four decay rates one can extract four
quantities, which are
$\arg{\lambda_{D_s^-K^+}}= -\gamma-\phi_{CP,B_s}^{\rm mix}+\delta $,
$\arg{\lambda_{D_s^+K^-}}= -\gamma-\phi_{CP,B_s}^{\rm mix} -\delta$,
$|\lambda_{D_s^-K^+}|=1/|\lambda_{D_s^+K^-}|$, and the overall
normalization
\cite{Aleksan:1991nh,Fleischer:2003yb,Nandi:2008rg,DeBruyn:2012jp}.
The LHCb analysis of $\BsorBsbar (t)\to D_s^{\mp} K^{\pm}$  
has found \cite{LHCb:2024xyw}
\begin{align}
  \gamma+ \phi_{CP,B_s}^{\rm mix} &=\; \lt( 79 \epm{12}{11} \rt)^\circ \label{gphisexp}
\end{align}
With $\phi_{CP,B_s}^{\rm mix}$ from \eq{acpmexps} this complies with the
result found for $\gamma$ from other measurements, but the large error in
\eq{gphisexp} currently limits the information on $\phi_{CP,B_s}^{\rm mix}$
or $\gamma$ from this measurement.
Unlike in $B_d\to J/\psi K_s$ and $B_s\to J/\psi \phi$, there is no
penguin pollution at all in decays like $B_s(t) \to D_s^- K^+$, in which
mixing-induced $CP$ violation stems from the interference of
$b\to c\bar u s$ and $\bar b\to c \bar u \bar s$ (or in the
corresponding decays with $s\to d$) amplitudes. 

To avoid penguin pollution, one can also exploit the interference of
$b\to c\bar u s$ and $b\to u \bar c s$ amplitudes in decays to
$\DorDbar$, in which the $D$ meson is identified in a $CP$ eigenstate
like $\pi^+\pi^-$ \cite{Gronau:1990ra}, which also works for direct $CP$
asymmetries \cite{Gronau:1991dp}. The penguin pollution in
$B_d\to J/\psi K_s$ and $B_s\to J/\psi \phi$ is discussed above after
\eqsand{betaex}{acpmexps}. In the future one could precisely determine
$ \phi_{CP,B_{d,s}}^{\rm mix}$ (together with $\gamma$) from
$B_d\to \DorDbar K_{\rm short}$ and $B_s\to \DorDbar \phi$, where no
penguin pollution is present.  Another issue concerns the neutral Kaon:
In future more precise measurements one may further wonder whether the
$CP$ violation in \kkm\ will lead to a bias in the extraction of
$2\beta$ from $B_d\to J/\psi K[\to\pi^+\pi^-]$.  If the $K$ meson does
not decay instantaneously, it  undergoes \kk\ oscillations, which
introduces some sensitivity to $\e_K$. This effect, however, is
calculable and one can correct for it \cite{Grossman:2011zk,Grossman:2025uwz}.

Flavor tagging at a hadron collider costs statistics, but one can
exploit the lifetime difference $\dg_s$ to determine the cosines of
$CP$-violating phases of mixing-induced $CP$ asymmetries from
\emph{untagged decays}, as pointed out by Dunietz
\cite{Dunietz:1995cp}. We can add \eqsand{gtfres}{gbtfres} to find the
untagged decay rate, which was defined in \eq{guntf}.
For clarity, I express the result in terms of
$\Gamma_{L,H}=\Gamma_s\pm \dg_s/2$:
\begin{align}
\guntf \, &=\; 
            A\,  e^{- \Gamma_L t} \,+\,  
          B \,  e^{- \Gamma_H t}  \nn
         \mbox{with }\quad A\,&=\, A(f)
  \;=\; \frac{|A_{f}|^2}{2} \lt( 1+ |\lambda_{f}|^2 \rt) \,
    \lt( 1-  A_{\dg} \rt) \;=\;  \frac{|A_{f}|^2}{2} \,\lt| 1+ \lambda_f \rt|^2 \nn
           B\,& =\, B(f) 
  \; =\; \frac{|A_{f}|^2}{2} \lt( 1+ |\lambda_{f}|^2 \rt) \,
    \lt( 1+  A_{\dg} \rt) \;=\;   \frac{|A_{f}|^2}{2}  \, \lt| 1- \lambda_f\rt|^2 
    \label{utadg}
\end{align}
where I have used \eq{dirmix}.
With the precisely measured $\Gamma_{L,H}$ we can fit the measured time evolution
of a chosen $\BsorBsbar\to f$ decay to \eq{utadg} to extract
$A(f)$ and $B(f)$. The ratio $A(f)/B(f)$, from which $|A_{f}|^2$ drops
out, provides information on $\lambda_f$ without determining it
completely.  If one fits the time evolution to a single
exponential $\exp(-i\Gamma_{\rm eff}t)$, one can calculate
$A(f)/B(f)$ from $\Gamma_{\rm eff}$
\cite{Hartkorn:1999ga,Dunietz:2000cr},
one therefore  often calls this  approach \emph{effective lifetime method}.
It is, however, safer to fit the time evolution to the two-exponential
formula than to use   $\Gamma_{\rm eff}$, which needs a good control of
detection efficiencies.

In gold-plated  $B_s \to f_{\rm CP}$ decays, we can
use $|\lambda_{f_{\rm CP}}|=1$ to infer $ A_{\dg} = -\real
\lambda_{f_{\rm CP}}$ from \eq{dirmix}, so that 
\begin{align}
A\,=\, A(f_{\rm CP})
  &=\;    |A_{f_{\rm CP}}|^2 \,
    (1+
    \real \lambda_{f_{\rm CP}})
    \; = \;  |A_{f_{\rm CP}}|^2  \lt( 1+\cos 
                                \phi_{CP,B_s\to f_{\rm CP}}^{\rm  mix}
    \rt) ,\nn
B\,=\, B(f_{\rm CP}) 
  &=\;  |A_{f_{\rm CP}}|^2 \,
    (1-
    \real \lambda_{f_{\rm CP}})
    \;=\; |A_{f_{\rm CP}}|^2 \, \lt( 1-\cos 
                                \phi_{CP,B_s\to f_{\rm CP}}^{\rm  mix}
   \rt) .
    \label{utadg2}
\end{align}
Here $ \phi_{CP,B_s\to f_{\rm CP}}^{\rm  mix}$ is the phase of
$\lambda_{ f_{\rm CP}}$, which quantifies the mixing-induced $CP$
violation in the studied decay. We have encountered the
special case $\phi_{CP,B_s}^{\rm  mix} \equiv \phi_{CP,B_s\to (J/\psi\phi)_{l=0,2}}^{\rm  mix}$
in our discussion of \bbms\ and quoted $A_{\dg}$ for this case in \eq{amixres6}.
Thus for the gold-plated decays into $CP$ eigenstates we can determine
$\cos \phi_{CP,B_s\to f_{\rm CP}}^{\rm  mix}$ from the ratio $A( f_{\rm
  CP})/B( f_{\rm CP})$. The method works as well for  gold-plated decays
into $CP$ non-eigenstates, in which case one needs the time evolution  for
both  $\BsorBsbar \to f $ and $\BsorBsbar \to \bar f $. 

Prominent applications of the lifetime method have addressed $B_s$
decays to $\rho_0 K_{\rm short}$, $D_s^{(*)\pm} K^{*\mp}$
  \cite{Dunietz:1995cp}, $D_s^{*+}D_s^{*-}$, $J/\psi \phi$ , $\rho^0 \phi$,
  $B_s\to K^* \bar K{}^*$ \cite{Fleischer:1996ai}, and the rare decay
  $B_s\to \mu^+\mu^-$, which is an important new-physics analyzer
  \cite{DeBruyn:2012wk}.  Another application is the measurement of
  $\dg_{\rm CP}$ through lifetime studies of $D_s^{(*)+}D_s^{(*)-}$
  \cite{Dunietz:2000cr}.

The ongoing search for $CP$ violation in \ddm\ is discussed in detail in
\cite{Friday:2025gpj}. The slow \dd\ oscillations make this difficult,
one can safely expand the time evolution formulae in
\eqsto{gtfres}{gbtfbres} to the second order in $t$; the physical
interpretation of the coefficients of the linear and quadratic terms were
derived in Ref.~\cite{Blaylock:1995ay}.  The prime effort in the search
for $CP$ violation in \ddm\ is devoted to the interference of
Cabibbo-favored (CF) and doubly Cabibbo-suppressed (DCS) decay
amplitudes, where a meson produced as $D$ decays through a
$c\to d u \bar s$ amplitude, which is proportional to $\lambda^2$. In
time the $D$ evolves into a superposition of $D$ and a tiny admixture of
$\bar D$, which decays with the CF $\bar c\to d \bar u \bar s$
amplitude. Interference is possible because of the valence $\bar u$
quark in $D$, so that the final state has the same quark content
$d u \bar s \bar u$ as the final state of the $\bar D$ decay.
The standard analysis studying $D\to K_{\rm short} \pi^+\pi^-$ 
gives both $a_{\rm fs}$ and $\imag \lambda_{K_{\rm short} \pi^+\pi^-}$,
From Ref.~\cite{Friday:2025gpj} one finds the 2025 world averages
\begin{align}
 a_{\rm fs}&=\; 2\,\lt(1- \lt| \frac{q}{p} \rt|\rt) \;=\;  0.008 \pm 0.104, \qquad\qquad
 \arg{\lambda_{K_{\rm short} \pi^+\pi^-}} \;=\; - 0.056 \epm{0.047}{0.051} .
\end{align}

In summary, mixing-induced $CP$ violation in gold-plated modes, which
are dominated by a single combination of CKM elements, can give access
to fundamental $CP$ phases, without the problem of penguin
pollution. Prime examples are the measurements of $2\beta$ and
$2\beta_s$ from $B_{d,s}$ decays to charmonium. There are gold-plated
modes with no penguin pollution at all like the decay
$\BsorBsbar (t)\to D_s^{\mp} K^{\pm}$ involving $CP$ non-eigenstates.
In \kkm\ $\imag \epsilon_K$ quantifies mixing-induced $CP$ violation in
$K\to \pi\pi$, while $\real \epsilon_K$ quantifies $CP$ violation in
mixing. In \ddm\ the most promising avenue to mixing-induced $CP$
violation and $CP$ violation in mixing utilizes the interference of DCS
$c\to d u \bar s $ and the CF $\bar c\to d \bar u \bar s $ decays.

\section{\mmmc\ in the Standard Model and beyond\label{sec:sm}}
The confirmation of the KM mechanism of $CP$ violation, which lead to
the 2008 Nobel Prize for Kobayashi and Maskawa, required the proof that
the $CP$ phase $\delta_{KM}\simeq \gamma$ extracted from the measured
$\epsilon_K$ correctly predicts $\phi_{CP,B_s}^{\rm mix}=2\beta$ as
measured by the asymmetric $B$ factories BaBar and Belle built for this
purpose. Both $\epsilon_K$ and $\phi_{CP,B_s}^{\rm mix}$ involve other
parameters beyond the $CP$ phases; one needs two quantities to construct
the UT in \fig{fig:ut}. Thus the prediction of $\phi_{CP,B_s}^{\rm mix}$
from $\epsilon_K$ also needed measurements and SM predictions of other,
$CP$-conserving quantities; altogether they constrain the allowed region
for the apex of the UT in the $\bar\rho$-$\bar\eta$ plane. From
\fig{fig:ut} one realizes that the measurement of
$ A_{CP}^{\rm mix}(B_d\to J/\psi K_{\rm short})=-\sin(2\beta)$ defines
an inclined line in the $\bar\rho$-$\bar\eta$ plane which intersects the
point $(1,0)$. The litmus test for the KM mechanism was the confirmation
that this line indeed intersects the previously determined allowed
region. The $CP$-conserving input for the described UT analysis were the
ratio of the semileptonic $b\to u$ and $b\to c$ branching ratios
determining one side of the UT trough
$|V_{ub}/V_{cb}|\simeq \lambda R_u $ and $\dm_d$ determining the other
side $R_t$ through $\dm_d\propto |V_{td}|^2 \propto |R_t|^2 |V_{cb}|^2$.
This procedure would not have been possible without the theory effort to
calculate these quantities, which include with $\epsilon_K$ and $\dm_d$
two different \mmm\ systems. At the beginning of the 1990s it was
unclear, whether prediction beyond semi-quantitive estimates were
possible at all.

As will be discussed in this section, predictions of flavor-changing
processes involve a perturbative piece, obtained by loop calculations in
perturbative QCD, and a non-perturbative calculation of hadronic matrix
elements.  The gold standard for the latter are computations with
\emph{lattice QCD}, in which the QCD path integral is discretized on a
space-time lattice and calculated with Monte-Carlo techniques.  In the
early 1990s lattice QCD studies were still in an exploratory stage and,
for example, did not include dynamical quarks (i.e.\ instead employed the ``quenched
approximation'') even at the time Belle and BaBar went into operation.
The field of \emph{precision flavor physics}, with the scope on FCNC
processes in $K$ and $B$ physics, was founded in the late 1980s by Buras
who initiated a program addressing the calculation of radiative
corrections to essentially all FCNC processes. The aim ---and result---
of this endeavour were robust predictions with theoretically
well-founded uncertainties, which, moreover, could be systematically
reduced with additional calculational effort to match the size of shrinking
error bars of modern experiments.  I mention a few milestones in this
paragraph (for overviews see Refs.~\cite{Buras:2011we,Buchalla:1995vs}).
To establish the field, conceptual problems had to be solved to define a
rigorous theoretical framework for the calculations. This progress
included the proofs of proper factorization of infrared singularities
and of the renormalization-scheme independence \cite{Buras:1989xd} of
the predicted observables, the development of the correct treatment of
bilocal matrix elements
\cite{Buchalla:1993wq,Herrlich:1993yv,Herrlich:1996vf}, and the
understanding of the renormalization of evanescent operators
\cite{Buras:1989xd,Herrlich:1994kh}.

Both $\dm_{d,s}$ in \bbm\ and the largest contribution to $\epsilon_K$
calculated from the \kkm\ amplitude involve box diagrams with top
quarks, the QCD corrections are proportional to $\alpha_s(m_t)\sim 0.1$
and perturbation theory was found to work well \cite{Buras:1990fn}. For
\kkm, however, also contributions with light $u,c$ quarks in the box
diagrams are relevant for $\e_K$ and are even dominant for
$\dm_K$. Since the leading-order prediction for $\dm_K$ fell short of
the experimental value by more than a factor of 2, there was doubt that
perturbation theory works for \kkm\ and uncontrolled additive hadronic
long-distance effects were invoked to explain the experimental value of
$\dm_K$. The issue was alleviated by QCD corrections and
Ref.~\cite{Herrlich:1993yv} established short-distance dominance of
$\dm_K$ in agreement with the expected suppression of additive long-distance
effects by a factor of $\lqcd^2/m_c^2$. By 1995 all QCD contributions to
$|\Delta F|=2$ transitions with heavy or light internal quarks had been
calculated at the two-loop level
\cite{Buras:1990fn,Herrlich:1993yv,Herrlich:1995hh,Herrlich:1996vf} and
a complete QCD-corrected prediction of $\e_K$ in terms of
$(\bar\rho,\bar\eta)$ became possible \cite{Herrlich:1995hh}. The
corrections to $\dm_K$ were disturbingly large and called for a
calculation of one higher order in $\alpha_s$, requiring a three-loop
calculation \cite{Brod:2011ty}.  With the corresponding result of
Ref.~\cite{Brod:2010mj} $\e_K$ has become a high-precision observable
\cite{Brod:2019rzc}.

The path to precision required a parallel effort on hadronic matrix
elements. One result of the definition of the perturbative framework was
the observation that a certain arbitrary element, the dependence on the
renormalization scheme, must cancel between perturbative pieces and
hadronic matrix elements. This criterion eliminated hadronic models and,
more generally, any approach without control over the renormalization
scheme, from the theorists' toolbox and strengthened the case for
lattice-QCD computations. There were also early analytical methods
compatible with precision perturbative calculations, most prominently
QCD sum rules \cite{Shifman:1978bx,Shifman:1978by}. In Kaon physics the
large-$N_c$ framework (a systematic expansion in terms of the inverse
number of colors) of \emph{Dual QCD} gave the correct value for the
$\Delta S=2$ hadronic matrix element of \kkm\
\cite{Bardeen:1987vg,Buras:2014maa} and an estimate of $\dm_K$ in the
right ballpark \cite{Gerard:1990dx,Bijnens:1990mz,Buras:2014maa}.

In the remainder of this section I will first discuss the Yukawa
interaction in the SM, which is the origin of flavor mixing.  Then, in
Sec.~\ref{sec:eff}, I will explain the concept of an effective hamiltonian,
exemplified for $|\Delta B|=2$ transitions and applied to $\dm_{d,s}$. Here also
precise SM predictions for these quantities and the associated
phenomenology will be presented. In Sec.~\ref{sec:effdg} the effective
 $|\Delta B|=1$ will be introduced and applied to mixing-induced $CP$
 asymmetries and $\Gamma_{12}^{d,s}$, with numerical predictions for
 $\dg_{d,s}$ and $a_{\rm fs}^{d,s}$. Sec.~\ref{sec:kaon} covers \kkm\
 with predictions for $\e_K$ and $\dm_K$ and presents the overall
 picture on the UT from all quantities discussed in this section.

\subsection{Yukawa interaction as the origin of flavor violation}
The $W$ boson is the gauge boson
related to the quantum numbers $(I,I_3)$ of the \emph{weak isospin}.
The SM implements maximal parity violation by placing right-handed quark
field into singlets of the weak gauge group SU(2), while left-handed
quark fields reside in doublets%
\begin{align}
  Q_1 &=\, \sv{u_L^\prime\\ d_L^\prime }, \qquad
  Q_2 =\, \sv{c_L^\prime\\ s_L^\prime }, \qquad 
  Q_3 =\, \sv{t_L^\prime\\ b_L^\prime } . \label{wes} 
\end{align}
SU(2) doublets have weak isospin quantum number $I=1/2$ with
$I_3=\pm 1/2$ for the up-type and down-type component, respectively.
The prime at the quark fields indicates that these fields are \emph{weak
  eigenstates}\ (a.k.a.\ as \emph{gauge} or \emph{interaction
  eigenstates}). SU(2) gauge symmetry dictates that the weak
interaction is built from the doublets $Q_j$ which leaves no room for
the CKM matrix $V$ at this stage. Thus we conclude that $V$ must
stem from a transformation of the weak eigenstates in \eq{wes} to
the physical quark fields $d_L,\ldots t_L$ in \eq{wex} and that
$(u_L,d_L)$, $(c_L,s_L)$, and $(t_L,b_L)$ are \emph{not}\ SU(2)
doublets.

To understand the relation between weak and physical quark eigenstate
fields we must study the mechanism to generate fermion masses in the SM.
Mass terms in the lagrangian involve quark fields of both chiralities,
for example $m_t \bar t_R t_L + m_t \bar t_L t_R$ for the top quark.
Such a term cannot be simply added to the lagrangian, because it violates
SU(2) symmetry. But it is possible to give masses to fermions by
employing gauge-invariant terms with the Higgs doublet field
$H=(G^+, v+(h^0+iG^0)/\sqrt2)^T$ by introducing the
\emph{Yukawa interaction}. The corresponding  lagrangian for quarks reads
\begin{align}
  L_Y^q &=\;  - \sum_{j,k=1,2,3} \lt[ Y_{jk}^d\,
          \bar Q_j \,H\, d_{k,R}^\prime \;+\; 
  Y_{jk}^u\, \bar Q_j \, \,\widetilde H \,u_{k,R}^\prime\rt]
  \;+\; \mbox{H.c.}, \label{yuq}
\end{align}
where $\widetilde H=v + \frac{h^0-i G^0}{\sqrt2}, -G^-)^T$ is the
charge-conjugate Higgs doublet. The Yukawa interaction involves two
complex $3\times 3$ matrices $Y^d$ and  $Y^u$ with
row and column indices $j$ and $k$ referring to the three fermion
generations.  We can readily identify the terms in
\eq{yuq} which are proportional to the Higgs vacuum expectation value
$v=174\gev$:
\begin{align}
  L_Y^q &\supset \; L_m^q \;=\; -  \sum_{j,k=1,2,3}
          \lt[ (d_L^\prime, s_L^\prime, b_L^\prime) \, M^d\, 
           \sv{d_R^\prime\\ s_R^\prime \\  b_R^\prime}
          \; +\;   (u_L^\prime, c_L^\prime, t_L^\prime) \, M^u\, 
           \sv{u_R^\prime\\ c_R^\prime \\  t_R^\prime} \rt]  
\label{yuk}
\end{align} 
with the \emph{quark mass matrices}
\begin{align}
  M^d &= \; Y^d v \qquad \mbox{and} \quad  M^u = \; Y^u v. \label{qmm}
\end{align}  
With four unitary rotations we can diagonalize the two mass matrices.
To this end we rotate the quark fields as
\begin{align}
  \sv{d_{L,R}^\prime \\ s_{L,R}^\prime \\ b_{L,R}^\prime} 
  &=\;  S_{L,R}^d   \sv{d_{L,R} \\ s_{L,R} \\ b_{L,R}},
  \qquad\qquad
  \sv{u_{L,R}^\prime \\ c_{L,R}^\prime \\ t_{L,R}^\prime}
  =\;  S_{L,R}^u   \sv{u_{L,R} \\ c_{L,R} \\ t_{L,R}}
\end{align} 
and choose the four unitary matrices $S_{L,R}^{d,u}$ such that  
the mass matrices in the new basis of physical quark fields are
diagonal,
\begin{align}
  \widehat M^u &\equiv  \; \begin{pmatrix} m_u & 0  &0 \\
    & m_c & 0 \\ 0&0& m_t  \end{pmatrix} \; =\;
    S_L^{u\dagger} M^u S_R^u\qquad
  \widehat M^d \equiv  \; \begin{pmatrix} m_d & 0  &0 \\
    & m_s & 0 \\ 0&0& m_b  \end{pmatrix} \;=\;
    S_L^{d\dagger} M^d S_R^d . \label{defsmat}
\end{align}  
One cannot choose non-unitary matrices for this purpose, because this
would destroy the kinetic term of the lagrangian. The physical quark
field eigenstates are also called \emph{mass eigenstates}.
Next we observe that all  unitary
rotations drop out in the flavour conserving couplings of the gauge
bosons, because they appear as $S_L^{d\dagger}S_L^d =\ldots
S_R^{u\dagger}S_R^u =1$ in the vertices. This is the origin of the
important feature of the SM that there are no \emph{flavor-changing
  neutral currents}\ at tree-level, meaning that $Z$ boson, photon, and
gluon all couple flavor-diagonal! The $W$ vertex, however, involves
\begin{align} 
  L_W &=\;
\frac{g_w}{\sqrt{2}} \lt[ 
\bar{u}_{jL}^\prime \, \gamma^{\mu} d_{kL}^\prime \,  +  \,
\bar{d}_{kL}^\prime \, \gamma^{\mu} u_{jL}^\prime \, W^{-}_{\mu} \rt]  
  \,=\,
    \frac{g_w}{\sqrt{2}} \sum_{j,k=1,2,3}
    \lt[\lt(S_L^{u\dagger} S_L^d\rt)_{jk} \,
    \bar{u}_{jL} \, \gamma^{\mu} d_{kL} \, W^{+}_{\mu} +
    \lt(S_L^{d\dagger} S_L^u\rt)_{kj} \,
    \bar{d}_{kL} \, \gamma^{\mu} u_{jL} \, W^{-}_{\mu} \rt], \no
\end{align}  
which coincides with \eq{wex} for $V=S_L^{u\dagger} S_L^d$. Thus the CKM
matrix is indeed unitary and stems from the mismatch of rotations of
left-handed up-type and down-type quark fields from the weak basis to
the physical basis. The unitarity of $V$ is automatic in the SM, it is
not possible to ``test CKM unitarity'' by comparing SM predictions
against a theory with SM particles but non-unitary $V$. Such a theory
is inconsistent and renders FCNC loops divergent which impedes any
experimentally testable prediction.

When the Yukawa interaction in \eq{yuq} is expressed in terms of the
physical quark fields, $Y^d$ and $Y^u$ are diagonal and the SM Higgs
boson field $h^0$ couples also flavor-diagonally. Contrary to the case
of the neutral gauge bosons, the absence of flavour-changing Higgs
couplings is not a consequence of any symmetry, but originates from the
\emph{minimality of the Higgs sector}. This is an ad-hoc choice in the
construction of the SM, there is no reason why Nature does not provide
several Higgs doublets. Already in a two-Higgs-doublet model (2HDM) the
four Yukawa matrices cannot all be brought to diagonal form and one
expects flavour-changing couplings of the three neutral Higgs bosons of
the 2HDM. Thus one finds contributions to, say, \kkm\ at tree-level,
mediated by a new Higgs boson $H^0$ with $\bar s d H^0$ coupling. 
Meson-antimeson mixing is instrumental to constrain the
parameter spaces of multi-Higgs-doublet models. This also holds  true for
multi-Higgs boson models, in which the FCNC couplings are switched off
by invoking new symmetries, because a charged Higgs boson can contribute
to \mmm\ amplitudes at loop-level like the $W$ boson. 

The electroweak and strong interactions of the SM are constrained by a
powerful principle, gauge symmetry, which dictates that the
boson-fermion and boson-boson couplings of a given interaction involve
the same coupling constant. With the two parameters of the Higgs
potential, the three gauge couplings and $\theta_{\rm QCD}$ quantifying
strong $CP$ violation, this makes six parameters in total. By contrast,
the quark Yukawa sector with the matrices $Y^d$ and $Y^u$ involves 10
physical parameters, which determine six quark masses and the four
parameters of $V$. In the lepton sector there are 10 or 12 parameters,
depending on whether neutrinos are Majorana or Dirac fermions. There a
several hierarchies in the elements of $Y^{d,u}$, resulting in
$m_u < m_d \ll m_s\ll m_c \sim m_b \ll m_t$ and the pattern in $V$
described by the Wolfenstein parametrization in \eq{wolf}. An
explanation of these features is the subject of \emph{flavor model
  building}, which aims at reducing the number of parameters from
symmetry considerations (see e.g.\ Ref.~\cite{Linster:2018avp})
and finding dynamical explanations of their values \cite{Froggatt:1978nt}.

Another ad-hoc feature of the Yukawa sector is the number of fermion
generations. Why did Nature provide us with three fermion families (and
the $CP$ violation which came with it)? Could there be more?  The
information from \mmm\ helped us to constrain the parameter space of a
hypothetical fourth fermion generation by severely constraining the
mixing of the fourth family with the other generations
\cite{Frampton:1999xi,Eberhardt:2010bm,Eberhardt:2012sb}, paving the way
for the exclusion of a fourth fermion generation by more than 5$\sigma$
\cite{Eberhardt:2012gv}.

In summary, flavor violation encoded in $V$ \emph{appears}\ in the weak
interaction of $W$ bosons, but \emph{originates}\ from the Yukawa
interaction of the Higgs field. The diagonalization of $Y^d$ and $Y^u$
rotates the quark fields from eigenstates of the weak interaction to
mass eigenstates. $V$ is a remnant of these unitary rotations.  Thus
flavor physics probes the Yukawa sector of the SM, which is poorly
understood and involves 10 free parameters in the quark sector. The loop
suppression of FCNC processes like \mmm\ amplitudes results form the
minimality of the SM Higgs sector and is a priori absent in models with
more than one Higgs doublet, which involve neutral Higgs bosons with
FCNC couplings.

\boldmath
\subsection{Effective ${|\Delta B|=2}$ hamiltonian and Standard-Model
  prediction for $\dm_{d,s}$\label{sec:eff}}
\unboldmath
In the following I derive the formalism needed to calculate the mass
difference for the \bbd\ and \bbms\ system. The central element is an
\emph{effective hamiltonian}\ describing the  $\Delta B=2$ transition mediated 
by the box diagrams proportional to $(V_{tb}V_{tq}^*)^2$. We can cover
the cases $q=d$ and $q=s$ simultaneously, because the corresponding
effective hamiltonians only differ by the exchange $d\leftrightarrow
s$. 

So far we have applied perturbation theory to the electroweak
interaction only, while the strong interaction is fully contained in
$H_0$ in \eq{hint}. In order to apply perturbative methods to QCD as
well, we must first separate \emph{short-distance}\ and
\emph{long-distance}\ interactions from each other. Short-distance QCD
is associated with high energy and mass scales, far above the scale
$\lqcd \sim 400\mev$ determining the size of typical strong binding
energies. Due to the asymptotic freedom of QCD one can apply perturbation
theory to the short-distance piece of the studied process by  
calculating Feynman diagrams with quarks and gluons.
Long-distance QCD is non-perturbative and confines the external quarks
of our box diagrams in \fig{fig:boxes} into mesons. The theoretical tool
for the desired separation is the \emph{Operator Product Expansion
  (OPE)}, which expresses a hadronic amplitude as a sum of terms which
factorize into a short-distance \emph{Wilson coefficient}\ and a
\emph{hadronic matrix element} containing the long-distance QCD
effects. The Wilson coefficients are calculable in perturbation theory
and the contributions are categorized by the order of
$\alpha_s=g^2/(4\pi)$ to which they are calculated, where $g$ is the QCD
coupling constant. These coefficients depend on the heavy masses in the
problem, the dependence of the \mmm\ amplitudes on $M_W$ and $m_t$ is
fully contained in the Wilson coefficients.  The hadronic matrix
elements instead contain the dynamics associated with Compton wavelengths
of order $\lqcd$, which cannot resolve the $W$ propagation.
For instance, the box diagram of \bbm\ with two heavy top
quarks reduces to a point-like four-quark interaction for the
long-distance piece of the transition amplitude; pictorially we can
shrink the heavy box diagram to a point for the long-distance piece of
the \bbm\ amplitude.

The result of the OPE is an \emph{effective field theory}\ with a simpler
interaction, in our case  described by four-quark vertices whose
effective coupling constants are the Wilson coefficients.
Technically, one distinguishes light and heavy degrees of freedom,
the quantum fields describing the former are kept as dynamical degrees
of freedom, while the fields corresponding to the heavy particles
are removed from the theory, their effect is fully contained in the
Wilson coefficients. In the following  $m_{\rm heavy}$ represents $M_W$
or $m_t$ while $m_{\rm light}$ stands for $m_b$, $m_c$, 
or $\lqcd$, while smaller quark masses are set to zero. 
% We now address the strong interaction, which is the main obstacle on our
% way from quark diagrams to mesonic amplitudes like $M_{12}$ and $A(M\to
% f)$.  In Sect.~\ref{sec:mqb} we have seen that weak processes of mesons
% are multi-scale processes. For instance, \bbm\ involves three largely
% separated scales, since $m_t \sim M_W \gg m_b \gg \lqcd$. These scales
% must be disentangled to separate the short-distance QCD, which is
% described by the exchange of quarks and gluons, from the long-distance
% hadronic physics, whose characteristic property is the confinement of
% quarks into hadrons. The key tool to separate the physics associated
% with the scale $m_{\rm heavy}$ from the dynamics associated with $m_{\rm
%   light}\ll m_{\rm heavy}$ is the construction of an \emph{effective field
%   theory}.
The corresponding effective hamiltonian $H^{\rm eff} $
is designed to reproduce the S-matrix elements of the Standard Model 
up to corrections of order $(m_{\rm light}/m_{\rm heavy})^n$ where $n$
is a positive integer:
\bea%
 \bra{f} \mathbf{T} e^{-i \int d^4 x H_{\rm int}^{\rm SM}(x)}
   \ket{i} &=&  \bra{f} \mathbf{T} e^{-i \int d^4 x H^{\rm eff}(x)}
   \ket{i} \lt[1 + 
   {\cal O} \lt( \frac{m_{\rm light}}{m_{\rm heavy}} \rt)^n \, \rt] 
  \label{eft} 
\eea%
I explain the method with an effective hamiltonian which reproduces
the amplitude for \bbm\ up to corrections of order $m_b^2/M_W^2$. That
is, I employ \eq{eft} for the case $i=\Bbar$ and $f=B$ (where $B=B_d$
or $B_s$), $m_{\rm light}=m_b$ and $m_{\rm heavy}=M_W \sim m_t$. In the
effective theory $H_0$ and $H_1$ in \eq{hint} are replaced by 
\begin{align}%
H_0 &=\; H^{\rm QCD (f=5)},\qquad\qquad 
H_1 \;=\; H^{\rm eff} \;=\;   H^{\rm QED (f=5)} + H^{|\Delta
  B|=2}
\label{hf5} .%
\end{align}%
Here the first terms is the usual QED hamiltonians with 5 ``active
flavours'', meaning that there is no top quark included and
$H^{|\Delta B|=2}$ described the weak interaction mediated by the
$\Delta B=2$ box diagrams proportional to $(V_{tb}V_{tq}^*)^2$ and their
$\Delta B=-2$ counterparts with outgoing $b$ quark lines and CKM factor
$(V_{tb}^*V_{tq})^2$. Also in  $H^{\rm QCD (f=5)}$ there is no top quark field.

\begin{figure}[t]
\centering 
\includegraphics[scale=0.6]{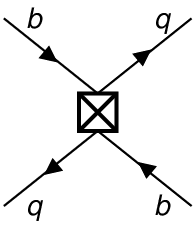}
\caption{The four-quark operator $Q$ for \bbmq\ with $q=d$ or $s$.
\label{fig:q}}
\end{figure}

Adapted to the process under study, $H^{|\Delta B|=2}$ only encodes the
physics related to \bbm, but does not describe other weak processes such
as meson decays.  The $\Delta B=2$ transition of the box diagram in
\fig{fig:boxes} is mediated by an effective four-quark coupling, the
four-quark operator
\begin{eqnarray}
  Q & =& 
    \ov{q}_L \gamma_{\nu} b_L \, \ov{q}_L \gamma^{\nu} b_L 
\qquad\qquad \mbox{with $q=d$ or $s$} ,\label{defQ}
\end{eqnarray}    
% For historical reasons $Q$ is called a \emph{four-quark operator}, but
% it is nothing but a point-like coupling of four quark fields as 
shown in \fig{fig:q}.
We have 
\beq%
H^{|\Delta B|=2} = \frac{G_F^2}{4 \pi^2}\, ( V_{tb} V_{tq}^* )^2 \,
    C^{|\Delta B|=2}( m_t, M_W, \mu )\, Q (\mu) + \mbox{H.c.} ,
\label{ch1:h2}
\eeq%
where the lengthy prefactor of $Q$ is just the effective
coupling constant multiplying the four-quark interaction of \fig{fig:q}.
The Fermi constant $G_F$, which is proportional to $1/M_W^2$,
enters quadratically and thus contains four powers of $1/m_{\rm heavy}$. 

The CKM elements of the box diagram are factored out to get a real
Wilson coefficient $C^{|\Delta B|=2}( m_t, M_W, \mu )$ and has mass
dimension two.  $\mu$ is the renormalization scale, familiar from any
QCD calculation. Just as any other interaction term, also $Q$ must be
renormalized. The renormalized operator $Q$ depends on $\mu$ through the
renormalization constant $Z_Q(\mu)$ via $Q=Z_Q Q^{\rm bare}$ and (in a
mass-independent scheme like $\ov{\rm MS}$) the latter dependence is
only implicit through $g(\mu)$, where $g$ is the QCD coupling
constant.\footnote{The analogy with the renormalization of the QCD
  coupling constant is more obvious if one interprets the product
  $C Z_Q Q^{\rm bare}$ in a different way: By grouping $Z_Q$ with $C$
  rather than $Q$ one recognizes $C$ as a renormalized coupling
  constant. The notion of a ``renormalized'' operator instead of a
  ''renormalized Wilson coefficient'' has historical reasons.}
% With
% the decomposition in \eq{ch1:h2} $C^{|\Delta B|=2}$ has dimension two
% and is real.
The coefficient $C^{|\Delta B|=2}$ is calculated from the definition of
$H^{\rm eff}$ in \eq{eft}: We compute the $\Delta B=2$ process both in
the Standard Model and with the interactions of $H^{\rm eff}$ and adjust
$C^{|\Delta B|=2}$ such that the two results are the same, up to
corrections of order $m_b^2/M_W^2$. Obviously we cannot do this with
mesons as external states $i$ and $f$. But a crucial property of
$H^{\rm eff}$ is the independence of the Wilson coefficient on the
external states. We can calculate it for an arbitrary momentum
configuration for the external quarks as long as the external momenta
are of the order of $m_{\rm light}$. That is, we do not need to know the
complicated momentum configuration of quarks bound in a hadron. In this
step, we switch from the Heisenberg picture to the interaction picture
for $H^{\rm QCD (f=5)}$, because we perform the calculation in an
entirely perturbative world with external quark rather than hadron states. The
strong coupling is determined at a renormalization scale of the order of
$m_{\rm heavy}$, where $g$ is small enough to expand amplitudes in
$\alpha_s$.  Furthermore, we write $C^{|\Delta B|=2}$ as an expansion in
$\alpha_s$:
% all QCD
% effects in $C^{|\Delta B|=2}$ are purely perturbative:%
\bea%
C^{|\Delta B|=2} &=& C^{|\Delta B|=2,(0)} + \frac{\alpha_s(\mu)}{4\pi}
C^{|\Delta B|=2,(1)} + \ldots
\label{cpert} %
\eea%
Since we aim at an expansion of the \bbm\ amplitude in terms of
$m_{\rm light}/m_{\rm heavy}$, we can expand 
the 
box diagram of \fig{fig:boxes} in terms of the momenta of the external
quarks, which are at most of order $m_b$. Thus to leading order in
$m_b/M_W$ (``leading power'') we can simply set 
the external momenta to zero.
Now the ``effective theory side''
of \eq{eft} involves the tree-level diagram corresponding to% 
\bea%
\bra{f} \mathbf{T} e^{-i \int d^4 x H^{\rm eff}(x)} \ket{i}^{(0)} 
  &\simeq & -i \int d^4 x
  \bra{f} H^{\rm eff}(x) \ket{i}^{(0)} 
  \; =\; -i \int d^4 x \bra{f} H^{|\Delta B|=2} (x) \ket{i}^{(0)} 
  \nn 
  & =& -i (2\pi)^4 \delta^{(4)}(p_f - p_i) \; \frac{G_F^2}{4
  \pi^2}\, ( V_{tb} V_{tq}^* )^2 \, C^{|\Delta B|=2,(0)} \, 
 \bra{f} Q \ket{i}^{(0)}  \no %
\eea%
where $\ket i=\ket{p_b,s_b; p_{\ov q}, s_{\ov q }}$ and $\ket f=
\ket{p_q,s_q; p_{\ov b}, s_{\ov b} } $ are the external states
characterized by the momenta and spins of the quarks. The superscript 
``$(0)$'' indicates the lowest order of QCD everywhere.  
Since $\bra{f} Q
\ket{i}$ reproduces the spinor structure (``Dirac algebra'') of the box
diagram, the coefficient $C^{|\Delta B|=2,(0)}$ inferred from this
\emph{matching calculation}\ is solely determined in terms of the loop
integral and therefore only depends on $M_W$ and $m_t$.

The matching calculation becomes more interesting at the \emph{next-to-leading
  order (NLO)} of QCD. Now $H^{\rm QCD}$ enters the matching calculation
and we must dress both the box diagram and the effective diagram in
\fig{fig:q} with gluons in all possible ways. Denoting the SM amplitude
by%
\bea%
{\cal M} &=& {\cal M}^{(0)} + \frac{\alpha_s}{4\pi} {\cal M}^{(1)} +
\ldots, \label{sma2} \eea%
our NLO matching calculation amounts to the determination of $C^{|\Delta
  B|=2,(1)}$ from%
\begin{align}
- {\cal M}^{(0)} - \frac{\alpha_s}{4\pi} {\cal M}^{(1)} &=\; 
\frac{G_F^2}{4
  \pi^2}\, ( V_{tb} V_{tq}^* )^2 \, 
\lt[
C^{|\Delta B|=2,(0)} + \frac{\alpha_s }{4\pi} C^{|\Delta B|=2,(1)} \rt]\,
% \nn & \qquad\qquad\cdot 
 \lt[ \langle Q \rangle^{(0)} + \frac{\alpha_s}{4\pi} \langle Q
\rangle^{(1)} \rt] \,
\lt[ 1+ {\cal O}\lt( \frac{m_b^2}{M_W^2} \rt) \rt] \; +\; {\cal O} \lt(
\alpha_s^2 \rt) \quad \label{manlo} %
\end{align}
On the RHS the external states are omitted for simplicity of notation
and I have expanded $\langle Q \rangle\equiv \bra{f} Q\ket{i}$ in
$\alpha_s$ as well.  The QCD corrections to the box diagram in
${\cal M}^{(1)}$ not only depend on the light scales, i.e.\ external
momenta and light quark masses, they also suffer from infrared (IR)
divergences. These divergences signal the breakdown of QCD perturbation
theory at low energies. However, the gluonic corrections to \fig{fig:q},
which are comprised in $\langle Q \rangle^{(1)}$, exactly reproduce the
infrared structure of the SM diagrams, with the same IR divergences and
the same dependence on the light mass scales. Collecting the
${\cal O}(\alpha_s)$ terms from \eq{manlo},%
\bea%
- {\cal M}^{(1)} &=& \frac{G_F^2}{4 \pi^2}\, ( V_{tb} V_{tq}^* )^2 \,
\lt[ C^{|\Delta B|=2,(0)} \langle Q \rangle^{(1)} + C^{|\Delta B|=2,(1)}
\langle Q \rangle^{(0)} \rt],
\label{manlocoll} %
\eea% 
one finds identical IR structures on the LHS and in the first term in the
square brackets, while $ C^{|\Delta B|=2,(1)}$  only contains heavy masses and
no IR divergences.  We conclude that the IR structure of the SM amplitude
properly factorizes with an ``infrared-safe'' $ C^{|\Delta B|=2}$. % This
The reason for the successful IR factorization is the fact that for a
soft gluon connecting two external quark lines the box diagram and the
four-quark operator look the same, so that the region of the loop
integral with small loop momentum gives the same result for
$- {\cal M}^{(1)}$ and $C^{|\Delta B|=2,(0)} \langle Q
\rangle^{(1)}$. There is only a mismatch stemming from the hard loop
momenta, which are not IR sensitive and feed into the other term in
\eq{manlocoll}, namely $C^{|\Delta B|=2,(1)} \langle Q \rangle^{(0)}$.

Our quark-level calculation is meaningful for $C^{|\Delta B|=2} $, but not for
$\langle Q \rangle$. In order to make a theoretical prediction for the \bbm\ 
amplitude, we must compute $\bra{B} Q \ket{\,\bar{B}}$ with nonperturbative
methods such as lattice QCD. 

Next I derive the result for the leading-order (LO) Wilson coefficient
$C^{|\Delta B|=2,(0)} $. In a first step let us decompose
${\cal M}^{(0)}$ as%
\beq%
{\cal M}^{(0)} = \sum_{j,k=u,c,t} V_{jb}V_{jq}^*\, V_{kb}V_{kq}^*\,
{\cal M }^{(0)}_{jk} \langle Q \rangle ^{(0)}, \qquad\qquad \mbox{$q=d$
  or $s$,} \label{loij}%
\eeq%
where ${\cal M }^{(0)}_{jk} \langle Q \rangle ^{(0)} $ is the result of
the box diagram containing internal quark flavours $(j,k)$ with the CKM
elements factored out.  We then write%
\bea%
{\cal M }^{(0)}_{jk} &=& - \frac{G_F^2}{4 \pi^2}\, M_W^2\, \widetilde S
(x_j, x_k ) \label{stil} %
\eea%
with $x_j=m_j^2/M_W^2$. The function $ \widetilde S (x_j, x_k )$ is
symmetric, $ \widetilde S (x_j, x_k )= \widetilde S (x_k, x_j )$.  Using
CKM unitarity to eliminate
$V_{ub}V_{uq}^*= - V_{tb}V_{tq}^* - V_{cb}V_{cq}^*$ one finds
\eq{loij}:%
\beq%
  - {\cal M}^{(0)} =  \frac{G_F^2}{4 \pi^2}\, M_W^2\, 
   \lt[  
     \lt( V_{tb}V_{tq}^* \rt)^2 S(x_t) \,+\, 
      2 V_{tb}V_{tq}^*\, V_{cb}V_{cq}^* S(x_c,x_t) \, + \,    
     \lt( V_{cb}V_{cq}^* \rt)^2 S(x_c) \rt] \, 
    \langle Q \rangle ^{(0)}. \label{sij}%
\eeq%
$S$ and $\widetilde S$ are related as% 
\bea%  
  S( x_j, x_k ) &=& 
             \widetilde  S (x_j,x_k ) - \widetilde S (x_j,0 ) 
              - \widetilde S (0,x_k ) + \widetilde S (0, 0),
 \qquad \mbox{for $j,k=c,t$},\nn
   S(x) &\equiv &S(x,x) , \label{inali}%  
\eea%
for zero  up-quark mass. In \eq{sij} the last two
terms are tiny, because $x_c \sim 10^{-4}$ and% 
\beq%
  S(x_c)= {\cal O} (x_c), \qquad\qquad  
  S(x_c,x_t)= {\cal O} (x_c \ln x_c). \label{ilexp}% 
\eeq%
where we recognize the GIM suppression discussed after \eq{brcpv}, as
four $\mathcal{O} (1)$ loop functions combine to something much smaller. 
There is no GIM suppression in top loops, because 
$x_t \sim 4 $. The dominant contribution to \eq{loij} involves% 
\bea%
S(x_t) &=& x_t \left[ \frac{1}{4} + \frac{9}{4}
    \frac{1}{1-x_t} - \frac{3}{2} \frac{1}{(1-x_t)^2} \right]
    - \frac{3}{2} \left[ \frac{x_t}{1-x_t} \right]^3 \ln x_t 
 \;\approx \; 2.3  .
 \label{sxt} %
\eea%
The tiny charm contribution does not contribute to 
$C^{|\Delta  B|=2,(0)}$ at all; to accommodate for it we must refine our 
operator product expansion to include higher powers of $(m_{\rm
  light}/m_{\rm heavy})$ in \eq{eft}. We can read off $C^{|\Delta
  B|=2,(0)}$ from \eqsand{manlo}{sij}:%
\bea%
C^{|\Delta B|=2,(0)} ( m_t, M_W, \mu ) = M_W^2\,
    S\, ( x_t ) . \label{wcini}
\eea% 
The functions $S(x)$ and $S(x_c,x_t)$ are called \emph{Inami-Lim}\ 
functions \cite{Inami:1980fz}.

The factorization in \eqsand{eft}{manlo} also solves another problem: %
No largely separated scales appear in
$C^{|\Delta B|=2}( m_t, M_W, \mu )$ provided that we take
$\mu = {\cal O} (M_W,m_t)$, so that no large logarithms can spoil the
convergence of the perturbative series. 
$\mu$  enters our matching
calculation at NLO as $\ln (\mu/M_W)$ 
in $C^{|\Delta B|=2\,(1)}$ and though $\alpha_s(\mu)$.  While no
explicit $\mu$-dependence is present in our LO result in \eq{wcini},
there is an implicit $\mu$-dependence through $m_t(\mu)$, which is a
running quark mass (typically defined in the $\ov{\rm MS}$ scheme).  The
mentioned  $\ln (\mu/M_W)$ term in $C^{|\Delta B|=2,(1)}$ has
two sources: Firstly, there is already a $\ln (\mu/M_W)$ term in
$ {\cal M}^{(1)}$, familiar to us from matrix elements with
$\ov{\rm MS}$-renormalized UV
divergences. Secondly, $ {\cal M}^{(1)}$ contains the large logarithm
$\ln(m_b/M_W)$ which is split between matrix elements and Wilson
coefficients as
\begin{align}
\ln \frac{m_b}{M_W} &=\; \ln \frac{m_b}{\mu} + \ln \frac{\mu}{M_W}
                      . \label{logs}
\end{align}
This feature is clear from \eq{manlocoll}, because 
$\langle Q \rangle ^{(1)}$ can only contain $\ln (m_b/\mu)$, as it is
independent of $M_W$, and conversely $C^{|\Delta
  B|=2}$ is independent of light scales like $m_b$.

The scale $\mu_{tW}= {\cal O} (M_W,m_t)$ at which we invoke \eq{manlo}
to find $ C^{|\Delta B|=2}$ is called the \emph{matching scale} (or
\emph{factorization scale}) and $C^{|\Delta B|=2}( m_t, M_W, \mu_{tW} )$
has a good perturbative behaviour. Similarly, no large logarithms occur
in $\langle Q (\mu_b) \rangle$, if we choose a scale $\mu_b\sim m_b$ in
the matrix element.  
Since $H^{|\Delta B|=2}$ does not depend on the unphysical scale $\mu$, 
we can choose any value for $\mu$, but this value must be
the same in $C(\mu)$ and $\langle Q (\mu) \rangle$. That forces us to
either relate $C(\mu_{tW})$ to $C(\mu_b)$ or to express
$\langle Q (\mu_b) \rangle $ in terms of $\langle Q (\mu_{tW}) \rangle $
in such a way that large logarithms%
\beq%
\alpha_s^n \ln^n \frac{\mu_{tW}}{\mu_b}%
\eeq%
are summed to all orders $n=0,1,2\ldots$ in perturbation theory.  This
can be achieved by solving the \emph{renormalization group (RG)
  equation} for either $C(\mu)$ or $\langle Q (\mu) \rangle$. All steps
of this procedure are analogous to the calculation of the running quark
mass, which can be found in any textbook on QCD. RG-improvement promotes
our LO result to a \emph{leading-log}\ quantity:%
\bea%
C^{|\Delta B|=2,(0)} ( m_t, M_W, \mu_b ) 
  &=& u^{(0)}(\mu_b, \mu_{tW}) C^{|\Delta B|=2,(0)} ( m_t, M_W, \mu_{tW} )     
  \label{rg1} \\
  \langle Q (\mu_{tW})\rangle &=& u^{(0)}(\mu_b, \mu_{tW})
  \langle Q (\mu_b) \rangle \label{rg2} \\
  u^{(0)}(\mu_b, \mu_{tW}) &=& \lt(
  \frac{\alpha_s(\mu_{tW})}{\alpha_s(\mu_b)} \rt)^{
    \frac{\gamma_+^{(0)}}{2\beta_0^{(5)}}} \qquad\qquad \mbox{with }
  \gamma_+^{(0)} = 4 .\label{rg3} %
\eea%
In flavor physics expressions like ``leading-order (LO)'' and
``next-to-leading order (NLO)'' are meant to include the RG resummation,
because fixed-order calculations are not common. I.e.\ ``(N)LO'' really means
``(next-to-)leading log''. 

The evolution factor $ u^{(0)}(\mu_b, \mu_{tW}) $ depends on the
\emph{anomalous dimension}\ of $Q$, which equals $(\alpha_s/(4\pi))
\gamma_+^{(0)}$ to leading-log accuracy.
$\beta_0^{(f)}=11-2f/3$ is the first term
  of the QCD $\beta$ function. One usually writes 
\bea%
C^{|\Delta B|=2} ( m_t, M_W, \mu_b )
 &=& \eta_B b_B(\mu_b) C^{|\Delta B|=2,(0)} ( m_t, M_W, \mu_{tW} )
\label{ceb}
\eea%
where all dependence on $\mu_b$ is absorbed into $b_B(\mu_b)$ and all
heavy scales reside in $\eta_B$.  This factorization is possible to all
orders in $\alpha_s$. It is trivially verified in the LO approximation
of \eq{rg3}, where one simply has
$u^{(0)} (\mu_b, \mu_{tW}) =\eta_B b_B(\mu_b)$. The anomalous dimension
$\gamma_+^{(0)}$ is calculated from the UV-divergent pieces of the
one-loop diagrams found by dressing $Q$ in \fig{fig:q} with a gluon.  In
\eq{ceb} $m_t$ is understood as $m_t(m_t)$ (and not as
$m_t(\mu_{tW})$). In this way $\eta_B$ is independent of $\mu_{tW}$ to
the calculated order; the residual $\mu_{tW}$ dependence is already tiny
in the NLO result. The NLO result for $C^{|\Delta B|=2}$ comprises all terms of order
$\alpha_s^{n+1} \ln^n (\mu_{tW}/\mu_b)$ and includes two ingredients:
first, the two-loop diagrams in which the box is dressed with an
additional gluon in all possible ways \cite{Buras:1990fn} and second, the NLO evolution factor
$ u^{(1)}(\mu_b, \mu_{tW}) $ refining $u^{(0)}$ in \eq{rg3} by 
corrections of order $\alpha_s$ found by calculating the two-loop contribution
$\gamma_+^{(1)}$ to the anomalous dimension of $Q$ \cite{Buras:1989xd}.

$\eta_B$ mildly depends on $x_t=m_t^2/M_W^2$ and in practice one
can treat it as a constant number \cite{Buras:1990fn}: %
\bea%
\eta_B = 0.55,\qquad\qquad b_B(\mu_b=m_b=4.2\,\gev) = 1.5 .
   \label{ebnum}%
\eea%
The dependences of $b_B$ on $\mu_b$ and the chosen renormalization
scheme cancel in the product $ b_B (\mu_b) \langle Q (\mu_b) \rangle$.
The quoted number is for the $\ov{\rm MS}$--NDR scheme, where ``NDR''
refers to the treatment of the Dirac matrix $\gamma_5$. Details on this
topic can be found in \cite{Buras:1989xd}.  We see that the impact of
short-distance QCD corrections is moderate, since
$\eta_B\,b_B(\mu_b)=0.84$. The NLO calculation of Ref.~\cite{Buras:1990fn} has
found only small two-loop corrections and the remaining uncertainty
of $\eta_B$ is around 2\%. 
Combining
Eqs.~(\ref{ch1:h2}), (\ref{wcini}) and (\ref{ceb}) we arrive at our final
result for the $|\Delta B|=2$ hamiltonian:
\begin{equation}
H^{|\Delta B|=2} \; = \; \frac{G_F^2}{4 \pi^2}\, M_W^2\, 
( V_{tb} V_{tq}^* )^2 \, \eta_B \, S (x_t) 
   b_B(\mu_b) Q(\mu_b) 
  \; + \; \mbox{H.c.} \label{desh2}
\end{equation}
   
Turning to the non-perturbative piece of the \bbm\ amplitude, we first
introduce the conventional parameterization of the hadronic matrix
element,%
\begin{align}
  \bra{B_q} Q(\mu_b) \ket{\bar{B}_q}
  & \equiv\;     \frac{2}{3} M_{B_q}^2 \,
    f_{B_q}^2 \, B_{B_q} (\mu_b)
\; \equiv \; 
    \frac{2}{3} M_{B_q}^2 \,
    f_{B_q}^2 \, \frac{\widehat B_{B_q}}{ b_B(\mu_b)} \label{mel}%
\end{align}
with the $B_q$ meson decay constant $f_{B_q}$ and the \emph{bag factor},
which is sometimes chosen as $B_{B_q} (\mu_b)$ and in other occasions as
$\widehat B_{B_q}=B_{B_q} (\mu_b) b_B(\mu_b)$. The second
parameterization incorporates the feature that the dependence on
renormalization scheme and scale must cancel between $ b_B(\mu_b) $ and
$ B_{B_q} (\mu_b)$, so that one can quote numbers for the scheme and
scale independent quantity $\widehat B_{B_q}$ without referring to
details of the renormalization.  The parameterization in \eq{mel} is
chosen in such a way that
$ B_{B_q} (\mu_b)= \widehat B_{B_q}/b_B(\mu_b)$ is close to one for
$\mu_b \sim m_b$.  With the help of our effective field theory we have
reduced the problem of long-distance QCD in \bbm\ to the calculation of
a single number. Lattice gauge theory computations cover the ranges
\cite{Hughes:2017spc,Dowdall:2019bea}%
\bea%
% alt: f_{B_d} \sqrt{\widehat B_{B_d}} & =& (225 \pm 35)\, \mev, \qquad\qquad
% alt: f_{B_s} \sqrt{\widehat B_{B_s}} \; =\; (270 \pm 45)\, \mev .
% B_d-hat = 1.222(61) Bs-hat= 1.232(53)
% f_{B_d} = 0.190 +/- 0.004 GeV f_{B_s} + 0.229 +/- 0.005 GeV 
f_{B_d} \sqrt{\widehat B_{B_d}} & =& (210 \pm 11)\, \mev, \qquad\qquad
f_{B_s} \sqrt{\widehat B_{B_s}} \; =\; (254 \pm 12)\, \mev, \qquad\qquad
\xi \: =\: \frac{f_{B_s} \sqrt{\widehat B_{B_s}}}{f_{B_s} \sqrt{\widehat
    B_{B_s}}} \;=\;  1.216\pm 0.016 .
\label{latt} %
\eea%
The quoted hadronic uncertainties are the dominant source of uncertainty
for the extraction of $|V_{tb}V_{tq}|$ from the measured $\dm_{B_q}$.

Putting \eqsand{desh2}{mel} together we find the desired 
element of the \bb\ mass matrix: % 
\begin{align}
M_{12}^q & = \; 
   \frac{\bra{B_q} H^{|\Delta B|=2}  \ket{\,\bar{B}{}_q} }{2 M_{B_q}} 
\; =\;   \frac{G_F^2}{12 \pi^2}\, \eta_B\, M_{B_q} \,
     \widehat{B}_{B_q} f_{B_q}^2 \,
    M_W^2\, S \bigg( \frac{m_t^2}{M_W^2} \bigg)
    \left( V_{tb} V_{tq}^* \right)^2 .
    \label{m12b}%
\end{align}%
We can now use $ \dm_{d}= 2 |M_{12}^d$ to determine $|V_{td}|$:
\begin{align}
 \dm_{d}  = 
(0.51\pm 0.02) \, \mbox{ps}^{-1}
          {\lt( \frac{|V_{td}|}{0.0086}\rt)^2 } \;
      \lt( \frac{f_{B_d} \sqrt{\widehat B_{B_d}}}{210 \, \mbox{MeV}} \rt)^2
\label{smdmd} .%
\end{align}
For  $ \dm_{s}= 2 |M_{12}^s|$ one finds  
\begin{align}
 \dm_{s}  = 
(16.3 \pm 0.6) \, \mbox{ps}^{-1}
          {\lt( \frac{|V_{ts}|}{0.04}\rt)^2 } \;
      \lt( \frac{f_{B_d} \sqrt{\widehat B_{B_d}}}{254 \, \mbox{MeV}} \rt)^2
\label{smdms} .%
\end{align}
$ \dm_{s}$ involves $|V_{ts}|$ which is fixed by CKM unitarity to 
$|V_{ts}|=0.98 |V_{cb}|$. Thus if one uses $|V_{cb}|$ as input, $
\dm_{s}$ is a direct test of the SM without sensitivity to
$(\bar\rho,\bar\eta)$. \eq{smdms} reproduces the experimental value in
\eq{wavdms} for $|V_{cb}|=0.0426 \epm{0.0022}{0.0019}$.

As mentioned after \eq{wavdms}, we can determine the side $R_t$ from
the ratio $\dm_d/\dm_s$ which is proportional to 
$|V_{td}/V_{ts}|=\, 1.02 \,\lambda \, R_t $, where the factor 1.02 subsumes
the higher-order terms in $\lambda$ mentioned in the text after \eq{wavdms}. 
With $\xi$ from \eq{latt} and $M_{B_s}/M_{B_d}= 1.017$  we find
the ``pocket calculator formula'':
\begin{align}
  R_t \; =\;
  \frac{1}{1.02 \,\lambda}
   \sqrt{\frac{\dm_{d}}{\dm_{s}}} \,
  \sqrt{\frac{M_{s}}{M_{d}}} \, \xi \; =\; 0.905 \, 
  \frac{0.225}{\lambda}\, \sqrt{\frac{\dm_{d}}{0.507\,\mbox{ps}^{-1}}}\,
  \sqrt{\frac{17.77 \,\mbox{ps}^{-1}}{\dm_{s}}} \frac{\xi}{1.216} .
  \label{rtpcf}
\end{align}  

In summary, $M_{12}^{q}$ can be calculated with the help of an OPE with
a particularly simple result, involving only a single Wilson coefficient
$C^{|\Delta B|=2}$ and a single hadronic $\Delta B=2$ matrix element
$ \bra{B_q} Q \ket{\bar{B}_q}$.  The former is calculated to NLO in
QCD perturbation theory; the missing three-loop contribution inflicts an
error of 4\% on this coefficient. The hadronic matrix elements in
\eq{latt} are determined from lattice QCD with a current accuracy of
slightly less than 11\%.
$\dm_q \simeq 2 |M_{12}^{q}| \propto |V_{tb} V_{tq}^*|^2$ determines
$|V_{tq}|$ with a current precision of 11\%, with the theoretical
uncertainty dominated by the lattice calculations of the hadronic
parameters in \eq{latt}. Since CKM unitarity fixes
$|V_{ts}|=0.98 |V_{cb}|$, measurements of $\dm_s$ directly probe the SM,
but this test currently suffers from the controversy on the value of
$|V_{cb}|$.  The ratio
$\dm_d/\dm_s \propto |V_{td}/V_{ts}|^2\propto R_t^2$ provides a precise
determination of the side $R_t$ of the UT, because the perturbative
coefficient and the implicit dependence on $|V_{cb}|$ drops out and the
uncertainty of the hadronic parameter $\xi$ in \eq{latt} is below
1.5\%. Easy-to-use formulae for phenomenological analyses are given in
\eqsto{smdmd}{rtpcf}.

\boldmath
\subsection{Effective $|\Delta B|=1$ hamiltonian and Standard-Model
  predictions for ${\dg_{d,s}}$ and ${a_{\rm fs}^{d,s}}$\label{sec:effdg}}
\begin{figure}
  \centering
   \includegraphics[width=0.75\textwidth]{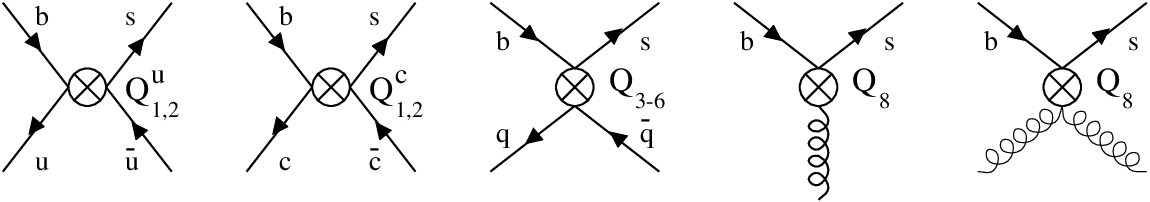} \\[2mm]
  \centering
  \includegraphics[width=0.13\textwidth,angle=-90]{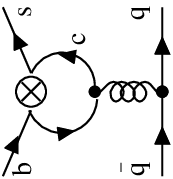} \hspace{3cm}
  \includegraphics[width=0.13\textwidth,angle=-90]{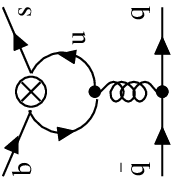} 
  \caption{Upper row: Operators in $H^{\Delta B=1}$ of
    \eqsto{eq::HamDB1}{operators}. The operators of  $H^{\Delta
      B=-1}=H^{\Delta B=1\, \dagger}$ are found by reversing the
    direction of the quark lines. Lower row:  
    Penguin diagrams describing the mixing of $Q_{2}^{(u)}$
    in \eqsand{q12}{qu12}
    into the penguin operators $Q_{3-6}$ in \eqsand{pengop1}{pengop2} 
    \label{fig:effp}}
\end{figure}

\unboldmath%
$B$ decays are processes in which the beauty quantum number
changes by one unit. The tree-level contribution to such decays involves
the exchange of one $W$ boson and the corresponding effective four-quark
operator is obtained by contracting the $W$ line to a point.  Thus the
$|\Delta B|=1$ hamiltonian found in this way is modeled after the Fermi
theory of beta decay.  $H^{|\Delta B|=1}$ comprises many operators,
because there are many non-leptonic decay channels; one further
categorizes the different terms by the other flavor quantum numbers.
For example, $b\to c \bar u d$ decays are described by
$H^{b\to c\bar u d}=H^{\Delta B=\Delta C=\Delta D =1} + \mbox{H.c.}$ and
when mentioning $H^{|\Delta B|=1}$ one usually only refers to the piece
which applies to the studied decay. Another reason for the proliferation
of operators compared to the Fermi theory is QCD: When we include diagrams
with gluons we find new contributions in which color indices are
contracted in a different way compared to the original operator
found from the tree diagram with $W$ exchange. For the description of $b\to s$
decays one needs
\begin{align}
        H^{|\Delta B|=1} 
        =&  \frac{4G_F}{\sqrt{2}}  \left[
        -\, \lambda^s_t \Big( \sum_{i=1}^6 C_i Q_i + C_8 Q_8 \Big) 
        - \lambda^s_u \sum_{i=1}^2 C_i (Q_i - Q_i^u) % \right.  \\
        % & \phantom{\frac{4G_F}{\sqrt{2}} \Big[}
        % \left.
        \, +\, V_{us}^\ast V_{cb} \, \sum_{i=1}^2 C_i Q_i^{cu} 
        + V_{cs}^\ast V_{ub} \, \sum_{i=1}^2 C_i Q_i^{uc} 
        \right]
        + \mbox{H.c.}\,,
        \label{eq::HamDB1}
\end{align}
with 
\begin{equation}
  \lambda^s_a = V_{as}^* V_{ab}\,, 
\end{equation}
where $a=u,c,t$ and $\lambda_t^s=-\lambda_c^s-\lambda_u^s$.  $G_F$ stands for the
Fermi constant. The operators are \cite{Gilman:1982ap}
\begin{align}
  Q_1
  &=\; Q_1^c\;=\; \bar{s}_L^\alpha \gamma_{\mu}  c_L^\beta \;
    \bar{c}_L^\beta \gamma^{\mu}  b_L^\alpha \,,
  &
    Q_2
  &=\; Q_2^c\;=\; \bar{s}_L^\alpha \gamma_{\mu}     c_L^\alpha\;
    \bar{c}_L^\beta     \gamma^{\mu}     b_L^\beta \,, \label{q12}\\
  Q^u_1
  &=\;
    \bar{s}_L^\alpha \gamma_{\mu} u_L^\beta\;
    \bar{u}_L^\beta \gamma^{\mu} b_L^\alpha \,,
  &  
  Q^u_2
  &=\;   \bar{s}_L^\alpha \gamma_{\mu}     u_L^\alpha \;
          \bar{u}_L^\beta     \gamma^{\mu}     b_L^\beta\,,\label{qu12}\\
  Q^{cu}_1
  &=\;  \bar{s}_L^\alpha \gamma_{\mu} u_L^\beta\;
         \bar{c}_L^\beta \gamma^{\mu}b_L^\alpha\,,
  &
 Q^{cu}_2
  &=\;  \bar{s}_L^\alpha \gamma_{\mu} u_L^\alpha\;
         \bar{c}_L^\beta \gamma^{\mu}b_L^\beta\,, \label{qcu12}\\
    Q^{uc}_1
  &=\;
    \bar{s}_L^\alpha \gamma_{\mu} c_L^\beta\;
    \bar{u}_L^\beta \gamma^{\mu} b_L^\alpha\,,
  &
 Q^{uc}_2
  &= \bar{s}_L^\alpha \gamma_{\mu}c_L^\alpha \;
      \bar{u}_L^\beta     \gamma^{\mu}     b_L^\beta \,. \label{quc12}\\
  Q_3
  &=\; \bar{s}_L^\alpha \gamma_{\mu}     b_L^\alpha
     \sum_q \bar{q}_L^\beta\gamma^{\mu}     q_L^\beta\,,
  &
  Q_4
  &=\; \bar{s}_L^\alpha \gamma_{\mu} b_L^\beta
    \sum_q \bar{q}_L^\beta\gamma^{\mu}  q_L^\alpha\,,
     \label{pengop1} \\ 
  Q_5
  &=\;     \bar{s}_L^\alpha \gamma_{\mu} b_L^\alpha
    \sum_q \bar{q}_R^\beta\gamma^{\mu}  q_R^\beta\,,
  &
    Q_6
  &=\; \bar{s}_L^\alpha \gamma_{\mu}     b_L^\beta
     \sum_q \bar{q}_R^\beta\gamma^{\mu}     q_R^\alpha\,,
   \label{pengop2} \\ 
    Q_8
  &=\; \frac{g_s}{16\pi^2} m_b \, \bar{s}_L \sigma^{\mu \nu} T^a b_R \,
    G_{\mu\nu}^a\, 
        \label{operators}
\end{align}
with color indices $\alpha$ and $\beta$. The operators 
are depicted in \fig{fig:effp}.  Those in \eqsto{q12}{quc12}
are called \emph{current-current}\ operators and are generated by tree
diagrams. For the current-current operators $Q_{1,2}$ and $Q_{1,2}^u$ we
can draw the penguin diagrams of \fig{fig:effp}, which require a
counterterm proportional to a linear combination of the four-quark
\emph{penguin operators}\ $Q_{3-6}$.  This feature is called
\emph{operator mixing}. Other diagrams giving rise to operator mixing
are those found by dressing the diagrams in the first row of
\fig{fig:effp} with a gluon. Thus e.g.\ also $Q_1^{cu}$ and $Q_2^{cu}$ mix
with each other.  Furthermore, diagrams like the one in \fig{fig:tpeng}
match onto penguin operators, so that their Wilson coefficients depend on
$m_t/M_W$.
The corresponding hamiltonian for $b\to d$ transitions is found from
\eq{eq::HamDB1} by changing the CKM factors to
$  \lambda^d_a = V_{ad}^* V_{ab}$ and replacing the $s$ field
by $d$ in the definitions of the operators. The Wilson coefficients are
the same. 

Operator mixing implies that the renormalization group equations for the
Wilson coefficients $C_{1-6}$ are coupled, so that the low-energy values
$C_j(\mu_b)$ also depend on the initial conditions $C_k(\mu_{tW})$ of
other coefficients. At two-loop level and beyond, the four-quark
operators also mix into the \emph{chromomagnetic penguin operator}. The
RG evolution factor $u^{(0)} (\mu_b, \mu_{tW})$ of \eq{rg3} is replaced
by matrices in the case of $H^{|\Delta B|=1}$, the term proportional to
$ \lambda^s_t$ involves a $7\times 7$ RG evolution matrix and the terms
with the other three CKM factors instead involve (the same) $2\times 2$
matrix. This $2\times 2$ matrix is diagonal if one switches from the
operator basis $(Q_1^{x},Q_2^{x})$ (with $x=u,c,cu,uc$) to
$(Q_+^{x},Q_-^{x})$ where $Q_{\pm}^x=(Q_2^x\pm Q_1^x)/2$. The first diagonal
element, i.e.\ the anomalous dimension of $Q_+^x$, is the sames as for $Q$
in \eq{defQ}, which explains the notation in \eq{rg3}, and
$\gamma_-^{(0)}=-8$.  Details on $H^{|\Delta B|=1}$ and the RG evolution
of its coefficients can be found in
Refs.~\cite{Buras:1989xd,Buras:1991jm,Buras:1992tc,
  Buras:1992zv,Ciuchini:1993vr,Buchalla:1995vs}, which report the two-loop
results for the NLO anomalous dimension matrix and the NLO initial  conditions for
the Wilson coefficients, $C_k(\mu_{tW})$. The calculation of the latter
involve the one-loop QCD corrections to the SM $b$
decay amplitude and the corresponding corrections to the four quark
operators. The NNLO result for  $ H^{|\Delta B|=1} $, which required
a three-loop calculation, has been presented in Ref.~\cite{Gorbahn:2004my}.

As a first application, I discuss the mixing-induced $CP$ asymmetries,
for which we do not need to know the values of the Wilson coefficients. 
The amplitudes $A_f$ and $\bar A_f$ of \eq{defaf} read, with
$H_{\rm int}$ of \eq{sma} replaced by $H^{|\Delta B|=1}=
       H^{\Delta B=1}+H^{\Delta B=-1}$, 
\begin{align}
  A_f &=\; \bra{f} H^{\Delta B=1} \ket{B}, \qquad\qquad
        \bar A_f \;=\; \bra{f} H^{\Delta B=-1} \ket{\bar B}
\end{align}
where I have expanded the S-matrix to the lowest order in $G_F$.

Taking the $CP$ phase $\phi_{CP,B_s}^{\rm  mix}$ (equal to
$-2\beta_s$ in the SM) as an example, we identify the terms with
$Q_{1,2}$ in \eq{eq::HamDB1} as responsible for the dominant tree
amplitude. Neglecting penguin contributions we set $\lambda_u^s=0$
and replace $\lambda_t^s$ by $-\lambda_c^s$. Then
\begin{align}
  A_{(J/\psi \phi)_l}
  &=\;  \frac{4 G_F}{\sqrt2}\,
    \lambda^{s*}_c \sum_j C_j \, \bra{(J/\psi \phi)_l}
     Q_j^\dagger 
    \ket{B_s}, \qquad\qquad 
    \bar A_{(J/\psi \phi)_l} \;=\; \frac{4 G_F}{\sqrt2}\,
    \lambda^s_c \, \sum_j C_j \,
    \bra{(J/\psi \phi)_l} Q_j
    \ket{\Bbar_s}.  
\end{align}  
Applying the $CP$ transform of \eq{end} to the quark currents in $Q_j$
on finds $CP Q_j \mathcal (CP)^\dagger = Q_j^\dagger$. Then we
insert $(CP)^\dagger CP=1$ into our matrix element and use 
\eq{defcandcp} and $\eta_{{\rm CP}, (J/\psi \phi)_l}=(-1)^l$ to 
find
\begin{align}
  \bar A_{(J/\psi \phi)_l}
  &=\; \frac{4 G_F}{\sqrt2}\,
    \lambda^s_c \, \sum_j C_j \, 
    \bra{(J/\psi \phi)_l} (CP)^\dagger CP
    Q_j  (CP)^\dagger CP    \ket{\Bbar_s} 
  \;=\; -  \frac{4 G_F}{\sqrt2}\, 
    (-1)^l\, \lambda^s_c \, \sum_j C_j \bra{(J/\psi \phi)_l} Q_j^\dagger
    \ket{B_s} \;=\; 
    - (-1)^l\, \frac{\lambda^s_c}{\lambda^{s*}_c} \, A_{(J/\psi \phi)_l}
\end{align}
which fills in the missing details of the derivation of
  $\bar A_{(J/\psi \phi)_l}/A_{(J/\psi \phi)_l} $ in \eq{lajpsph}.

\begin{figure}[t]
\centering 
\includegraphics*[scale=0.7]{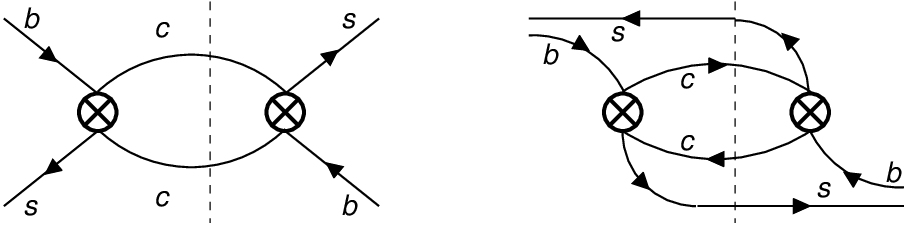}
\caption{Second-order contribution of $H^{|\Delta B|=1}$ 
  to \bbms. The diagrams with two charm quarks,
   found by contracting the $W$ lines in the corresponding box
   diagrams to a point,
   constitute the dominant contribution to 
   $\Gamma_{12}^s$ and involve two insertions of $Q_2$ of
   \eq{q12}.}\label{fig:dega}
\end{figure}

Next I discuss the calculation of $\Gamma_{12}^{d,s}$.  There is no
contribution of $H^{|\Delta B|=2}$ to $\Gamma_{12}^{d,s}$, because
$\bra{B_q} H^{|\Delta B|=2} \ket{\Bbar_q}$ has no absorptive part.
The relevant contributions instead come from two transitions
mediated by $H^{|\Delta B|=1}$. We expanding the S-matrix % in \eq{sma}
to second order:
\begin{align}
S&=\;
                       \mathrm{T}\!\exp \lt[-i \!
     \int \! d^4 x \, \lt(H^{|\Delta B|=2} (x)+H^{|\Delta B|=1}(x)\rt)   \rt] 
   \;=\; -i \int d^4 x\,  H^{|\Delta B|=2} (x) \,- \,
     \frac12 \int d^4 x\,d^4 y\, \mathrm{T}\lt[  H^{|\Delta B|=1} (x) H^{|\Delta
   B|=1} (y)\rt]
   \;+ \ldots
\end{align}
where the dots represent terms which do not contribute to $|\Delta B|=2$
transitions. 
The $\Delta B=2$ matrix element in the effective theory then reads
\begin{align}
  % -{\cal M}^{\rm eff}
  \Sigma_{12}\;=\; M_{12} -\frac{i}{2} \Gamma_{12}
   &=\; \bra{B} 
                     H^{|\Delta B|=2}(0)    \ket{\Bbar}  \; - \; 
                     \frac{i}{2} \bra{B}
                     \int d^4 x\, \mathrm{T}\lt[ H^{|\Delta B|=1}(x) H^{|\Delta
                     B|=1}(0) \rt]
                     \ket{\Bbar}.\label{s12}
\end{align}
The LO contributions to $\Gamma_{12} $, which stem from the second term
in \eq{s12}, are shown in \fig{fig:dega} for the case of \bbms.  The
contribution from the second, bilocal term \eq{s12} to \bbm\ is much
smaller than the one from $H^{|\Delta B|=2}$, which is enhanced due to
the heavy top mass entering \eq{sxt} while $\Gamma_{12}$ scales like
$m_b^2$.  Therefore we can neglect the bilocal contribution in $M_{12}$
and only need to consider it for $\Gamma_{12}$. From this observation we
also verify that $|\Gamma_{12}/M_{12}|= {\cal O}(m_b^2/m_t^2)$, which
I used in the discussion after \eq{mgqp:e}.

\eqsto{agmb}{mgsol:b} relate $\Gamma_{12}^q$ and $M_{12}^q $ to $\dg_q$ and
$a_{\rm fs}^q$ for $q=d$ or $s$:
\begin{align}
  \frac{\dg_q}{\dm_q}
  &\simeq \; -\real
                        \frac{\Gamma_{12}^q}{M_{12}^q},\qquad\qquad
  a_{\rm fs}^q
  \;\simeq \; \imag
                        \frac{\Gamma_{12}^q}{M_{12}^q}. \label{dgdm}
\end{align}
The calculation of the ratio $\Gamma_{12}^q/M_{12}^q$ has two
advantages over the calculation of $\Gamma_{12}^q$: The dependence
on $V_{cb}=A\lambda^2$, which normalizes the CKM parameter $V_{tq}$,
drops out from the ratio. Furthermore, the dominant contribution to
$\Gamma_{12}^q$ is proportional to the hadronic matrix element in
\eq{mel} which also enters $M_{12}^q$ and  therefore cancels 
to a large extent from the ratio.

$\Gamma_{12}^q$ involves two novel features compared to $M_{12}^q$:
Firstly, it is calculated in a power series in $\lqcd/m_b$ which results
from an OPE with $m_b$ as the hard scale called \emph{Heavy Quark
  Expansion (HQE)} \cite{Shifman:1984wx,Khoze:1986fa}. Technically, one
expands the loop diagrams of \fig{fig:dega} in the inverse of the
external $b$-quark momentum and matches the different terms onto matrix
elements of local $\Delta B=2$ operators. These operators are
pictorially found by contracting the hard loop momentum in bilocal
diagrams like those in \fig{fig:dega} to a point, leading to the
effective interaction of \fig{fig:q}.  Higher powers of $1/m_b$ in the
coefficients come with higher dimensions of the corresponding operators,
thus the $1/m_b$ corrections to the leading-power result involve
dimension-7 operators, whose matrix elements have an extra power of
$\lqcd$ compared to \eq{mel} (entering e.g.\ as a power of
$M_B-m_b$). Thus the prediction of $\Gamma_{12}^q$ is a double expansion
in the two parameters $\lqcd/m_b$ and $\alpha_s(m_b)$.  Secondly, the
leading-power contribution involves two operators, apart from $Q$ in
\eq{defq} this is \cite{Lenz:2006hd}
\begin{align}
  \widetilde{Q}_S &= \bar{q}_L^\alpha b_R^\beta \;
                    \bar{q}_L^\beta b_R^\alpha \label{defqq}
\end{align}
with matrix element 
\begin{align}
 \bra{B_q} \widetilde Q_S (\mu_2)\ket{\overline B_q} &= \frac{1}{3}  M^2_{B_q}\,
  f^2_{B_q} \widetilde  B_{S,B_q}^\prime (\mu_2) \label{defbs}
\end{align}
With $B_{S,B_s}^\prime (m_b)=1.31\pm 0.09$ and
$B_{S,B_d}^\prime (m_b)=1.20\pm 0.09$ \cite{Dowdall:2019bea} we find
this matrix element smaller than
$\bra{B_q}  Q \ket{\overline B_q}$ in \eq{mel} and the uncertainty
in $\Gamma_{12}^q/M_{12}^q$ from these matrix element is not an issue
in the predictions of the quantities in \eq{dgdm}.

For the discussion of $ a_{\rm fs}^q$ it helps to decompose $
\Gamma_{12}^q$ as \cite{Beneke:1998sy}
\begin{align}
        \Gamma_{12}^q &= - (\lambda_c^q)^2\Gamma^{cc}_{12} 
        - 2\lambda_c^q\lambda_u^q \Gamma_{12}^{uc} 
        - (\lambda_u^q)^2\Gamma^{uu}_{12} \\
                      &= -(\lambda_t^q)^2 \left[
                          \Gamma_{12}^{cc} 
                          + 2 \frac{\lambda_u^q}{\lambda_t^q}\left(\Gamma_{12}^{cc}-\Gamma_{12}^{uc}\right)
                          + \left(\frac{\lambda_u^q}{\lambda_t^q}\right)^2 
                          \left(\Gamma_{12}^{uu}+\Gamma_{12}^{cc}-2\Gamma_{12}^{uc}\right)
                          \right]
        .
        \label{eq::Gam12}
\end{align}
Here the superscript labels the quark flavors on the two internal lines,
$- (\lambda_c^s)^2 \Gamma^{cc}_{12}$ is shown in \fig{fig:dega}. To
prepare for the normalization to $M_{12}^q\propto \lambda_t^{q\,2}$ I
have traded $\lambda_c^q=-\lambda_t^q-\lambda_u^q$ for $\lambda_t^q$ in
the second line of \eq{eq::Gam12}. We observe that the terms
proportional to $\lambda_u^q/\lambda_t^q$ and
$(\lambda_u^q/\lambda_t^q)^2$ are GIM-suppressed, since they vanish for
$m_c=m_u$.

To discuss $\dg_q/\dm_q $ and $ a_{\rm fs}^d$ it is useful to define real
parameters $a$, $b$ and $c$ through \cite{Beneke:2003az}
\begin{eqnarray}
\frac{\Gamma_{12}^d}{M_{12}^d} 
        &=&  \frac{\lambda_t^{d\,2}}{M_{12}} \lt[ - \Gamma_{12}^{cc} \; +\;
        \,2\, 
                \lt( \Gamma_{12}^{uc} \, - \, \Gamma_{12}^{cc} \rt)
                \,  \frac{\lambda_u^{d}}{\lambda_t^{d\,2}} \; 
        +\;  
           \lt( 2\, \Gamma_{12}^{uc}  \, - \, \Gamma_{12}^{cc}   
                \, - \, \Gamma_{12}^{uu}  \rt)  
        \,  \frac{\lambda_u^{d\,2}}{\lambda_t^{d\,2}} \rt]\no\\[1mm]
        &\equiv&  
            10^{-4} \lt[ c + a 
                \,  \frac{\lambda_u^{d}}{\lambda_t^{d}}  \; + \;  
            b 
                \,  \frac{\lambda_u^{d\,2}}{\lambda_t^{d\,2}}\, \rt]. 
  \label{a14}
\end{eqnarray}
$a$, $b$, and $c$ depend on the particle masses and, in particular, 
are functions of 
\begin{align}
  z&=\, \frac{m_c^2}{m_b^2}  \label{defz}
\end{align}     
By calculating $\Gamma_{12}^{ab}$ one finds that $a$ is linear in $z$.
$b$ is proportional to $(\lambda_u^q/\lambda_t^q)^2$ and negligible,
because it is proportional to $z^3$ at LO and $\alpha_s z^2$ at
NLO. Furthermore we verify from \eq{eq:beta} that both
$|\lambda_u^s/\lambda_t^s|\propto \lambda^2$ and
$\real (\lambda_u^d/\lambda_t^d) \propto \cos\alpha$ are small; in the
latter case this stems from the fact that $\alpha$ happens to be close
to 90$^\circ$. Thus $\dg_q/\dm_q=-\real(\Gamma_{12}^q/M_{12}^q) $ is
dominated by $c$ for both $q=d$ and $q=s$.

For  $a_{\rm fs}^q$, however, one observes
\begin{align}
  a_{\rm fs}^q &= \;
  \left[a\, \imag \frac{\lambda_u^d}{\lambda_t^d} +
    b\, \imag \frac{(\lambda_u^d)^2}{(\lambda_t^d)^2} \right]\cdot 10^{-4}
                \; \propto \;   
                 z 
                 \label{eq:afsgaab} 
\end{align}
from \eq{a14}, so that $ a_{\rm fs}^d$ is suppressed w.r.t.\
$\dg_d/\dm_d$ by a factor of $z$. Furthermore, in $ a_{\rm fs}^s$ there
is an additional CKM suppression from
$\imag (\lambda_u^s/\lambda_t^s) \propto \lambda^2\simeq 0.05$.

\begin{figure}[t]
\centering 
\includegraphics*[width=0.6\textwidth]{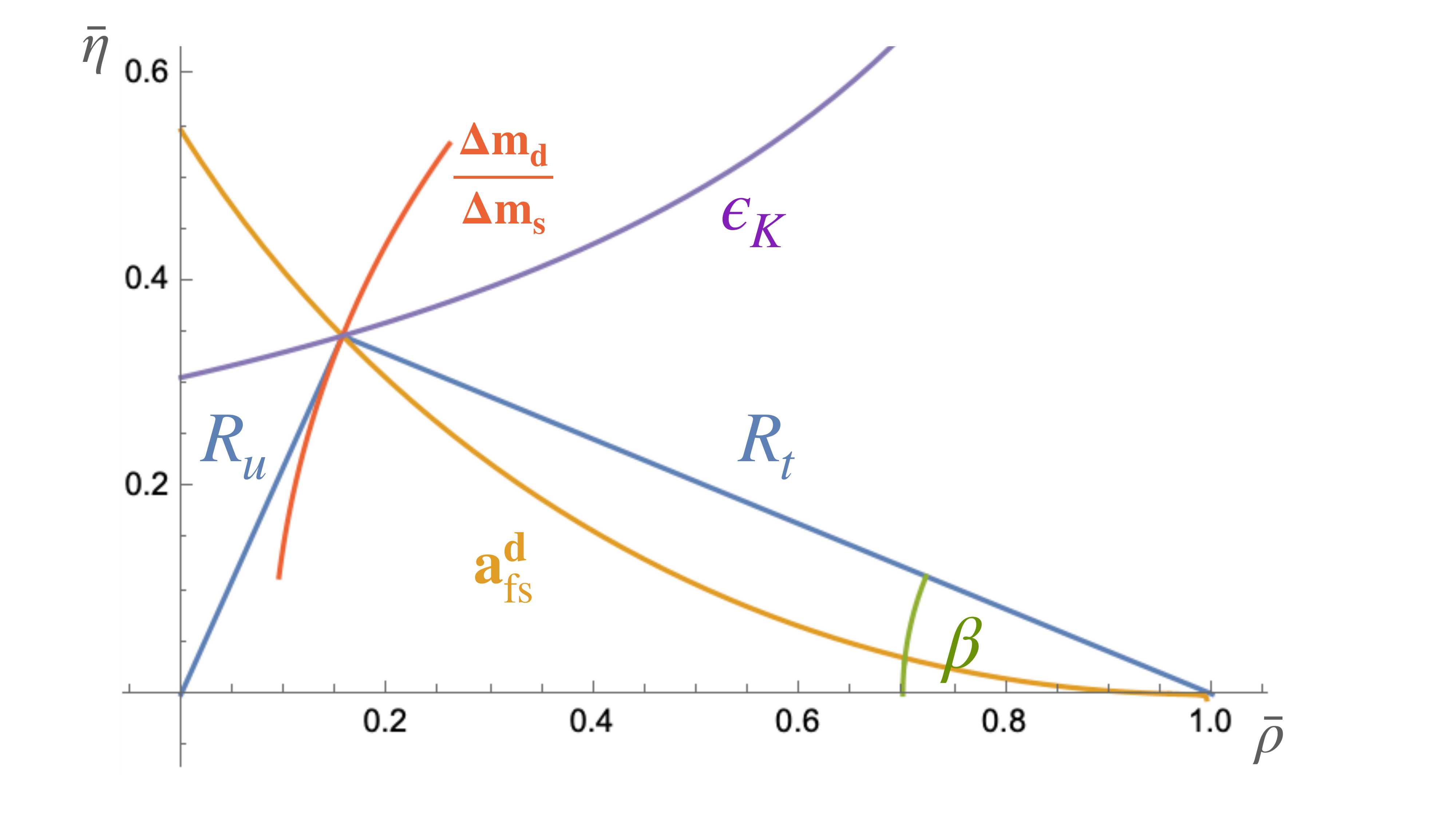}
\caption{Contraints on the unitarity triangle from
      $\dm_d/\dm_s$, $\epsilon_K$, $a_{\rm fs}^{d}$, and 
      $ A_{CP}^{\rm mix}(B_d\to J/\psi K_{\rm short})=-\sin(2\beta)$,
      schematically for a hypothetical perfect agreement with the SM and no uncertainties. 
  \label{fig:utall}}
\end{figure}

With
\begin{equation}
  \frac{\lambda_u^d}{\lambda_t^d} =
  \frac{1-\bar\rho - i \bar\eta}{(1-\bar\rho)^2+\bar\eta^2} - 1
  \,.\label{eq:lambdUoT_param}
\end{equation}
we can express
$\imag (\lambda_u^d/\lambda_t^d) $ in terms of UT parameters. Neglecting
the small term with $b$ one finds
\begin{align}
  a_{\rm fs}^d & =\;  \imag \frac{\lambda_u^q}{\lambda_t^q}  \, a \cdot
                 10^{-4} \;=\; - \frac{\bar\eta}{{ (1 - \bar\rho{}
                 )^2+\bar\eta{}^2}} \, a \cdot
                 10^{-4} \;= \; -  \frac{\sin\beta}{R_t} \, a \cdot
                 10^{-4}
                 \label{afsert}
\end{align}
from \eq{eq:afsgaab}.  From the third expression one realizes that a
future improved measurement of $a_{\rm fs}^d$ in \eq{eq:expafsd} will
define a circle in the $\bar\rho$-$\bar\eta$ plane which gives
complementary information to the circle found from $\dm_d$, the line
from $ A_{CP}^{\rm mix}(B_d\to J/\psi K_{\rm short})=-\sin(2\beta)$, and
other standard observables used in UT phenomenology, see
\fig{fig:utall}. From \eq{afsert} one easily finds the equation for the
desired new circle:
\begin{eqnarray}
  (\ov \eta - R_{\rm fs})^2 + (1 - \ov \rho)^2 \; =\; R_{\rm fs}^2
  && \qquad\qquad \mbox{with }\qquad 
      R_{\rm fs} =\,  - \frac{a}{2 a_{\rm fs}^{d\, \rm exp}}, \label{circ} 
\end{eqnarray}
where $ a_{\rm fs}^{d\, \rm exp}$ denotes the experimental value of
$ a_{\rm fs}^{d}$.  Thus the circle is centered on the vertical line
with $\bar\rho=1$ and intersects the point $(\bar\rho,\bar\eta)=(1,0)$;
its radius $R_{\rm fs}$ is slightly larger than 1 if the SM describes
$ a_{\rm fs}^{d}$ correctly.

The last expression in \eq{afsert} involves the two
quantities inferred from measurements of
$A_{CP}^{\rm mix}(B_d\to J/\psi K_{\rm short})$ and $\dm_d$. Thus
measuring all of $A_{CP}^{\rm mix}(B_d\to J/\psi K_{\rm short})$,
$\dm_d$, and $ a_{\rm fs}^d$ over-constrains $\beta$ and $R_t$ and thus constitutes
a probe of BSM physics in \bbmd\ \emph{alone}, without the need of input
from other quantities. Pictorially, BSM physics in  \bbmd\ will reveal
itself in this way if the intersection of the circles from $\dm_d$ and
 $ a_{\rm fs}^d$ will not be spiked by the line inferred from $A_{CP}^{\rm
   mix}(B_d\to J/\psi K_{\rm short})$.

The OPE matches the result of the diagrams in \fig{fig:dega} (and the
corresponding ones with one or two up quarks) onto local $\Delta B=2$
operators to yield an expression of the form
\begin{align}
        \Gamma_{12}^{ab} 
        &=\;  \frac{G_F^2m_b^2}{24\pi M_{B_q}} \left[ 
        H^{ab}(z)   \langle B_q|Q|\bar{B}_q \rangle
        + \widetilde{H}^{ab}_S(z)  \langle B_q|\widetilde{Q}_S|\bar{B}_q \rangle
        \right]
        + \mathcal{O}\lt( \frac{\Lambda_{\rm QCD}}{m_b} \rt) \,,
        \label{eq::Gam^ab}
\end{align}
with new Wilson coefficients $H^{ab}(z)$ and
$\widetilde{H}^{ab}_S(z)$. The diagrams of \fig{fig:dega} determine them
to LO, for the NLO and NNLO results one had to calculate diagrams with
one and two extra gluons, respectively, as well as the corresponding
diagrams for the $\Delta B=2$ operators.

$\Gamma_{12}^q$ has been calculated to LO in
Refs.~\cite{Hagelin:1981zk,Franco:1981ea,
  Chau:1982da,Buras:1984pq,Khoze:1986fa,Datta:1987gw,Datta:1988um} with
focus on the predictions of $\dg_{s}$ and $a_{\rm fs}^d$. The NLO
prediction of the contribution with the large Wilson coefficients $C_1$,
$C_2$, and $C_8$ to $\dg_s$ was presented in
Refs.~\cite{Beneke:1998sy,Ciuchini:2003ww,Lenz:2006hd}, the NLO results
for $a_{\rm fs}^{d,s}$ and $\dg_d$ were derived in
Refs.~\cite{Ciuchini:2003ww,Lenz:2006hd,Beneke:2003az}.  The NLO
contribution with the small (of order 0.05) four-quark penguin
coefficients $C_3\ldots C_6$ was obtained in Ref.~\cite{Gerlach:2021xtb,
  Gerlach:2022wgb}.  The NNLO calculation of the three-loop diagrams has
been tackled in terms of an expansion in $z$. The contribution
calculated first only involved three-loop diagrams with fermion loops
\cite{Asatrian:2017qaz,Asatrian:2020zxa,Hovhannisyan:2022miy}.
Ref.~\cite{Gerlach:2022wgb} contains the NNLO results with one $C_8$ and
one four-quark coefficient, which only involve two-loop diagrams. The
numerically most important piece of the NNLO prediction stems from
three-loop diagrams with two current-current operator coefficients
$C_{1,2}$, found in an expansion to order $z$ in
Ref.~\cite{Gerlach:2022hoj} and to order $z^{50}$ in
Ref.~\cite{Gerlach:2025tcx}, except for the charm mass effects in the
tiny gluon self-energy diagrams, which are only known to order $z^6$.
The result of Ref.~\cite{Gerlach:2022hoj} is satisfactory for $\dg_q$,
but not for $a_{\rm fs}^q\propto z$, for which the calculation in
Ref.~\cite{Gerlach:2025tcx} was needed.  Ref.~\cite{Gerlach:2022wgb}
also contains a first step towards NNNLO, with the calculation of two-loop
diagrams proportional to $C_8^2$.

The cited calculations are all for the leading-power term, i.e.\ they
address QCD corrections to $H^{ab}(z)$ and $\widetilde{H}^{ab}_S(z)$ in
\eq{eq::Gam^ab}.   The NNLO predictions are \cite{Gerlach:2025tcx}
\begin{align}
        \frac{\Delta \Gamma_s}{\Delta M_s} 
  & =\, (
      4.37_{-0.44}^{0.23}{}_{\rm scale}  \pm  0.12_{\rm matrix el.}
\pm 0.79_{1/m_b} \pm 0.05_{\rm input}) \times 10^{-3}\;
    (\overline{\textrm{MS}}) \nn
         \frac{\Delta \Gamma_s}{\Delta M_s} 
  &=\, (
      4.27_{-0.37}^{0.36}{}_{\rm scale}  \pm  0.12_{\rm matrix el.}
\pm 0.79_{1/m_b} \pm 0.05_{\rm input}) \times 10^{-3}\; ({\textrm{PS}})
\label{dgdmres}
\end{align} 
The two results correspond to the modified minimal subtraction
($\overline{\textrm{MS}}$) and potential-subtracted ($\textrm{PS}$)
renormalization schemes. The dependence on renormalization scheme and
scale diminishes order-by-order of $\alpha_s$; it is commonly used as an
estimate of the uncertainty related to the omission of unknown higher
orders of $\alpha_s$.  The indicated scale dependence of the NNLO result
in \eq{dgdmres} is slightly below 9\%, which is larger, but close to the
experimental error in \eq{dgsexp}.  The second uncertainty stems from
the ratio of the hadronic matrix elements, i.e.\ from
$B_{S,B_s}^\prime/B_{B_s}$. The dominant source of uncertainty are the
poorly know matrix elements of the dimension-7 operators entering the
$\lqcd/m_b$ corrections. Moreover, the coefficients of the dimension-7
operators are only known to LO \cite{Beneke:1996gn}, and an NLO
calculation is necessary for a meaningful lattice-continuum matching of
the matrix elements.  The last uncertainty in \eq{dgdmres} stems from the physical
parameters, mostly from CKM elements. One should keep in mind that their
values and uncertainties are found under the assumption that there is no
new physics in the quantities entering the global fit of $\bar\rho$ and
$\bar\eta$ from data. Thus future measurements in tension with the
presented SM predictions may not necessarily be related to new physics
in the quoted quantities but instead in the quantities from which
$\bar\rho$ and $\bar\eta$ are determined.

Currently, we are far away from a measurement of $\dg_d$ with a 
precision comparable to $\dg_s$ in \eq{dgsexp}. The SM predictions 
for $\dg_d/\dm_d$ and  $\dg_s/\dm_s$ are almost equal and, for the time
being, one can use
\begin{align}
  \frac{\Delta \Gamma_d}{\Delta M_d} &=\; 0.963  \frac{\Delta \Gamma_s}{\Delta M_s}
\label{eq:canuse}
\end{align}                    
for the results in both renormalization schemes. Once better measurements
than those summarized in \eq{dgdexp} will be available, one should
use the precise formulae of Ref.~\cite{Gerlach:2025tcx}.

One can use the experimental values in \eqsand{wavdmd}{wavdms} to predict
\cite{Gerlach:2025tcx}
\begin{align}
  \Delta \Gamma_d &=\, \frac{\Delta \Gamma_d}{\Delta M_d}
                    \dm_d^{\rm exp} \,=\, 
                    {{(0.00211 \pm 0.00045)}} ~\mbox{ps}^{-1}, \qquad\qquad 
  \Delta \Gamma_s\,=\, \frac{\Delta \Gamma_s}{\Delta M_s} \dm_s^{\rm exp}
  \, =\, ({0.077}\pm 0.016)\,\mbox{ps}^{-1}
\label{dgres}
\end{align} 
from \eq{dgdmres}, with the numbers being the averages of the 
$\ov{\rm MS}$ and PS schemes. The benefit of a future measurement
of $\Delta \Gamma_d$ will be a precise test of BSM physics in
$\Gamma_{12}^{d,s}$  through the ratio $\dg_d/\dg_s$, because most of
the uncertainties in \eq{dgdmres} drop out from this ratio. 

The NNLO predictions  for the coefficients in \eq{a14} are \cite{Gerlach:2025tcx}
\begin{align} %NNLO
  a &=\, 12.2 \pm 0.6 , \qquad\qquad
   b\,=\, 0.23 \pm 0.06    
\end{align}
and  a breakdown of the different sources of uncertainties can be found
in Ref.~\cite{Gerlach:2025tcx}. $c$ in \eq{a14} essentially equals
$10^4\cdot \real (\Gamma_{12}^d/M_{12}^d)= - 10^4\cdot \dg_d/\dm_d$, which
amounts to
\begin{align} %NNLO
 c&=\, -42\pm 9 .\label{resc}
\end{align}
Using the values from a recent global fit \cite{CKMfitter},
\begin{align}
    \frac{\lambda^d_u}{\lambda^d_t} &=\,   (0.0105 \pm 0.0107) - (0.4259
                                      \pm 0.0091){\rm i},
                                      \qquad\qquad
  \frac{\lambda^s_u}{\lambda^s_t} \,=\, -(0.00877 \pm 0.00043) +(0.01858
                                      \pm 0.00038){\rm i}, \label{eq::ckm_input}
\end{align}
one finds \cite{Gerlach:2025tcx}
\begin{align*}
    a_{\rm fs}^d &=\,  -\left({5.21 \pm 0.32} \right) \times 10^{-4}, \qquad\qquad
    a_{\rm fs}^s \,=\, \phantom{-} \left({2.28 \pm 0.14} \right) \times
                   10^{-5}\,.
                   \label{afsres}
\end{align*}
Note that the parameters $a$, $b$, $c$ for \bbms\ are slightly different
from those in \bbmd, similar to the situation in \eq{eq:canuse}. 
The smallness of $ a_{\rm fs}^s $ compared to $a_{\rm fs}^d$ stems from
the small imaginary part of the CKM factor in \eq{eq::ckm_input},
$\imag (\lambda^s_u/\lambda^s_t) \simeq \bar\eta \lambda^2 $. 

In Ref.~\cite{Laplace:2002ik} it was pointed out that $ a_{\rm fs}^d$ is
very sensitive to new physics in $M_{12}^d$, as a small BSM $CP$ phase
spoils the approximate phase alignment of $\Gamma_{12}^d$ and
$M_{12}^d$, so that $ a_{\rm fs}^d$ picks up a term with the large
coefficient $c$ in \eq{resc} which is not suppressed by $z$. This
analysis was later extended to BSM physics in $\Gamma_{12}^d$
\cite{Bobeth:2011st,Bobeth:2014rda}. 
If one parameterizes BSM physics as
\begin{align}
 \frac{\Gamma_{12}^q}{M_{12}^q} &= \,  \frac{\Gamma_{12,\rm
                                  SM}^q}{M_{12,\rm SM}^q} \, d_{\rm NP}^q e^{i\phi_{\rm NP}^q},
\end{align}  
with $d_{\rm NP}^q>0$ and a BSM $CP$ phase $\phi_{\rm NP}^q$, 
one finds \cite{Laplace:2002ik,Beneke:2003az,Bobeth:2011st}
\begin{align}
   a_{\rm fs}^q &=\,  a_{\rm fs, SM}^q \, d_{\rm NP}^q  \cos \phi_{\rm
                  NP}^q \, -\, \frac{\dg_{q,\rm SM}}{\dm_{q,\rm SM}} \,
                   d_{\rm NP}^q  \sin \phi_{\rm NP}^q . 
\end{align}
Thus new physics enhances $ a_{\rm fs}^q$ with a lever arm of
$(\dg_{q,\rm SM}/\dm_{q,\rm SM})/ a_{\rm fs, SM}^q$ multiplying
$ \sin \phi_{\rm NP}^q$. This enhancement factor equals 8 for $q=d$ and
190 for $q=s$ when using the central values in \eqsand{dgdmres}{afsres}.
While the effect of BSM physics is spectacular for $ a_{\rm fs}^s$, the
experimental value in \eq{acpmexps} permits only BSM contributions of a
few degrees in $M_{12}^s$. $\Gamma_{12}^s$ cannot receive sizable BSM
contributions either. Still, for $\phi_{\rm NP}^ s=-5^\circ$ and $d_{\rm NP}^s=1$ we find $
a_{\rm fs}^q$ enhanced by more than a factor of  15 to $a_{\rm fs}^s=4
\cdot 10^{-4} $. The room for BSM physics in $ a_{\rm fs}^d$ is larger,
because a precise SM prediction for $M_{12}^d$ suffers from the unclear
situation with $V_{cb}$ and $V_{ub}$ and the doubly Cabibbo-suppressed
$\Gamma_{12}^d$  can receive BSM contributions as well, so that $|a_{\rm
  fs}^d |$ above $10^{-3}$ cannot be excluded now.

In summary, the decay matrix element $\Gamma_{12}^{d,s}$ determines
$\dg_{d,s}$ and $a_{\rm fs}^{d,s}$. While $\Gamma_{12}^{d,s}$ originates from
a $\Delta B=2$ transition, there is no contribution from the
$|\Delta B|=2$ hamiltonian to this quantity, which instead involves two
$|\Delta B|=1$ interactions as shown in \fig{fig:dega}. The
corresponding $|\Delta B|=1$ hamiltonian in \eq{eq::HamDB1} describes
decays of $b$-flavored hadrons; $\Gamma_{12}^q$ receives
contributions from all decays to final states which are common to $B_q$
and $\Bbar_q$, as shown in \eq{ga12af}.  $\dg_s$ is precisely measured
and a calculation of QCD corrections to $\Gamma_{12}^q$ at NNLO was
necessary to bring the perturbative uncertainty to a level which is
similar to the experimental error.  A future measurement of $a_{\rm
  fs}^{d,s}$ will define the circle of \eq{circ} in the $\bar\rho$-$\bar\eta$ plane,
which gives new information on $(\bar\rho,\bar\eta)$, complementary
to other constraints. In particular, in combination with $\dm_d$ and
$A_{CP}^{\rm mix}(B_d\to J/\psi K_{\rm short})$ a measurement of the $CP$ asymmetry 
$a_{\rm  fs}^{d}$ will over-constrain the UT and provide a SM test of BSM physics
in \bbmd\ without the need of external input from other observables
entering the global UT fit.  $\dg_d$ and  $a_{\rm  fs}^{s}$ play roles
as BSM physics tests; in the case of $a_{\rm  fs}^{s}$ already small BSM
CP phases can enhance this quantity from its tiny SM prediction to a value
accessible at experiment.  

\boldmath
\subsection{Effective hamiltonians for \kkm\ with predictions for ${\e_K}$
  and ${\dm_K}$, overall picture of the UT\label{sec:kaon}}
\unboldmath%
The formalism to describe $CP$ violation in \kkm\ and the relation of
$CP$ asymmetries to the %fundamental
$CP$ phase $\phi_K=\arg (-M_{12}^K/\Gamma_{12} ^K)$ has been presented
in \eqsto{siso}{phikres2}. The quantity $\e_K$ introduced in \eq{ek}
encodes both $CP$ violation in mixing and mixing-induced $CP$ violation,
while the semileptonic $CP$ asymmetry $A_L=a_{\rm fs}^K/2$ of \eq{aphi}
is a measure of only the former type of $CP$ violation. $A_L$ and
$\real \epsilon_K$ provide the same information on $\phi_K$. In contrast
to \bb\ and \ddm\ $\imag \epsilon_K$ encoding mixing-induced $CP$
violation in the decay $K\to (\pi\pi)_{I=0}$ does not provide new
information compared to $a_{\rm fs}^K$ and also determines $\phi_K$, see
\eq{ephi}.  I will now show how $\e_K$ is calculated in the SM.

In \kkm\ it is common practice to adopt the standard phase convention
for the CKM matrix in which $V_{us}V_{ud}^*$ is real and
positive. Starting from from \eq{defphi}, we write
\begin{eqnarray}
   \phi &=& \arg \left( -\frac{M_{12}^K}{\Gamma_{12}^K} \right) 
       \; \simeq \; \frac{\imag M_{12}^K}{|M_{12}^K|} - 
                    \arg(-\Gamma_{12}^K) 
       \; =\; 2 \, \lt[ \frac{\imag M_{12}^K}{\dm_K^{\rm exp}} + \xi_K \rt] 
\label{phimg}
\end{eqnarray}
where 
\begin{eqnarray}
  2 \xi_K &\equiv & - \arg(-\Gamma_{12}^K) 
          \;\simeq \; 
                    - \arg \lt( - \frac{\ov A_0}{A_0} \rt)
                    . \label{defxik}
\end{eqnarray}
In \eq{phimg} I have used that the phases of both $M_{12}^K$ and $-\Gamma_{12}^K$
are small in the standard CKM phase convention and further 
traded $|M_{12}^K|$ for the experimental $\dm_K/2$. Furthermore, in \eq{defxik} 
the saturation of $\Gamma_{12}$ by $A_0^*\, \ov{A}_0$ in \eq{g12k} 
has been used. Inserting \eq{phimg} into our result for $\e_K$ in
\eq{ephi} gives
\begin{align}
  \e_K &\simeq\,    \sin (\phi_\e) e^{i \phi_\e} \, 
          \lt[ \frac{\imag M_{12}^K}{\dm_K^{\rm exp}} + \xi_K \rt] .\label{ekm12}
\end{align}
The term with $\xi_K$ encodes the CP-odd phase in
$\ov K \to (\pi\pi)_{I=0}$, which we expect to appear in a quantity
measuring mixing-induced $CP$ violation. It only contributes $-6$\% to
$\e_K$ \cite{Buras:2008nn} (a recent lattice determination
finds $|\xi_K|$ slightly larger \cite{Jwa:2025fon}) and will be briefly discussed below. The
ballpark contribution to $\e_K$ stems from $\imag M_{12}^K$, i.e.\ from the
familiar \kk\ box diagram.

The LO $\Delta S=2$ transition amplitude ${\cal M}^{(0)}$ corresponding to the \kkm\ box
diagram in \fig{fig:boxes} is found from the corresponding $\Delta B=2$
expression in \eq{sij} by the substitutions $b\to s$ and $q\to d$ in the
CKM elements. Everything else is unchanged, in particular we encounter
the same Inami-Lim functions as in \bbm. An important difference is the
hierarchy among the three CKM combinations in \eq{sij}, the smallness of
$ ( V_{ts}V_{td}^* )^2 \simeq A^4 \lambda^{10} (1-\ov\rho -i \ov \eta)^2$
compensates the large size of $S(x_t)$ and the terms with $S(x_c,x_t)$
and $S(x_c)$ become important, where I recall the definition $x_q\equiv
m_q^2/M_W^2$.  For $\e_K$ in \eq{ekm12} we need 
the imaginary parts of the CKM factors. To lowest order in the
Wolfenstein expansion one finds
\begin{align}
    \imag  ( V_{ts} V_{td}^* )^2 &\simeq\; 2 (A \lambda^2)^4 \lambda^2
                                     \, \ov\eta \, (1-\ov \rho ), \qquad\qquad
2\,  \imag  (V_{ts} V_{td}^* V_{cs} V_{cd}^*) \;\simeq \;
  - \imag  ( V_{cs} V_{cd}^* )^2  \simeq 2 (A \lambda^2)^2 \lambda^2
  \, \ov \eta  \label{theimags}
\end{align}
and the numerical hierarchy becomes evident from $A \lambda^2=|V_{cb}|=0.04$.
Thus, in addition to $S(x_t)$ in \eq{sxt} we also need $S(x_c)= x_c
+{\cal O}(x_c^2)$ and  
\begin{align}
S(x_c,x_t) &=\; - x_c \ln x_c 
+x_c \lt[ \frac{x_t^2-8 x_t+4 }{4 (1-x_t)^2} \ln x_t 
          + \frac{3}{4} \frac{x_t}{x_t-1}   \rt] + O(x_c^2 \ln x_c) 
\label{sxcxt}   .
\end{align}  
These three contributions require a very different treatment of their
QCD corrections. The term with $S(x_t)$ follows the pattern which we
discussed for \bbm, leading to the effective hamiltonian in \eq{desh2}.
There is only one difference between the QCD correction factor
$\eta_B b_B(\mu_b)$ in \eq{desh2} and its counterpart $\eta_{tt}
b_K(\mu)$ in the $|\Delta S|=2$ hamiltonian: The RG evolution of the
coefficient of the local operator
\begin{align}
  Q & =\;  
    \ov{d}_L \gamma_{\nu} s_L \, \ov{d}_L \gamma^{\nu} s_L .  
\label{defQk}
\end{align}   
must be carried to a lower scale $\mu_K={\cal O}(1\gev)$ and to this end
one must match $Q$ from five-flavour QCD to three-flavour QCD
at an intermediate scale $\mu_{bc}={\cal O}(m_c)$. This requires
to change $f=5$ to $f=3$ in the evolution of  
\eq{rg3} for $\mu_K \leq \mu \leq \mu_{bc}$. Beyond LO there is also a threshold
correction $C^{|\Delta S|=2, f=3}(\mu_{bc})/C^{|\Delta S|=2,
  f=5}(\mu_{bc}) $, which is numerically very small. 

The other two contributions, with light charm and up quarks on the
internal lines,  require the consideration of terms with two
$|\Delta S|=1$ hamiltonians. We have encountered this piece as the
second term of $\Sigma_{12}$ in \eq{s12} in the discussion of $|\Delta
B|=2$ transitions. This term does not only contribute to $\Gamma_{12}$,
but also to $M_{12}$ but is negligible in the case of $|\Delta
B|=2$ transitions. In \kkm\ the transition amplitude at intermediate
scales  $\mu_{bc} \leq
\mu \leq \mu_{tW}$ involves both a local $|\Delta S|=2$ hamiltonian
and the bilocal matrix element with two copies of the
$|\Delta S|=1$ hamiltonian, both of which are obtained from their
$|\Delta B|=1,2$ counterparts by appropriately replacing the quark
fields in the operators and changing the CKM elements.
When we arrive at the renormalization scale $\mu_{bc}$, we must match
our $\Delta S=2$ amplitude of the five-flavour theory to an amplitude
in a theory which only has $u$, $d$, and $s$ as dynamical quark fields.
For $\mu \leq \mu_{bc}$ the $\Delta S=2$ transition solely involves the local
$\Delta S=2$ operator of \eq{defQk}, just as in the case of the 
top quark contribution proportional to $(V_{ts}V_{td}^*)^2$ discussed first. 
The two-step matching of the contributions with  $(V_{cs}V_{cd}^*)^2$
and $V_{cs}V_{cd}^*V_{ts}V_{td}^*$ with the RG evolution between the
scales $\mu_{tW}$ and  $ \mu_{bc}$ leads to a result in which the
product of $\alpha_s$ and the large
logarithm $\ln x_c$ is summed to all orders in perturbation theory.
The final result for the effective hamiltonian reads
\begin{eqnarray}
H^{|\Delta S|=2} & = & \frac{G_F^2}{4 \pi^2}\, M_W^2 \, 
  \lt[ ( V_{ts} V_{td}^* )^2 \, \eta_{tt} \, S ( x_t ) \, + \,
      2 V_{ts} V_{td}^* V_{cs} V_{cd}^*  \, \eta_{ct} \, 
                                      S (x_c, x_t ) % \rt. \nn 
                                      % && \lt. \qquad \qquad \qquad \qquad
                                           \,+ \,
       ( V_{cs} V_{cd}^* )^2 \, \eta_{cc} \, x_c  \rt] \,
   b_K(\mu_K) Q(\mu_K) 
  \; + \; h.c. \label{deshs2}
\end{eqnarray}
with the QCD corrections encoded in $\eta_{tt}$, $\eta_{ct}$, and
$\eta_{cc}$, with a common factor   $b_K(\mu_K)$ factored out. 
The hadronic matrix elements is parametrized as
\begin{align}
 \bra{K} Q(\mu_K) \ket{\,\ov{\!K}} & =\; 
  \frac{2}{3} M_{K}^2 \, f_K^2 \, 
  \frac{\widehat B_K}{ b_K(\mu_K)} , \label{melk}% 
\end{align}
where $M_K=497.6\mev$ \cite{ParticleDataGroup:2024cfk} and $f_K= 156\mev$
\cite{FlavourLatticeAveragingGroupFLAG:2024oxs}
are Kaon mass and decay constant, respectively.
With $M_{12}^K= \bra{K} H^{|\Delta S|=2} \ket{\Kbar}$ we can determine
$\imag M_{12}^{K}$ in terms of  $\widehat B_K$ and write for $\e_K$ in
\eq{ekm12}:
\begin{align}
1.21 \cdot 10^{-7} &=\; \widehat B_K 
\left[ 
       \imag (V_{ts}V_{td}^*)^2 \, \eta_{tt} \, S(x_t) +
        2 \, \imag 
        \left( V_{ts}V_{td}^* V_{cs}V_{cd}^* \right) \, \eta_{ct} \, 
          S(x_c,x_t) + \imag  (V_{cs}V_{cd}^*)^2 \, \eta_{cc} \, x_c \right] . \label{cons} 
\end{align}
Here the number on the LHS originates from 
\begin{align}
  % M_W= 80.370 
  % Gf=1.16637 10^(-5)
  % M_K=0.4976
  % fk= 0.156
  % dm_K = 3.476 \pm 0.006 *10^{-15} GeV
  % phi_epsilon = 0.97 \pm 0.01  Pi/4
  % |eps_K| =  2.228 \pm 0.011) \times 10^{-3}
  1.21 \cdot 10^{-7} 
&=\; 
 \frac{12   \pi^2\, \dm_K^{\rm exp}}{G_F^2 \, f_K^2\, M_K \, M_W^2\sin\phi_{\e}}
  \left( |\epsilon_K^{\rm exp}|   - \tilde{\xi}_K \sin \phi_{\e} \right)
     \label{numbek}
\end{align}
where I have used the numbers quoted in this report as well as $G_F=1.16637
10^{-5}\gev^{-2}$ and $M_W=80.370\gev$.
In \eq{numbek} $\tilde\xi_K \sim 0.6 \xi_K$ subsumes $\xi_K$ in \eq{ekm12} and
the bilocal contribution to $\imag M_{12}^K$ from two insertions of the
$|\Delta S|=1$ hamiltonian \cite{Buras:2010pza,Bai:2023lkr,Jwa:2025fon},
i.e.\ the bilocal ``long-distance'' matrix element
$\bra{K} \int d^4 x\, \mathrm{T}\lt[ H^{|\Delta S|=1}(x) H^{|\Delta
                     S|=1}(0) \rt] \ket{\Kbar}$ contributes to both
                   $\Gamma_{12}^K$ and $M_{12}^K$.
The effect of the term with $\tilde\xi_K$ in \eq{numbek} can be implemented as
a 3\% reduction of the RHS via  $ |\epsilon_K^{\rm exp}|   - \tilde{\xi}_K
\sin \phi_{\e} \simeq 0.97 |\epsilon_K^{\rm exp}|$.  We can use    \eq{theimags}             
to express \eq{cons} in terms of $\bar\rho$ and $\bar\eta$. Dividing
\eq{cons} by $2\lambda^2$ with $\lambda=0.225$ and trading $A$ for
$|V_{cb}| = A \lambda^2$ one finds
\begin{align}
1.20 \cdot 10^{-6} &=\; \widehat B_K \,  |V_{cb}|^2\, \, \ov \eta \, 
\left[ |V_{cb}|^2\, (1-\ov \rho )\eta_{tt} S(x_t)
       \, +  
                     \, \eta_{ct} \, 
          S(x_c,x_t) \, -  \, \eta_{cc} \,  x_c \right] , \label{cons2} 
\end{align}
where a tiny term of order $\lambda^2$ (to be found in
Ref.~\cite{Nierste:1996zu})
has been neglected.

Since the perturbative calculation of $\eta_{ct}$, and $\eta_{cc}$ involves an expansion in
$\alpha_s(\mu_{bc}) \sim 0.3$, we must calculate these coefficients to
higher order than in the case of $\eta_{tt}$. The development of the
effective-hamiltonian  framework and the 
LO calculation of $H^{|\Delta S|=2}$ was performed by Gilman and Wise
\cite{Gilman:1982ap} for the case of a top quark mass far below
$M_W$. The calculation confirmed the results for $\eta_{tt}$ and
$\eta_{cc}$ found before with other methods
\cite{Vainshtein:1975xw,Vysotsky:1979tu}. Later the LO calculation was
extended to the case of a heavy top quark \cite{Flynn:1989cf}.

The motivation for the NLO calculation of $H^{|\Delta S|=2}$ was, in the
first place, to find out whether RG-improved perturbation theory works
at all in a non-leptonic process involving low scales with
$\alpha_s\gtrsim 0.3$. While the NLO calculation of $\eta_{tt}$, which
only involves $\alpha_s(\mu_{tW})$, showed a good behaviour of the
perturbative series \cite{Buras:1990fn}, the fate of a reliable
prediction of $\e_K$ depended on the control over $\eta_{ct}$, and
$\eta_{cc}$. The NLO calculation of $\eta_{cc}$ indeed showed a
disturbingly large positive correction of 65\%, which, however, could be traced back to
an accidental cancellation  among different terms in the LO result which
was weakened at NLO \cite{Herrlich:1993yv}. The NLO correction to
$\eta_{ct}$ implied an upward shift of 31\% of this quantity
\cite{Herrlich:1995hh,Herrlich:1996vf},
which is of the expected size of a correction proportional to $\alpha_s(m_c)$. 
The NLO calculation further permits control over the definition of the
quark masses, the numbers quoted below for $\eta_{tt}$, $\eta_{ct}$,
and $\eta_{cc}$ correspond to the use of the  $\ov{\rm MS}$ scheme, 
i.e.\ $x_q\equiv \lt(\bar m_q(\bar m_q)/M_W\rt)^2$ is to be used in
\eqsand{cons}{cons2}. 

To calculate $\dm_K = 2 |M_{12}^K| \cos \phi_K \simeq 2 \,\real M_{12}^K$
we decompose $\dm_K$ into a short-distance piece from the local
contribution to $M_{12}^K$ and a long-distance piece from the bilocal
contribution with two $|\Delta S|=1$ hamiltonians:
\begin{align}
\dm_K &=\; \dm_K^{\rm SD} + \dm_K^{\rm LD} \nn 
\dm_K^{\rm SD} \equiv \frac{1}{m_K}\, \real \bra{K} H^{|\Delta S|=2}
        \ket{\bar{K}},
  & \qquad\qquad
 \dm_K^{\rm LD} \equiv    - \real\, \frac{i}{2 m_K} \int \! d^4 x \,
    \bra{K} H^{|\Delta S|=1} (x)\, H^{|\Delta S|=1} (0) \ket{\bar{K}} ,
\label{defsdld}
\end{align}  
which again is specific to the standard CKM phase convention, for which
the dominant CKM combination $(V_{us}V_{ud}^*)^2$ is real. While the
tiny contribution proportional to $(V_{ts}V_{td}^*)^2$ is relevant 
for $\tilde\xi_K$ in \eq{numbek}, it is negligible for $\dm_K$ and omitted in
\eq{defsdld}. Since the CKM factor is real, the dispersive part of the
bilocal matrix element entering $ \dm_K^{\rm LD}$ is identical to the
real part. The dispersive part of the mixing amplitude has been
explained after \eq{absdisp}.

The  values for the QCD coefficients in \eq{deshs2} are
\begin{align}
  \eta_{tt} &=\; 0.5765 \pm 0.0065\quad \mbox{(NLO) \cite{Buras:1990fn}},
              \qquad            \eta_{ct} \;=\;
               0.496 \pm 0.047\quad \mbox{(NNLO) \cite{Brod:2010mj}},  %mc= 1.286(13) GeV 
\qquad            \eta_{cc} \;=\;   1.87 \pm 0.76 \quad
              \mbox{(NNLO) \cite{Brod:2011ty}}. % mc= 1.279(13) GeV
              \label{etanum}
\end{align}  
The value for $\eta_{tt}$ is an update taken from
Ref.~\cite{Brod:2011ty} using $\bar m_t(\bar m_t)= (163.7 \pm 1.1)
\gev$. The $\ov{\rm MS}$ top mass is smaller than the pole mass quoted
in the context of collider physics by roughly $8\gev$. 
$\eta_{cc}$ depends steeply on $m_c$, the quoted value
is for $\bar m_c(\bar m_c)= (1.279 \pm 0.0013) \gev$, which should
be kept in mind when comparing the NNLO value in \eq{etanum} with the
NLO values in Refs.~\cite{Herrlich:1993yv,Herrlich:1995hh,Herrlich:1996vf}. 
With today's precise value of $m_c$, the uncertainties of $m_c$ and other
input parameters has no relevance. The uncertainties of $\eta_{cc}$ and
$\eta_{ct}$ are dominated by the scale uncertainty, estimated by varying
$\mu_{bc}$ around $\bar m_c$. This uncertainty will diminish once
perturbative calculations beyond NNLO will be performed. The NNLO result
for $\eta_{ct}$ in \eq{etanum} is larger than the NLO result of
Ref.~\cite{Herrlich:1996vf} by just 8.5\%, which testifies to
a good behavior of the perturbative series. The NNLO result for
$\eta_{cc}$, however, shows the same pathological situation as the NLO
result of Ref.~\cite{Herrlich:1993yv,Herrlich:1996vf}. The correction is  
large, increasing $\eta_{cc}$ by 36\%\   over the NLO
value \cite{Brod:2011ty}, which exceeds the LO value by 65\%, so that
$\eta_{cc}$ more than doubled due to two-loop and three-loop QCD
corrections.

This development has immediate consequences for $\dm_K$, as
with \eqsand{deshs2}{melk} we find from \eq{defsdld}: 
\begin{align}
  \dm_K^{\rm SD}  &=\; \frac{G_F^2}{6\pi^2} f_K^2 M_K \widehat{B}_K  \lt( \real
          (V_{cs}V_{cd}^*) \rt)^2 \, \eta_{cc} m_c^2 \label{dmksd}
\end{align}
where $M_W^2 x_c = m_c^2$ has been used. The contributions with other
CKM factors amount to 1\% and are negligible.

Using the values for $M_K$ and $f_K$ quoted below \eq{melk},
%$M_K=497.6\mev$ \cite{ParticleDataGroup:2024cfk}, $f_K= 156\mev$
$\bar m_c(\bar m_c)=  1.279 \gev$, and $ \real
(V_{cs}V_{cd}^*)=-0.218$ we obtain
\begin{align}
  \frac{\dm_K^{\rm SD}}{\dm_K^{\rm exp}}
    &=\;  (1.16 \pm 0.47) \, \widehat{B}_K \label{sdexp}
\end{align}  
where I further used the experimental value of \eq{dmkexp}.
% dm_K^exp = 3.476 \pm 0.006 *10^{-15} GeV
% The lattice calculation of Ref.~\cite{Carrasco:2015pra}
% has computed $\widehat B_K = 0.717 \pm 0.024$. 
% B_K = 0.717(18)(16) for n_f= 2+1+1, ETM Coll.  
A full NNLO prediction further requires a two-loop
lattice-continuum matching of the hadronic matrix element,
i.e.\ of $\widehat B_K$, which involves the matching of the $\ov{\rm
  MS}$ result to a different renormalization scheme suited for
a non-perturbative calculation. Ref.~\cite{Gorbahn:2024qpe} presents
this calculation for  $\widehat B_K$ and the application of  the result to an
average of different lattice computations to find
$\widehat B_K= 0.7627\pm 0.0060$. Thus \eq{sdexp} boils down to 
\begin{align}
  \frac{\dm_K^{\rm SD}}{\dm_K^{\rm exp}}
    &=\;  (0.89 \pm 0.36), \,  \label{sdexp2}
\end{align} 
implying short-distance dominance of $\dm_K$. The long-distance
contribution lacks the prefactor $m_c^2$  of \eq{dmksd} and should
therefore be smaller by a factor of $\lqcd^2/m_c^2 \sim 0.1$. 
There are exploratory lattice calculations of  $\dm_K^{\rm LD}$ plus certain terms
contained in $\dm_K^{\rm SD}$, employing the feature that this
calculation is easier in QCD with four active flavours because the GIM
cancellation beween charm and up contributions improves the UV behaviour
of the calculated quantity \cite{Christ:2012se,Bai:2014cva}. 

$\epsilon_K$ in \eq{cons2} defines a hyperbola in the 
$\bar\rho$-$\bar\eta$ plane characterized by $\eta_{tt} S(x_t)$ and
\begin{align}
  \eta_{ct} \, S(x_c,x_t) \, -  \, \eta_{cc} \,  x_c
  % \;& \equiv\; -\eta_{ut} {\cal S}(x_c,x_t)   .
      \label{etaut}
\end{align}
shown in \fig{fig:utall}. In Ref.~\cite{Brod:2019rzc} it has been
observed that the pathological term in $ \eta_{cc} \, x_c$ drops out
from this combination, so that the prediction of $\epsilon_K$ does not
inherit the uncertainty from the poor convergence of the perturbative
series of $\eta_{cc}$. The first term $\eta_{ct} \, S(x_c,x_t) $ is
parametrically larger by a factor of the large logarithm $\ln x_c$, so
that the term cancelling with $\eta_{cc} \, x_c$ is numerically
sub-leading and therefore has a smaller impact on $\eta_{ct}
S(x_c,x_t)$. The cancellation can be understood by switching to a
different form of \eq{deshs2} via the replacement
$V_{cs}V_{cd}^* \to -V_{us}V_{ud}^*-V_{ts}V_{td}^*$
\cite{Brod:2019rzc}. The resulting prediction for $\epsilon_K$ reads
\cite{Brod:2019rzc}
\begin{align}
   |\epsilon_K| &= \; (2.16 \pm 0.18) \times 10^{-3} \, \frac{\widehat B_K}{0.7625}
 \label{epskbgs}
\end{align}  
for $\bar\rho=0.16$, $\bar\eta= 0.38$, $|V_{cb}|=(42.2 \pm 0.8) \cdot
10^{-3}$, and the quark masses quoted above. The largest uncertainty in
\eq{epskbgs} stems from the input parameters, the perturbative
uncertainty is down to 3\%.  The hyperbola of \eq{cons2} reads with
$\eta_{tt}$ in \eq{etanum} and $ \eta_{ct} \, S(x_c,x_t) \, -  \,
\eta_{cc} \,  x_c = (7.98\pm 0.18) \cdot 10^{-4} $ \cite{Brod:2019rzc}:
\begin{align}
 1.20 \cdot 10^{-6} &=\; \widehat B_K \,  |V_{cb}|^2\, \, \ov \eta \, 
\left[ |V_{cb}|^2\, (1-\ov \rho ) (1.36 \pm 0.02) 
       \, +  \,  (7.98\pm 0.18) \cdot 10^{-4}
                      \right] . \label{cons3} 
\end{align}  
The first term in the square bracket stemming from $\eta_{tt} S(x_t)$
contributes about 3/4 to $|\epsilon_K|$, so that
$ |V_{cb}|$ essentially contributes to $|\epsilon_K|$ with the fourth
power. The  value of  $ |V_{cb}|$ is controversial, the determinations
from exclusive and inclusive semileptonic $B$ decays do not agree.
For the lower value, found from   exclusive decays, the hyperbola from
$\epsilon_K$ is not consistent with other constraints on the apex
$(\bar\rho,\bar\eta)$ of the UT.

Finally, $\dg_K$ in \eq{dgkexp} has defied any calculation from first
principles.   For this we need $\Gamma_{12}^K$, which is completely
dominated by the isospin-0 amplitude $A_0$, see \eq{g12k}.
The experimental fact $|A_0|\simeq 22 |A_2|$ is called ``$\Delta I=1/2$
rule'' and no analytical calculation could reproduce the value of
$|A_0/A_2|$.

\begin{figure}[tp]
\centering 
\includegraphics*[width=0.7\textwidth]{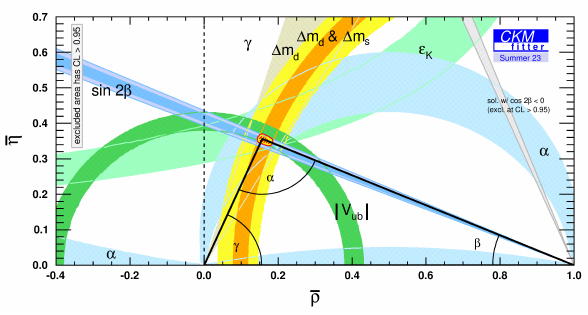}\\ %[-30mm]
\includegraphics*[width=0.5\textwidth]{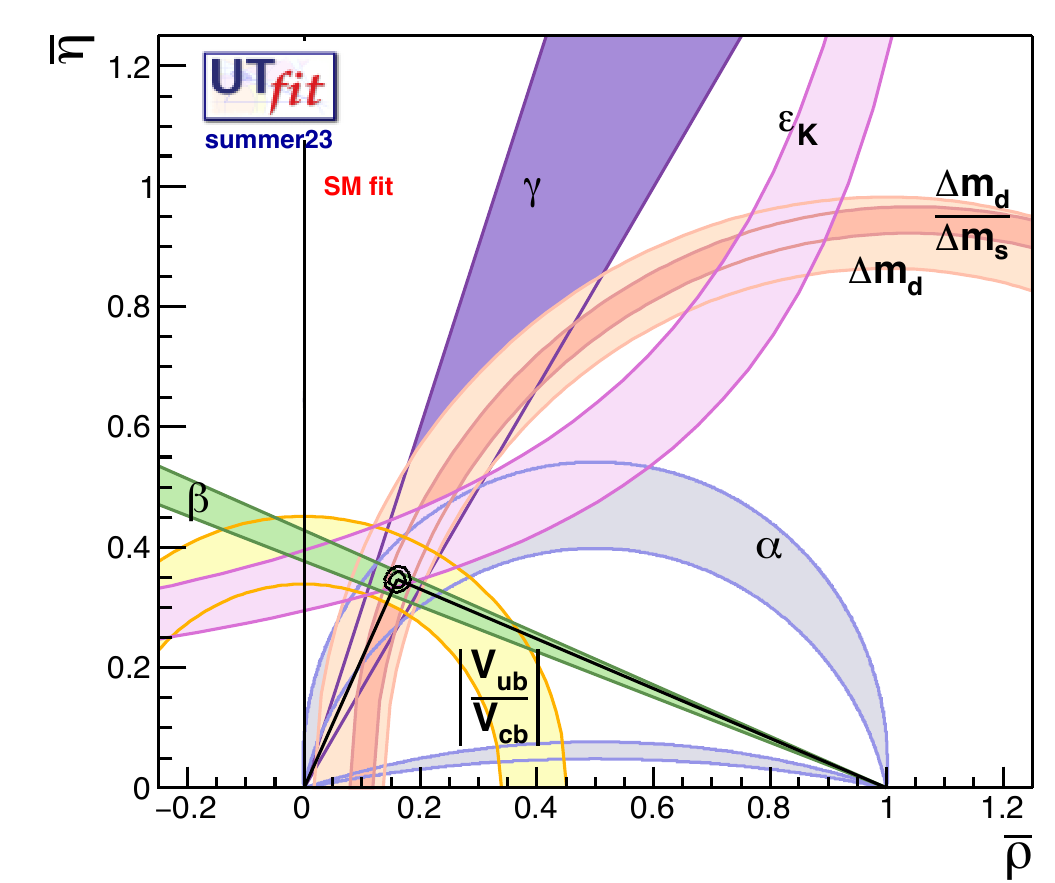}
\caption{Contraints on the unitarity triangle from global fits
  to all quantities constraining  $(\bar\rho,\bar\eta)$.
  The upper and lower plots are from CKMfitter \cite{Charles:2004jd}
  and UTfit \cite{UTfit:2022hsi}, respectively. The 68\% C.L.\ contours
  for $(\bar\rho,\bar\eta)$ are indicated.  CKMfitter uses a Frequentist
  approach in which systematic errors and theoretical uncertainties are
  scanned over, which results in a larger 68\% C.L.\ region than the
  Bayesian approach of UTfit. The 3$\sigma$ regions found by both groups
  are similar.\label{fig:utall_fitter}}
\end{figure}

\fig{fig:utall} shows the constraints on the apex of the unitarity
triangle from the mixing-related observables discussed in this review.
The actual situation from global fits to all measured quantities sensitive
to $\bar\rho$, $\bar\eta$ is illustrated in \fig{fig:utall_fitter},
which shows the results from the two major groups performing such
analyses, CKMfitter and UTfit.  

To study BSM physics one may parameterize the BSM contributions in a
model-independent way by foreseeing parameters modifying magnitude and
phase of $M_{12}^{d,s,K}$ and constrain these in conjunction with
$\bar\rho$ and $\bar\eta$ in a global fit
\cite{Lenz:2010gu,Lenz:2012az,UTfit:2022hsi}.  Ref.~\cite{UTfit:2022hsi}
find that ${\cal O}(20\%)$ BSM effects are allowed in $|M_{12}^{d,s}|$
and $\imag M_{12}^K$, while there is less space for new physics in the
phases of $M_{12}^{d,s}$. With the result of the model-independent fit
one can constrain any BSM model of interest, unless the model correlates
different mixing observables.

BSM models addressing the gauge sector or Dark Matter are usually
agnostic about the flavor structure, so that predictions rely on
additional assumptions on the flavor sector. A widely studied approach
is \emph{minimal flavor violation (MFV)}, which organizes the flavour
pattern of the SM Yukawa interactions in \eq{yuk} in terms of small
symmetry-breaking parameters called spurions
\cite{Chivukula:1987py,DAmbrosio:2002vsn}.  The MFV approach assumes
that the same spurions governing \eq{yuk} also determine the flavor
structure of the studied BSM model, so that the new interaction involve
the same CKM elements as the SM contributions. This reduces the BSM
sensitivity of FCNC processes drastically and was originally motivated
to permit lighter BSM particles in the reach of contemporary
experiments.  However, for instance in models with more than one Higgs
doublet, one even finds imprints on \mmm\ in the MFV case
\cite{Buras:1989ui,Buras:2002wq,Isidori:2006pk,Gorbahn:2009pp,Buras:2013raa,
  Buras:2013rqa,Eberhardt:2013uba,Enomoto:2015wbn,Atkinson:2021eox}.
Some of the cited papers have considered the special case of the
Minimals Supersymmetric Standard Model with decoupled heavy
superpartners, resulting in a 2HDM in which a neutral heavy Higgs boson
affects \bbms\ in a critical way. If one relaxes the MFV hypothesis by
adding a third spurion to the two spurions of the SM, one also finds
large effects in \mmm\ observables \cite{Lang:2022mxu}.

%%%%%%%%%%%%%%%%%%%%%%%%%%%%%%%%%%%%%%%%%
%% Mandatory: A concluding paragraph summing up your main points in the chapter
%% Optional: Also include big questions in the field that are still to be answered. What topics/methods/questions are researchers like to focus on next?
\section{Conclusions\label{sec:con}}
This review article summarizes the theoretical formalism and the
phenomenological methodology of \kk, \dd, \bbd, and \bbms.
All these \mmm\ systems involve $|\Delta F|=2$ transitions, in which the
flavor quantum number $F=S,C$ or $B$ characterizing the meson changes 
by two units. The relevant $2\times 2$ matrix $M-i \Gamma/2$
is composed of the hermitian mass and decay matrix and \mmm\ occurs
because  $M_{12}-i \Gamma_{12}/2 \neq 0$. Upon diagonalization of
$M-i \Gamma/2$ one finds the two mass eigenstates, which are
superpositions of the particle state $\ket{M}$ and the antiparticle
state $\ket{\bar M}$ and follow exponential decay laws.  
\mmmc\  is characterized by four quantities, the mass and width
differences between the two mass eigenstates as well as the $CP$ asymmetries in
flavor-specific decays and a chosen exclusive decay. I have presented
the formulae connecting these quantities to $M_{12}$ and $\Gamma_{12}$
and the Standard-Model (SM) predictions for them. While all four mixing
complexes follow the same pattern, the numerical values of the
corresponding quantities are very different with, for example, 
$\BsorBsbar$ mesons oscillating very rapidly while the 
very slow \dd\ oscillations had impeded their discovery for a long
time. 

Another objective has been the recapitulation of the historical
development of the field since the early 1950s, highlighting how the
study of \mmm\ helped to shape the SM. The confirmation of
the Kobayashi-Maskawa mechanism of $CP$ violation needed a firm
prediction of the size of mixing-induced $CP$ violation in \bbmd\ from
the $CP$-violating quantity $\e_K$ in \kkm, the \bbd\ oscillation
frequency, and the semileptonic $B$ branching ratios for decays with and
without charm in the final state. The needed predictions involved a rigorous
theoretical basis and sophisticated calculational tools, which were
developed in the late 1980s and early 1990s and proceeded 
along three avenues: (i) the establishment of a framework for
perturbative calculations of short-distance QCD corrections, (ii) the
advancement of lattice-QCD computations as a first-principle method
to tackle long-distance QCD, and (iii) the identification of sizable 
theoretically clean mixing-induced $CP$ asymmetries such as
$ A_{CP}^{\rm mix}(B_d\to J/\psi K_{\rm short})$. The theoretical
progress was not only instrumental  for the precise determination of
CKM elements, which are fundamental parameters of the SM Yukawa sector,
but also lead to the identification of ``gold-mines'' for the search for
BSM physics, namely theoretically clean observables with sensitivity to
new physics.   With the exception of the width difference in the \kk\
system, all above-mentioned observables could be calculated in \kk, \bbd,
and \bbms\ with good  accuracy. \ddm, however, has defied
any calculation from first principles, with theory failing even at order-of-magnitude
predictions for mass and width differences.   

\mmmc\ processes are highly sensitive to BSM physics, with the potential
to reveal virtual effects of  new particles with masses far 
above  100\tev. To disentangle BSM physics from SM contributions one
must determine the CKM elements together with the parameters of the
studied BSM model; especially   $(\bar\rho, \bar\eta)$  determined from
the global fit of \fig{fig:utall_fitter} will be ``contaminated'' by BSM
physics in the \mmm\ observables. At present, the identification of
BSM physics in this way is impeded by the controversies on the values of
$|V_{cb}|$ and $|V_{ub}|$ found from inclusive and exclusive decays.
With better experimental possibilities for $B$ physics  at future runs of the
LHC, with higher luminosity at Belle II, and further at the FCC-ee,
mixing-induced $CP$ asymmetries will become an important tool to
discover or constrain BSM physics in  rare decays. For example, if the flavor anomalies
observed today in angular observables and branching fractions of
$B\to K \ell^+\ell^-$ and $B_s \to \phi \ell^+\ell^-$ decays will
manifest themselves also in the corresponding mixing-induced $CP$ asymmetries, this will
constitute an unambigous discovery of BSM physics. 
On the theoretical side, continued effort is needed to keep up with the
ever decreasing experimental error bars, especially better lattice-QCD
predictions are highly desirable. Lattice QCD is further likely to
emerge as the best avenue to address \ddm.  Finally, in the field of BSM
physics flavor and collider observables will always go hand-in-hand
to identify allowed parameter spaces and \mmm\  observables play an important
role to this end. 

\begin{ack}[Acknowledgments]%
  This research was supported by the Deutsche Forschungsgemeinschaft
  (DFG, German Research Foundation) under grant 396021762 - TRR 257 through
  project C1b of the Collaborative Research Center \emph{Particle Physics after the
    Higgs Discovery (P3H)}.
\end{ack}

%%%%%%%%%%%%%%%%%%%%%%%%%%%%%%%%%%%%%%%%%%%%
%% Optional: A list of references to other relevant works/articles/websites which are not cited in the text but that would further enhance a readers understanding of this topic
%\seealso{article title article title}

%%%%%%%%%%%%%%%%%%%%%%%%%%%%%%%%%%%%%%%%%
%% Mandatory: Bibliography using bibtex 
\bibliographystyle{Numbered-Style} %% for Numbered Reference Style
\bibliography{meson_mix_nierste}

\end{document}